\documentclass[useAMS,usenatbib]{mn2e}
\usepackage{graphicx}
\usepackage{epsfig}
\usepackage{multirow}

\title[A Spectral Study of Unobscured Type 1 AGN - I]{A Combined Optical and X-ray Study of Unobscured Type 1 AGN. I. Optical Spectra and SED Modeling}

\author[C. Jin, M. Ward, C. Done and J. M. Gelbord]
{Chichuan Jin\thanks{E-mail: chichuan.jin@durham.ac.uk}$^{1}$,
Martin Ward$^{1}$, Chris Done$^{1}$, Jonathan Gelbord$^{1,2}$\\
$^{1}$Department of Physics, University of Durham, South Road, Durham, DH1 3LE, UK\\
$^{2}$Department of Astronomy $\&$ Astrophysics, Penn State University, PA 16801, USA}

\begin{document}

\date{Accepted by MNRAS}

\maketitle

\label{firstpage}

\begin{abstract}

We present modeling and interpretation of the continuum and emission lines for a sample of 
51 unobscured Type 1 active galactic nuclei (AGN). All of these AGNs 
have high quality spectra from both XMM-Newton and Sloan
Digital Sky Survey (SDSS). We extend the wavelength coverage
where possible by adding simultaneous UV data from the OM onboard XMM-Newton. Our sample
is selected based on low reddening in the optical and low gas columns implied by their X-ray
spectra, except for one case, the BAL-quasar PG 1004+130.
They also lack clear signatures for the presence of a warm absorber. Therefore the
observed characteristics of this sample are likely to be directly related to the intrinsic
properties of the central engine.

To determine the intrinsic optical continuum we subtract the Balmer continuum and all major 
emission lines (including FeII). We also consider possible effects of contamination from 
the host galaxy. The resulting continuum is then used to derive the properties of the underlying
accretion disc. We constrain the black hole masses from spectral fits of the Balmer emission lines and
determine the best fit value from the modeling of broadband spectral energy distributions (SED). 
In addition to the disc component, many of these SEDs also exhibit a strong soft X-ray excess, 
plus a power law extending to higher X-ray energies. 
We fit these SEDs by applying a new broadband SED model
which comprises the accretion disc emission, low temperature optically thick Comptonisation
and a hard X-ray tail by introducing the concept of a corona radius (\citealt{Done11}).
We find that in order to fit the data, the model often requires  
an additional long wavelength optical continuum component, whose origin
is discussed in this paper. We also find that the Photo-recombination edge of Balmer continuum 
shifts and broadens beyond the standard limit of 3646{\AA}, implying an
electron number density which is far higher than that in the broad line region clouds.

Our results indicate that the Narrow Line Seyfert 1s in this sample tend to have 
lower black hole masses, higher Eddington ratios, softer 2-10~keV band spectra,
lower 2-10 keV luminosities and higher $\alpha_{ox}$, compared with
typical broad line Seyfert 1s (BLS1), although their bolometric
luminosities are similar. We illustrate these differences in
properties by forming an average SED for three subsamples, based on
the FWHM velocity width of the H$\beta$ emission line.

\end{abstract}

\begin{keywords}
accretion, broadband SED modeling, active-galaxies: nuclei
\end{keywords}

\section{Introduction}

The spectral energy distribution (SED) of AGN
has been modeled for several decades. Initial studies
focused on the infrared, optical and ultraviolet continuum
(e.g. \citealt{Wills85}; \citealt{Canalizo01}; \citealt{Lacy07}). With
the inclusion of X-ray data, it was possible to define the continuum 
on both sides of the ultraviolet/X-ray gap (imposed by galactic
photoelectric absorption), and so constrain the properties of the
accretion disc (e.g. \citealt{Ward87}; \citealt{Elvis94}). 
Refinements to modeling the optical/UV continuum include
subtraction of the complex blended features arising from permitted
iron emission, the so-called small blue-bump from the Balmer
continuum, and contamination across the entire spectrum from a
stellar component (\citealt{Maoz93}; \citealt{Boisson00})

The observed spectral differences between various types of AGN are not
only due to selective absorption and orientation effects, as implied
by the simplest version of AGN unification model (\citealt{Antonucci93}),
but also result from a wide range in basic physical
parameters, such as black hole mass and accretion rate
(e.g. \citealt{Boroson92}; \citealt{Boller96}; \citealt{Done05};
\citealt{Zhou06}). To better understand the accretion
processes occurring close to the super massive black hole (SMBH),
we construct broadband SEDs. Galactic
dust reddening, together with the intrinsic reddening of the AGN itself,
attenuates the optical/UV band emission. Furthermore,
Photoelectric absorption from gas modifies the lower energy X-ray
continuum. But these factors can be quantified and corrected.
Thereby we can recover the intrinsic SED, except
for the unobservable far-UV region. If we have reliable data on both
sides of the energy gap between the UV and soft X-ray, we can apply 
a multi-component model which spans across it.

\begin{table*}
 \centering
  \begin{minipage}{170mm}
   \caption{The Seyfert 1 Galaxy Sample Set}
   \label{object list}
     \begin{tabular}{@{}clcccccc@{}}
\hline\hline
  & & & 2XMMi Catalog & XMM-Newton & SDSS DR7 & SDSS & EPIC\\
  ID & Common Name$^{a}$ & Redshift & IAU Name (2XMM$^{b}$)& Obs Date & MJD-Plate-Fibre & Obs Date & Counts$^{c}$ \\
\hline
 1 & UM 269 & 0.308 & J004319.7+005115 & 2002-01-04 & 51794-0393-407 & 2000-09-07 & 19126\\
 2 & MRK 1018 & 0.043 & J020615.9-001730 & 2005-01-15 & 51812-0404-141 & 2000-09-25 & 2056\\
 3 & NVSS J030639 & 0.107 & J030639.5+000343 & 2003-02-11 & 52205-0709-637 & 2001-10-23 & 35651\\
 4 & 2XMMi/DR7 & 0.145 & J074601.2+280732 & 2001-04-26 & 52618-1059-399 & 2002-12-10 & 9679\\
 5 & 2XMMi/DR7 & 0.358 & J080608.0+244421 & 2001-10-26 & 52705-1265-410 & 2003-03-07 & 2912\\
 6 & HS 0810+5157  & 0.377 & J081422.1+514839 & 2003-04-27 & 53297-1781-220 & 2004-10-19 & 4189\\
 7 & RBS 0769 & 0.160 & J092246.9+512037 & 2005-10-08 & 52247-0766-614 & 2001-12-04 & 32731\\
 8 & RBS 0770 & 0.033 & J092342.9+225433$^{*}$ & 2006-04-18 & 53727-2290-578 & 2005-12-23 & 104028\\
 9 & MRK 0110 & 0.035 & J092512.8+521711 & 2004-11-15 & 52252-0767-418 & 2001-12-09 & 515453\\
 10 & PG 0947+396 & 0.206 & J095048.3+392650 & 2001-11-03 & 52765-1277-332 & 2003-05-06 & 58555\\
 11 & 2XMMi/DR7 & 0.373 & J100025.2+015852 & 2003-12-10 & 52235-0501-277 & 2001-11-22 & 7187\\
 12 & 2XMMi/DR7 & 0.206 & J100523.9+410746 & 2004-04-20 & 52672-1217-010 & 2003-02-02 & 5437\\
 13 & PG 1004+130 & 0.241 & J100726.0+124856 & 2003-05-04 & 53055-1744-630 & 2004-02-20 & 3781\\
 14 & RBS 0875 & 0.178 & J103059.0+310255 & 2000-12-06 & 53440-1959-066 & 2005-03-11 & 69434\\
 15 & KUG 1031+398 & 0.043 & J103438.6+393828 & 2002-05-01 & 53002-1430-485 & 2003-12-29 & 63891\\
 16 & PG 1048+342 & 0.160 & J105143.8+335927 & 2002-05-13 & 53431-2025-637 & 2005-03-02 & 47858\\
 17 & 1RXS J111007 & 0.262 & J111006.8+612522$^{*}$ & 2006-11-25 & 52286-0774-600 & 2002-01-12 & 6147\\
 18 & PG 1115+407 & 0.155 & J111830.2+402554 & 2002-05-17 & 53084-1440-204 & 2004-03-20 & 64601\\
 19 & 2XMMi/DR7 & 0.101 & J112328.0+052823 & 2001-12-15 & 52376-0836-453 & 2002-04-12 & 10098\\
 20 & RX J1140.1+0307 & 0.081 & J114008.7+030710 & 2005-12-03 & 51994-0514-331 & 2001-03-26 & 35616\\
 21 & PG 1202+281 & 0.165 & J120442.1+275412 & 2002-05-30 & 53819-2226-585 & 2006-03-25 & 66550\\
 22 & 1AXG J121359+1404  & 0.154 & J121356.1+140431 & 2001-06-15 & 53466-1765-058 & 2005-04-06 & 12975\\
 23 & 2E 1216+0700 & 0.080 & J121930.9+064334 & 2002-12-18 & 53140-1625-134 & 2004-04-26 & 8028\\
 24 & 1RXS J122019 & 0.286 & J122018.4+064120 & 2002-07-05 & 53472-1626-292 & 2005-04-12 & 8338\\
 25 & LBQS 1228+1116 & 0.236 & J123054.1+110011 & 2005-12-17 & 52731-1232-417 & 2003-04-02 & 165823\\
 26 & 2XMMi/DR7 & 0.304 & J123126.4+105111 & 2005-12-17 & 52731-1232-452 & 2003-04-02 & 8816\\
 27 & MRK 0771 & 0.064 & J123203.6+200929 & 2005-07-09 & 54481-2613-342 & 2008-01-15 & 40705\\
 28 & RX J1233.9+0747 & 0.371 & J123356.1+074755 & 2004-06-05 & 53474-1628-394 & 2005-04-14 & 6041\\
 29 & RX J1236.0+2641 & 0.209 & J123604.0+264135$^{*}$ & 2006-06-24 & 53729-2236-255 & 2005-12-25 & 17744\\
 30 & PG 1244+026 & 0.048 & J124635.3+022209 & 2001-06-17 & 52024-0522-173 & 2001-04-25 & 8509\\
 31 & 2XMMi/DR7 & 0.316 & J125553.0+272405 & 2000-06-21 & 53823-2240-195 & 2006-03-26 & 7591\\
 32 & RBS 1201  & 0.091 & J130022.1+282402 & 2004-06-06 & 53499-2011-114 & 2005-05-09 & 209458\\
 33 & 2XMMi/DR7 & 0.334 & J132101.4+340658 & 2001-01-09 & 53851-2023-044 & 2006-04-26 & 4425\\
 34 & 1RXS J132447 & 0.306 & J132447.6+032431 & 2004-01-25 & 52342-0527-329 & 2002-03-09 & 6305\\
 35 & UM 602 & 0.237 & J134113.9-005314 & 2005-06-28 & 51671-0299-133 & 2000-05-07 & 18007\\
 36 & 1E 1346+26.7 & 0.059 & J134834.9+263109 & 2000-06-26 & 53848-2114-247 & 2006-04-23 & 71985\\
 37 & PG 1352+183 & 0.151 & J135435.6+180518 & 2002-07-20 & 54508-2756-228 & 2008-02-12 & 36171\\
 38 & MRK 0464 & 0.050 & J135553.4+383428 & 2002-12-10 & 53460-2014-616 & 2005-03-31 & 13974\\
 39 & 1RXS J135724 & 0.106 & J135724.5+652506 & 2005-04-04 & 51989-0497-014 & 2001-03-21 & 12081\\
 40 & PG 1415+451 & 0.114 & J141700.7+445606 & 2002-12-08 & 52728-1287-296 & 2003-03-30 & 55786\\
 41 & PG 1427+480 & 0.221 & J142943.0+474726 & 2002-05-31 & 53462-1673-108 & 2005-04-01 & 70995\\
 42 & NGC 5683 & 0.037 & J143452.4+483943 & 2002-12-09 & 52733-1047-300 & 2003-04-04 & 18885\\
 43 & RBS 1423 & 0.208 & J144414.6+063306 & 2005-02-11 & 53494-1829-464 & 2005-05-04 & 37568\\
 44 & PG 1448+273 & 0.065 & J145108.7+270926 & 2003-02-08 & 54208-2142-637 & 2007-04-18 & 134532\\
 45 & PG 1512+370 & 0.371 & J151443.0+365050 & 2002-08-25 & 53083-1353-580 & 2004-03-14 & 40432\\
 46 & Q 1529+050 & 0.218 & J153228.8+045358 & 2001-08-21 & 54563-1835-054 & 2008-04-07 & 10952\\
 47 & 1E 1556+27.4 & 0.090 & J155829.4+271715 & 2002-09-10 & 52817-1391-093 & 2003-06-27 & 6995\\
 48 & MRK 0493 & 0.031 & J155909.6+350147 & 2003-01-16 & 53141-1417-078 & 2004-05-14 & 124115\\
 49 & II Zw 177 & 0.081 & J221918.5+120753 & 2001-06-07 & 52221-0736-049 & 2001-11-08 & 36056\\
 50 & PG 2233+134 & 0.326 & J223607.6+134355 & 2003-05-28 & 52520-0739-388 & 2002-09-03 & 7853\\
 51 & MRK 0926 & 0.047 & J230443.3-084111 & 2000-12-01 & 52258-0725-510 & 2001-12-15 & 59513\\
\hline\hline
   \end{tabular}
  \\
  \\
 $^{a}$ for some targets without well-known names, we simply use `2XMMi/DR7'; \\
 $^{b}$ the full name should be `2XMM J...', but for those targets with * symbol, their full names should be `2XMMi J...'; \\
 $^{c}$ the total counts in all three EPIC monitors, namely pn, MOS1 and MOS2, and there are at least 2000 counts in at least one \\
 of these three monitors; \\
 \end{minipage}
\end{table*}

\subsection{Previous Work} 

Many multi-wavelength studies have been carried out
previously. \citet{Puchnarewicz92} studied the optical properties of
53 AGNs in \citet{Cordova92}'s sample with ultra soft X-ray
excesses, and found that they tend to have narrower permitted lines 
than optically selected samples. Supporting this finding, \citet{Boller96}
studied ROSAT selected AGN with extremely soft X-ray spectra, and
found that they tend to be Narrow-Line Seyfert 1s (NLS1s). Correspondingly
they found that optically selected NLS1s often have large soft X-ray
excesses. \citet{Walter93} combined soft X-ray and optical data for 58
Seyfert 1s, and showed that their broadband SED have a bump from UV to
soft X-rays, which is now refered to as the big blue bump (BBB). \citet{Grupe98}
and \citet{Grupe99} used a sample of 76 bright soft X-ray selected
Seyferts with infrared data, optical spectra and soft X-ray
spectra. Their results reinforced the connection between the optical and
soft X-ray spectra, and confirmed the existence of strong BBB emission
in these objects. \citet{Elvis94} studied 47 quasars in a UV-soft
X-ray sample, and derived the mean SEDs for 
radio-loud and radio-quiet sources. Recently, more detailed spectral
models have been applied to broadband SEDs including
simultaneous optical/UV and X-ray observations which avoid potential
problems caused by variability.
\citet{Vasudevan07} (hereafter VF07) combined a disc and broken
powerlaw model to fit optical, far UV and X-ray data for 54 AGN.
They found a well-defined relationship between the hard X-ray
bolometric correction and the Eddington ratio. \citet{Brocksopp06}
analysed the data from XMM-Newton's simultaneous EPIC (X-ray) and OM (optical/UV)
observations for 22 Palomar Green (PG) quasars. Another sample
consisting of 21 NLS1s and 13 broad line AGNs was also defined using simultaneous
data from XMM-Newton's EPIC and OM monitor (\citealt{Crummy06}).
The SEDs of this sample were then fitted using various broadband SED 
models such as disc plus powerlaw model,
disc reflection model and disc wind absorption model (\citealt{Middleton07}).
\citet{Vasudevan09} derived SEDs using
XMM-Newton's simultaneous X-ray and optical/UV observations for 29
AGNs selected from \citet{Peterson04}'s reverberation mapped
sample. The well constrained black hole masses available for this sample enabled
them to fit a better constrained accretion disc model, combined with a powerlaw, to
the source's broadband SEDs. Hence they derived more reliable Eddington
ratios.

\subsection{Our AGN Sample}

In this paper we define an X-ray/optically selected sample of 51 AGN, all
of which have low reddening (so excluding Seyfert 2s and 1.9/1.8s),
to construct SEDs ranging from about 0.9 microns to 10 keV.  We also
apply corrections for the permitted iron features, the Balmer continuum
and stellar contribution, in order
to model the non-stellar continuum free from emission line
effects. Included in this sample are a number of NLS1s, 
a subclass of AGN whose permitted
line widths are comparable to those of forbidden lines. Their
[OIII]$\lambda$5007/H$\beta$ ratio is also lower than the typical value of
broad line Seyfert 1s (BLS1s) (\citealt{Shuder81}; \citealt{Osterbrock85}).
For consistency with previous work, we classify
AGNs in our sample as NLS1s if they have ratios of
[OIII]$\lambda$5007/H$\beta <$ 3 and FWHM$_{H\beta} <$ 2000 km/s
(\citealt{Goodrich89}). We identify 10$\sim$12 NLS1s in our 
sample\footnote{Although 2XMM J112328.0+052823 and 1E 1346+26.7 have
H$\beta$ FWHMs of 2000 $km~s^{-1}$, 2050 $km~s^{-1}$ respectively,
they both have H$\alpha$ FWHM of 1700 $km~s^{-1}$, and also share
other NLS1's spectral characteristics. Thus they could both
potentially be classified as NLS1s, making a total of 12.}. 

All objects in our sample  have high quality
optical spectra taken from the Sloan Digital Sky Survey (SDSS) DR7,
X-ray spectra from the XMM-Newton EPIC cameras, and in some cases
simultaneous optical/UV photometric data points from the XMM-Newton OM
monitor. Combining these data reduces the impact of intrinsic
variability and provides a good estimate of the spectral shape in the
optical, near UV and X-ray regions.  In addition, by analyzing the
SDSS spectra, we can derive the parameters of the principal
optical emission lines and underlying continuum.
An important result from reverberation mapping study is 
the correlation between black hole mass, monochromatic
luminosity at 5100 {\AA} and H$\beta$ FWHM (e.g. \citealt{Kaspi00};
\citealt{Woo02}; \citealt{Peterson04}). We measure these
quantities from the SDSS spectra, and then estimate black hole masses using
this correlation.

Compared with previous work, a significant improvement of our 
study is that we employ a new broadband SED model which
combines disc emission, Comptonisation and a high energy powerlaw
component in the context of an energetically self-consistent model for
the accretion disc emission (\citealt{Done11}, 
also see Section~\ref{broadband-sed-model}).
By fitting this model to our data, we can
reproduce the whole broadband SED from the optical to X-ray. From
this detailed SED fitting, we derive a number of interesting AGN
properties such as: the bolometric luminosity, Eddington ratio, hard
X-ray slope, and the hard X-ray bolometric correction. Combining all
the broadband SED parameters with the optical parameters, we
can provide further evidence for many previously suggested correlations, including all
the correlations between optical and X-ray claimed in previous work, plus many others such
as the H$\beta$ FWHM versus X-ray slope, black hole mass versus Eddington
ratio, FeII luminosity versus [OIII]$\lambda$5007 emission line
luminosity and the high excitation lines (e.g. [FeVII]$\lambda$6087,
[FeX]$\lambda$6374) versus their ionizing flux
(e.g. \citealt{Boroson92}; \citealt{Boller96}; \citealt{Grupe98};
\citealt{Grupe99}; \citealt{Sulentic00}; \citealt{Mullaney09}).

This paper is organized as follows. Section 2 describes the sample
selection and data analysis procedures. The detailed spectral fitting
methods and results including Balmer line fitting, optical spectral
fitting and broadband SED fitting are each discussed in sections 3,
4 and 5, separately. We present the statistical properties of our 
sample in section 6. The summary and conclusions are given in section 7. 
A flat universe model with Hubble constant of $H_{0} = 72$ km s$^{-1}$ Mpc$^{-1}$,
$\Omega_{M} = 0.27$ and $\Omega_{\Lambda} = 0.73$ is adopted.
In another paper, we will present our analysis of correlations 
between selected optical/UV emission features and the SED components,
and discuss their physical implications (Jin et al. in prep.,
hereafter Paper II).

\section{Sample Selection and Data Assembly}

To identify a sample of Type 1 AGNs having both high quality X-ray and optical
spectra, we performed a cross-correlation between {\tt 2XMMi} catalog
and {\tt SDSS DR7} catalog.  We filtered the resulting large sample
as described below. Our final sample consists of 51 Type 1 AGNs
including 12 NLS1s,
all with high quality optical and X-ray spectra and low 
reddening/absorption, and with H${\beta}$ line widths 
ranging from 600 $km s^{-1}$ up to 13000 $km s^{-1}$.
All the sources are listed in Table~\ref{object list}.

\subsection{The Cross-correlation of 2XMMi \& SDSS DR7}

The first step was to cross-correlate between {\tt 2XMMi} and {\tt
SDSS DR7} catalogs. The {\tt 2XMMi} catalog contains 4117 XMM-Newton EPIC
camera observations obtained between 03-02-2000 and 28-03-2008,
and covering a sky area of $\sim$ 420 deg$^{2}$. The {\tt SDSS DR7} is the
seventh data release of the Sloan sky survey. The SDSS
spectroscopic data has sky coverage of $\sim$ 8200 deg$^{2}$, with
spectra from 3800 {\AA} to 9200 {\AA}, and spectral resolution between
1800 and 2200.\\
Our cross-correlation consisted of three steps:\\ 
1. We first
searched for all XMM/SDSS position pairs that lay within 20$\arcsec$
of each other, resulting in 5341 such cases.\\ 
2. For these 5341
unique X-ray sources, we imposed two further selection criteria:
that source positions be separated by less than 3$\arcsec$, or that sources
be separated by no more than 3 $\times$ the XMM-Newton position
uncertainty and no more than 7$\arcsec$. This filtering resulted in
3491 unique X-ray sources. The 3$\arcsec$ separation is chosen 
because we want to include all possible XMM/SDSS pairs during these
early filtering steps. From the 2XMMi and SDSS DR7 cross-correlation,
there are 114 XMM/SDSS pairs whose separations are less than
3$\arcsec$, but are still nevertheless greater than 3 $\times$ the XMM
position uncertainty.  We included all of these pairs. The 7$\arcsec$
separation upper limit mitigates spurious matches, especially
for fainter objects and/or those located far off-axis.\\ 
3. We selected only objects classified as extragalactic, giving a 
total of 3342 for further analysis.

\subsection{Selection of Seyfert 1 with High Quality Spectra}
Within these 3342 unique X-ray sources which satisfied all 
the above criteria, we applied further filtering to select
only Type 1 AGNs having both high quality optical and X-ray
spectra. The five steps in the filtering were as follows:\\ 
1. In order to obtain black hole mass estimates and also as 
reddening indicators,
we require $H\beta$ and $H\alpha$ emission lines to be measurable.
So we only selected sources with $H\beta$ in emission (as
indicated by the SDSS $H\beta$ line models with at least 3$\sigma$
significance and $EW > 0$) and redshift $z < 0.4$.  This selection
resulted in 802 unique X-ray sources, and 888 XMM/SDSS pairs (since
some X-ray objects were matched with more than one SDSS spectrum).\\
2. Then we searched for the Type 1 AGNs (including subtypes 1.0, 1.5, 1.8 and 1.9)
which have a minimum of 2000 counts in at
least one of the three EPIC cameras. Our search retrieved 96 such
broad line AGNs. We then inspected each of these
XMM/SDSS pairs, to confirm that all the matches were indeed genuine.\\
3. From inspection of the SDSS spectra, we excluded 22 sources whose
blueward part of the $H\beta$ line showed strong reddening or low S/N,
which would distort the $H\beta$ line profile. We also excluded one
object, RBS 0992, because its SDSS spectrum did not show an $H\beta$
line, due to a bad data gap. We ensured that the remaining 73 objects
all had good $H\beta$ line profiles.\\ 
4. As a simple method to assess the spectral quality of the X-ray data, we used {\it wabs*powerlaw\/} 
model in {\tt xspec11.3.2} to fit the rest-frame 2-10 keV X-ray spectra 
of all 73 objects. The $error$ command was used to
estimate the 90{\%} confidence region for the $photon~index$
parameter. Based on the results, 16 objects with $photon~index$ uncertainties greater than
0.5 were thereby excluded, leaving 57 Type 1 AGNs with relatively well
constrained 2-10 keV spectra.\\ 
5. By examining the 0.2-10 keV X-ray
spectra, we excluded another 6 objects (i.e. IRAS
F09159+2129, IRAS F12397+3333, PG 1114+445, PG 1307+085, PG 1309+355
and PG 1425+267) whose spectral shapes all showed clear evidence of an absorption 
edge at $\sim$0.7 keV (possibly originating from combined Fe I L-Shell 
and O VII K-Shell absorptions (\citealt{Lee01}; \citealt{Turner04})). This  
is a typical spectral signature of a warm absorber (e.g. \citealt{Nandra94}; 
\citealt{Crenshaw03}). By removing such objects with complex X-ray spectra,
our broadband SED fitting is simplified. Our final sample contains 51 Type 1 AGNs.

\label{aperture correction}
\begin{figure*}
\centering
   \includegraphics[clip=]{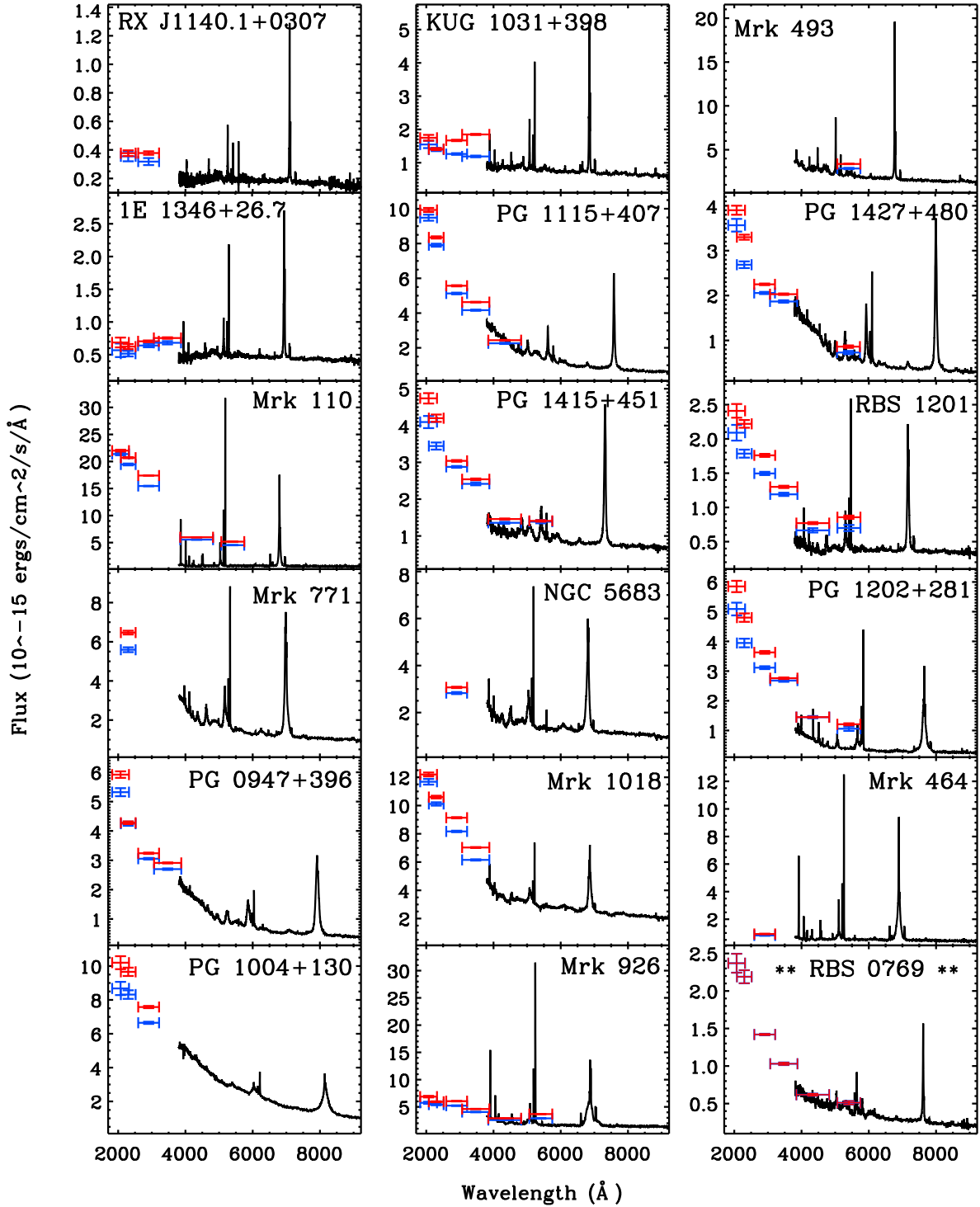}
   \caption{The aperture effect correction results for 17 extended
   sources in the sample.  The point like source RBS 0769 (the last
   figure marked by **) is also shown for comparison.  We over-plot OM
   data points on to the SDSS spectrum. Red OM points are data obtained
   directly from the OM PPS files. Blue OM points are the corresponding
   data after applying a smaller 6\arcsec aperture to all OM filters,
   and applying appropriate OM corrections to the flux eg.
   deadtime correction, coincidence loss correction and OM time
   sensitivity degradation correction.}
   \label{OMcorrection}
\end{figure*}

\subsection{Characteristics of the Sample}

The sample selection procedure described above ensures that every source
in our AGN sample has both high quality optical and X-ray
spectra. In addition, a large fraction of the sample have simultaneous
optical/UV photometric points from the OM monitor.
Such high quality data enables accurate spectral fitting.
In the optical band our sample is selected to have low reddening, since if present 
this would significantly modify the intrinsic continuum as well as
the optical emission lines. This requirement reduces the complexity
and uncertainty in our modeling of the intrinsic continuum, 
and also increases the overall quality of H$\beta$ and
H$\alpha$ line profiles useful for estimating the black hole masses.
Furthermore, low reddening is essential in the
UV band. The inclusion of OM-UV photometric data observed simultaneously
with the X-ray spectra provides a reliable link between these bands.
This helps to reduce fitting uncertainty of the SED resulting
from optical and X-ray variability. Besides, all
sources are well constrained in the 2-10 keV band, which is
directly associated with the compact emitting region of the AGN. 
Our exclusion of objects with evidence of a warm absorber means that
the 2-10 keV spectral index is likely to be intrinsic rather than hardened 
by absorption in the soft X-ray region.

In summary, compared with previous AGN samples used for 
broadband SED modelling, the spectrally
`cleaner' nature of our sample should make the reconstructed broadband
SEDs more reliable. Consequently, the parameters derived from the broadband
spectral fitting should be more accurate. This may reveal new
and potentially important broadband correlations, which we will discuss in
detail in paper II.

\subsection{Additional Data}

The 51 Type 1 AGNs all have SDSS survey-quality spectra (flagged as
``sciencePrimary'' in SDSS catalog), including 3 objects that have
multiple SDSS spectra (i.e. NVSS J030639, 1RXS J111007 and Mrk1018). 
In such cases we adopt the SDSS
spectrum which connects most smoothly with the OM data.

For each object, we used all available EPIC X-ray spectra
(i.e. pn, MOS1 and MOS2) for the broadband SED modeling, unless the
spectrum had few counts and low S/N. We also searched
through the {\tt XMM-OM SUSS} catalog for all data in the
OM bands (i.e. V, B, U, UVW2, UVM2 and UVW1), which are observed
simultaneously with the corresponding EPIC spectrum. Of our 51
sources, we have 14 sources with SDSS optical spectra and XMM EPIC
X-ray spectra, and 37 sources which in addition to this also have
XMM-OM photometry.

\begin{figure*}
\centering \includegraphics[bb=20 530 590 840,
scale=0.85,clip=]{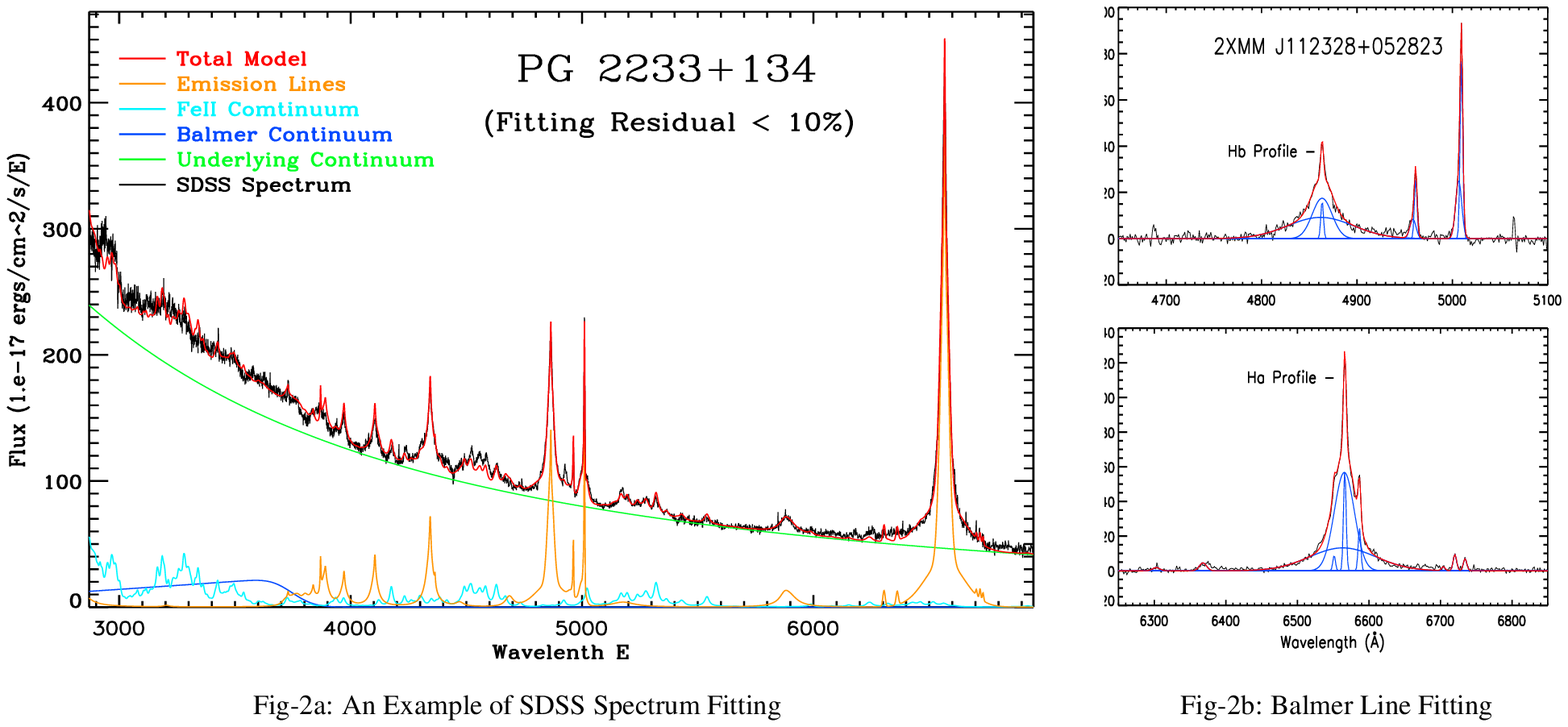}
\caption{An example of results from SDSS spectrum fitting. The left
panel shows a good fit for PG 2233+234. The black line is the observed
spectrum, the red line is the total model spectrum.  The green line
represents the observed underlying continuum. The Balmer continuum
(blue), FeII emission (light blue) and other strong emission lines
(orange) are shown underneath. The right panel shows an example of
detailed line profile fitting to the FeII subtracted region around the
H$\beta$ (upper) and H$\alpha$ lines (lower) including H$\alpha$,
H$\beta$, [OIII] $\lambda$5007/4959 doublets, [NII] $\lambda$6585/6548
doublets, Li $\lambda$6708, [SII] $\lambda$6717/6733 doublets, [OI]
$\lambda$6300/6363 doublets.  In our profile fitting, three Gaussian
components are used for H$\beta$ and H$\alpha$, two components for
[OIII] $\lambda$5007, and one Gaussian for all other lines.  The
various Gaussian profiles are shown in blue, the total model is shown
in red.}
\label{SDSS_fit}
\end{figure*}

\subsection{OM Data Corrections and Aperture Effects}
\label{OM:correction}
In the procedure of combining the SDSS spectra and OM data points, we
identified that in some objects there is a clear discrepancy between
these two data sets. The OM points often appear higher on the spectral 
plots (brigher) than 
is consistent from a smooth extrapolation of the SDSS spectral shape. In 
fewer cases this discrepancy appears
in the opposite sense, with the OM points apparently too low (fainter), see
Figure~\ref{OMcorrection} for some examples). 
This discrepancy may arise for several reasons, including a
simple aperture effect. Compared to 3{\arcsec} diameter 
for the SDSS spectroscopy fibres, the OM monitor has a much larger aperture, 
i.e. 12{\arcsec} and 35{\arcsec} diameter for the OM optical and
OM UV filters respectively (\citealt{Talavera09}). If the host galaxy is
sufficiently extended,
e.g. in the case of RE J1034+396, the larger aperture of the OM would
include more host galaxy emission than that in the SDSS
spectrum (see also section~\ref{discrepancy} for other
possible reasons to account for this discrepancy). To investigate
the aperture issue in more detail, we performed the following tests:\\
(1) We examined the combined SDSS and OM data plots, searching for those
objects with excess OM flux compared with that expected from the
extrapolated SDSS spectrum. We identified 27 such cases out of
the 51 sources;\\
(2) Within this sample of 27 sources, we checked the catalog flag
for an extended source in each OM filter. We noted those flagged as an
extended source in at least one OM filter. This yielded 13 sources out of the 27.\\
(3) We also extracted 
the SDSS CCD images for all 51 objects and visually checked whether they
appeared extended. As a result, we included another 4
objects for which their SDSS CCD images show that their host galaxy
is extended beyond the 3\arcsec diameter of the SDSS aperture. Either
they were not flagged as extended sources in any OM filter,
or they did not have any OM optical data. For these 17 objects,
an aperture effect could at least be 
partially responsible for an excess flux in the OM data.\\
(4) For these 17 objects we downloaded all available OM
image files. In each OM image, we applied a 6{\arcsec} diameter
aperture from which to extract the flux. We used the same
sized aperture placed on a blank region of sky close to the object, to
estimate the background. The quoted PSF FWHM of the OM for
the different filters are: V(1.35{\arcsec}), B(1.39{\arcsec}) ,
U(1.55{\arcsec}), UVW1(2.0{\arcsec}), UVM2(1.8{\arcsec}),
UVW2(1.98{\arcsec}). Thus in all cases 6{\arcsec} is at least 3$\times$PSF FWHM.
So this aperture includes effectively all optical flux for a point source,
and more than 90\% that from a UV point source detected by the OM.

Before subtracting the background flux from the source+background flux, we 
performed three count rate calibrations, according the method
described in the OM instrument document.\footnote{URL:
http://xmm2.esac.esa.int/docs/documents/CAL-TN-0019.ps.gz; Also see
the XMM-Newton User Handbook:
http://xmm.esac.esa.int/external/xmm\_user\_support/document-ation/uhb/index.html.}
The first is the deadtime correction, required because for a small fraction of
the exposure time the CCD is in readout mode, and so cannot record
events. The second calibration is for coincidence losses, which occur
when more than one photon arrives on the CCD at the same location and
within the same frame time, so results in under counting. The third
calibration is for the OM time sensitivity degradation correction.
We performed these calibrations, according to the algorithms set out in
the OM instrument document, separately for the background and source+background
count rates. We then subtracted the background count
rate from the source+background count rate to obtain the corrected source count
rate.

Figure~\ref{OMcorrection} shows the OM data points before and after
correction for aperture effects for the 17 objects.
The reduced OM aperture does improve the
alignment between the OM points and SDSS spectrum. This correction not
only lowers the OM flux, but also changes the continuum shape defined
by the OM points. Although choice of an aperture smaller than 6{\arcsec} will
lower the OM fluxes by a larger factor, it will also introduce uncertainties
and systematics caused by the PSF. Therefore we compromise by adopting
a 6{\arcsec} diameter aperture. In our subsequent SED modeling we use the
aperture corrected OM data.

\section{Optical Spectral Modeling: The Emission Lines}
\label{spectral modeling}

Our optical spectral modeling employs linked $H\alpha$ and $H\beta$
profile fitting and the complete optical spectral fitting.  We wrote the code in IDL 
(Interactive Data Language) v6.2, to perform all the optical spectral fitting. 
The `MPFITEXPR' program from the Markwardt IDL Library is incorporated 
within our code to perform the Levenberg-Marquardt least-squares algorithm 
used to obtain the best-fit parameters. The SDSS spectra
(stored in SDSS {\it spSpec\/} files) were extracted directly from the SDSS DR7
data archive and analyzed in IDL using our code.
A detailed description of our spectral modeling procedures is presented in the
following subsections.

\subsection{Profile Fitting of the H$\alpha$, H$\beta$ and [OIII]$\lambda$5007 Emission Lines }
\label{emission line fitting}
Based on current AGN emission line models, there are thought to be
stratified regions emitting different lines. These regions are divided
somewhat arbitrarily into a narrow line region (NLR), a broad line region
(BLR) and possibly an intermediate line region (ILR, e.g. \citealt{Grupe99};
\citealt{Hu08}; \citealt{Mei09}; \citealt{Zhu09}). Following previous
studies, we use several separate Gaussian profiles representing each
of these emitting regions to model the Balmer line profiles.

The H$\alpha$ and H$\beta$ line profiles each pose distinct difficulties 
for the spectral analysis. In the case of the $ H\beta$ line, the permitted FeII
emission features (which are often strong in NLS1s) and broad HeII
4686 line blended with the $H\beta$ line, which can affect the
determination of the underlying continuum and hence the
$H\beta$ line profile. For the $H\alpha$ line, there is the
problem of blending with the [NII] $\lambda$6584,6548 doublet,
improper subtraction of which may distort $H\alpha$'s intrinsic profile. Our
approach, therefore, is to fit $H\alpha$ and $H\beta$ simultaneously using the same
multi Gaussian components. The assumed similarity between the intrinsic
profiles of these two Balmer lines assists in deblending 
from other nearby emission lines, and should yield a more robust
deconvolution for the separate components of their profile.

\subsection{The FeII Problem}
\label{FeII:problem}
We use the theoretical FeII model templates of \citet{Verner09}. These
include 830 energy levels and 344,035 transitions between 2000{\AA}
and 12000 {\AA}, totaling 1059 emission lines.
The predicted FeII emission depends on physical
conditions such as microturbulence velocity and hardness of the
radiation field, but we use the template which best matches the
observed spectrum of I ZW 1 (\citealt{Boroson92},
\citealt{Veron-Cetty04}) i.e. the one with $n_{H}=10^{11}~cm^{-3}$,
$v_{turb}=30~kms^{-1}$, $F_{ionizing}=20.5~cm^{-2}s^{-1}$. Detailed
modelling of high signal-to-noise spectra shows that the FeII emission is
often complex, with four major line systems in the case of 1 Zw 1, (one broad line
system, two narrow high-excitation systems and one low-excitation
system \citealt{Veron-Cetty04}; \citealt{Zhou06}; \citealt{Mei09}). However,
for simplicity we will assume only one velocity structure and 
convolve this template with a single Lorentzian profile.

We fit this to the actual FeII emission line features between 5100
{\AA} and 5600 {\AA} (no other strong emission lines lie in this
wavelength range) of the de-redshifted SDSS spectra, leaving the
FWHM of the Lorentzian and the normalization of the FeII as free parameters.
The resulting best-fit FeII model to this restricted wavelength range, was 
then extrapolated and subtracted from the entire SDSS spectrum. A major
benefit from subtracting the FeII features is that the
profiles of the [OIII] $\lambda$5007 lines no
longer have apparent red-wings. This is particularly important for the
NLS1s, where the FeII emission is often strong. After subtracting FeII, we
used either 2 or 3 Gaussian components (depending on the profile complexity)
to fit the [OIII] $\lambda$5007 line.

\subsection{Deconvolution of the Balmer Lines} 

After fitting the [OIII] $\lambda$5007 line,
we start to fit the H$\alpha$ and H$\beta$ line profiles simultaneously.
Following previous studies we consider a simplified picture in which the
Balmer lines have three principal components, namely a
narrow component (from the NLR), an intermediate component (from a
transition region ILR between the NLR and BLR or from the inner edge of
dusty torus (\citealt{Zhu09})), and a broad component (from the BLR).
The intermediate and broad components are both represented
by a Gaussian profile, whereas the narrow component is assumed to
be similar to that of [OIII] $\lambda$5007. Since we do not know whether
or not the Balmer decrements are the same in these different emitting
zones, the relative strengths of different line components were not fixed,
but their FWHM and relative velocity were both kept the same.
The [OIII] $\lambda$4959 line was set at $1/3$ that of [OIII] $\lambda$5007
from atomic physics. The [NII] $\lambda$6584,6548 line doublet were also fixed to the [OIII]
$\lambda$5007 line profile. For simplicity, the [SII] $\lambda$6733,6717 doublet,
[OI] $\lambda$6300,6363 doublet and Li 6708 were all fitted with a single
Gaussian profile separately, because they are all relatively
weak lines and do not severely blend with Balmer lines.

In order to separate the narrow component of the Balmer lines from the
other components as accurately as possible, particularly for NLS1s and
some broad line objects which lack clear narrow line profiles, we
applied the following four different fitting methods: \\
1. The profile of the narrow component is held the same as the entire [OIII]
$\lambda$5007 profile; and the normalization of each component in
the $H\alpha$ and $H\beta$ lines are left as free parameters; \\
2. Only the central narrow component of the [OIII] $\lambda$5007 profile 
is used to define the profile of the Balmer narrow component, and of the
[NII] $\lambda$6585,6550 doublet; the normalization of each component
in the $H\alpha$ and $H\beta$ lines are free parameters; \\
3. The shape of the narrow component is held the same as the entire [OIII]
$\lambda$5007 profile, and also the normalization of the $H\beta$ line
narrow component is set to be 10\% of [OIII] $\lambda$5007, this ratio being
an average for the NLR in typical Seyfert 1s (\citealt{Osterbrock85}; \citealt{Leighly99});
all other components have their normalizations as free parameters; \\
4. All conditions are the same as in method 3, except that the
Balmer line narrow component and the [NII] $\lambda$6584,6548 doublet
adopt the central narrow Gaussian component of the [OIII] $\lambda$5007 line.

We applied each of the above fitting methods to every object in our
sample, and then compared the results. For those objects with
clear narrow components to their Balmer lines, we used the best
fitting result from method 1 and 2.  For the other objects whose
narrow components were not clearly defined or even visible, we adopted
method 3 and method 4, unless method 1 or 2 gave much better fitting
results. Figure~\ref{SDSS_fit} right panel shows an example of our
fitting. Results for the whole sample are shown in Figure~\ref{SED}.

After obtaining the best-fit parameters, we used the intermediate and
broad components to reconstruct the narrow-line subtracted $H\beta$
line profile, and then measured the FWHM from this model. 
The rationale for using this method, instead of directly measuring the FWHM of
the $H\beta$ line from the data, is because for low signal/noise line profiles direct
measurement of FWHM can lead to large uncertainties, whereas our
profile models are not prone to localized noise in the data.
The H$\beta$ FWHM measurements for each of the 51 sources, after
de-convolving using the instrumental resolution of 69 $km s^{-1}$, are
listed in Table~\ref{SED-key-parameters}.

\section{Optical Spectral Modeling}

In order to obtain the underlying continuum, we must model the entire
SDSS spectrum so that we can remove all the emission lines as well as 
the Balmer continuum and host galaxy contribution.
As we are now concerned with the broad continuum
shape, we choose to refit the FeII spectrum across the
entire SDSS range, rather than restricting the fit to the H$\alpha$
and H$\beta$ line regions as discussed in the previous section.

Figure~\ref{SDSS_fit} shows an illustrative example of our 
optical spectral fitting, and the results for each of the 51 sources
are presented in Figure~\ref{SED}. In the following
subsections we give further details of the components that make up
these modeled spectra.

\subsection{Emission Lines Including FeII}

We use the models for [OIII], $H\alpha$ and $H\beta$ as derived
above. We add to this a series of higher order Balmer lines: from
5$\rightarrow$2 ($H\gamma$) to 15$\rightarrow$2.
We fix the line profile of these to that of $H\beta$ up to 
9$\rightarrow$2, then simply use a single Lorentzian profile for the rest
weak higher order Balmer lines. We fix the line ratios for each 
Balmer line using the values in \citet{Osterbrock89}, Table 4.2, with
$T_{e}$ between 10,000 K and 20,000 K. We similarly use a single Lorentizan 
to model the series of Helium lines (HeI 3187, HeI 3889, HeI 4471, HeI 5876,
HeII 3204, HeII 4686) and some other emission lines
(MgII 2798, [NeIII] $\lambda$3346,4326, [OII]
$\lambda$3727,3730, [OI] $\lambda$6302,6366, [NII] $\lambda$6548,6584,
Li 6708, [SII] $\lambda$6717,6733).

We use the same model for the FeII emission as described in
Section~\ref{emission line fitting}. However we now fit this to the
entire SDSS wavelength range, rather than restricting the fit to
5100--5600 {\AA}. 

\subsection{The Balmer Continuum}

Another potentially significant contribution at shorter wavelengths is
from the Balmer continuum. \citet{Canfield81} and
\citet{Kwan81} predicted the optical depth at the Balmer continuum
edge to be less than 1, we use Equation~\ref{balmer-conti} to model
the Balmer continuum under the assumptions of the optically thin case
and a single-temperature electron population (also see
\citealt{Grandi82}; \citealt{Wills85}).
\begin{equation}
\label{balmer-conti}
F_{v}^{BC}=F_{v}^{BE}e^{-h(v-v_{BE})/(kT_{e})}~~~(v \geqslant v_{BE})
\end{equation}
where $F_{v}^{BE}$ is the flux at Balmer edge, $v_{BE}$ corresponds to the Balmer
edge frequency at 3646{\AA}. $T_{e}$ is the electron temperature. $h$ is the Planck's constant, 
$k$ is the Boltzmann's constant. This Balmer continuum equation is then
convolved with a Gaussian profile to represent the real Balmer 
bump in SDSS spectra.

There are several parameters that may slightly modify or significantly
change the shape of the Balmer continuum. It is already seen that the
electron temperature $T_{e}$ appearing in Equation~\ref{balmer-conti}
and the optical depth can both change the Balmer continuum shape, but
there are additional important factors. Any intrinsic velocity
dispersion will Doppler broaden all the Hydrogen emission
features. Therefore a better description of the Balmer continuum can
be obtained by convolving Equation~\ref{balmer-conti} with a
Gaussian profile, whose FWHM is determined by
the line width of $H\beta$ (or other broad lines), as shown by
Equation~\ref{balmer convolve}, where G(x) represents a Gaussian
profile with a specific FWHM.
\begin{equation}
\label{balmer convolve}
F_{\lambda}^{BC}=F_{\lambda}^{BE}e^{hc/({\lambda}_{BE}kT_{e})}{\int}_{0}^{+\infty}
e^{-hc/({\lambda}kT_{e})}G({\lambda}_{1}-\lambda)d{\lambda}_{1}
\end{equation}

\begin{figure}
\centering
\includegraphics[scale=0.48,clip=]{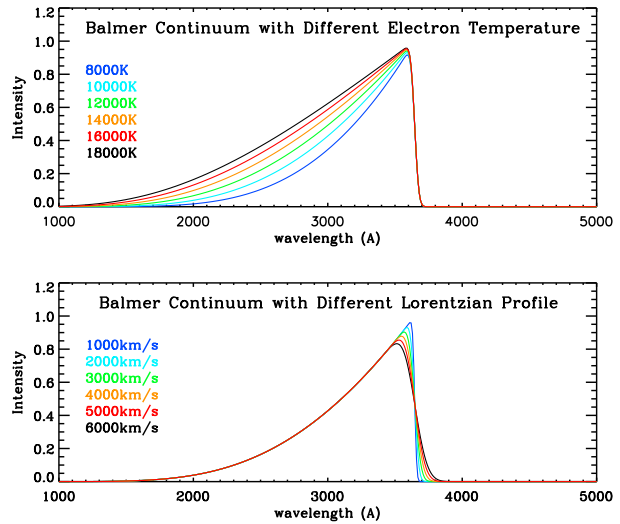}
\caption{The Balmer continuum models of \citet{Grandi82}.
The upper panel shows the dependence of the model on the electron temperature.
The lower panel shows the dependence of the model on the FWHM of the convolved Lorentzian profile.}
\label{balmerconti model}
\end{figure}

Figure~\ref{balmerconti model} shows how the Balmer continuum's shape
depends on the electron temperature and velocity broadening in
Equation~\ref{balmer convolve}. The electron temperature modifies the
decrease in the Balmer continuum towards shorter wavelengths, but has
little effect on the broadening of (Balmer Photo-recombination) BPR edge. On the contrary, velocity
broadening mainly affects the shape of the BPR edge, but the emission
longward of 3646{\AA} is still very weak compared to the emission
blueward of the BPR edge, i.e. the BPR edge is still sharp.

We initially applied Equation~\ref{balmer convolve} to fit the Balmer
continuum bump below 4000{\AA} in the SDSS spectra. We assumed the
velocity profile for the convolution was a Gaussian with its FWHM
determined from the $H\beta$ line profile, and the wavelength of the
position of the BPR edge was taken as the laboratory wavelength of
3646\AA. However, this model did not provide an acceptable fit, for
example see the model shown by the blue line in Figure~\ref{example of
Balmer continuum fit}. It appears that the observed spectrum requires
a model with either a more extended wing redward of the BPR edge, or
a BPR edge that shifts to longer wavelength than 3646\AA. However,
additional velocity broadening should affect both the Balmer continuum and
Balmer emission lines equally, as they are produced
from the same material, although the multiple components present in
the line make this difficult to constrain.

One way the wavelength of the edge may be shifted without affecting the lines is
via density (collisional, or Stark) broadening (e.g. \citealt{Pigarov98}).
Multiple collisions disturb the outer energy levels, leading to an effective
$n_{max}$ for the highest bound level $\ll \infty$, i.e. lowering the effective
ionization potential. We set the edge position and the FWHM as free parameters,
and let the observed spectral shape determine their
best fit values. The red line shown in Figure~\ref{example of Balmer
continuum fit} represents a good fit, obtained with FWHM of 6000 $km
s^{-1}$ and the BPR edge wavelength of 3746 \AA, which implies
$n_{max}\sim 12$. The theoretical $n_{max}$ can be determined by the plasma 
density $N_e$ and temperature $T_e$ as $n_{max}=2\times 10^4 (T_e/N_e)^{1/4}$
(\citealt{Mihalas78}), so for a
typical temperature of $10^{4-5}K$, the required density is $7\times
10^{16-17}$ cm$^{-3}$. Such high density is not generally associated
with the BLR clouds, and may give support to models where the low
ionization BLR is from the illuminated accretion disc
(e.g. \citealt{Collin-Souffrin90}). However, any reliable
estimation of the density would require more accurate subtraction of
other optical components such as the FeII line blends and many other
non-hydrogen emission lines, which is not the focus of this
paper. Nonetheless, this remains an interesting problem which is
worthy of further study.

\begin{figure}
\centering
\includegraphics[scale=0.48,clip=]{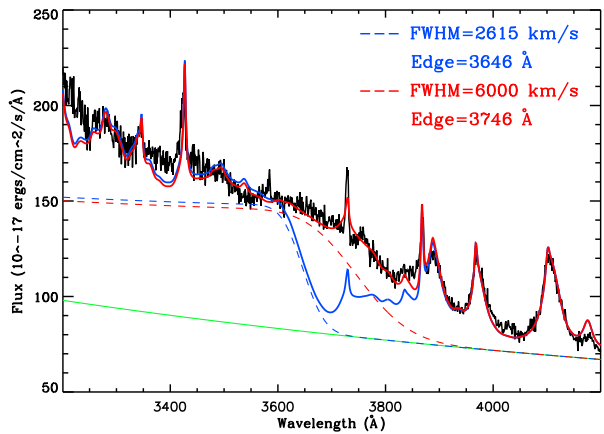}
\caption{An expanded view of the region around the BPR edge in PG 1427+480.
The blue and dashed lines represent the Balmer continuum model superposed on the underlying disc
continuum (green solid line) using standard parameters (blue dash), and also a set of best fit
parameters (red dash line). The red and blue solid lines are models of the total optical spectrum, including
the corresponding Balmer continuum components and plus other components described in the text.
The observed spectrum is shown in black.}
\label{example of Balmer continuum fit}
\end{figure}

Yet another issue in modeling the Balmer continuum is how to quantify the
the total intensity of this continuum component, especially when there
is limited spectral coverage bellow 4000{\AA}, which makes it difficult to
define the overall shape. The theoretical flux ratio between the
Balmer continuum and the $H\beta$ line under case B conditions can be
expressed by Equation~\ref{Bac-Hb relation} (\citealt{Wills85}),
\begin{equation}
\label{Bac-Hb relation}
I(Bac)/I(H\beta) = 3.95~T_{4}^{0.4}
\end{equation}
but other theoretical calculations of photonionization models show
that by varying the Balmer optical depth, electron temperature and
electron number density, this can result in very different values of
I(Bac)/I(H$\beta$). For example, \citet{Canfield81}'s calculation resulted in a
I(Bac)/H$\alpha$ range of 0.05$\sim$10, \citet{Kwan81} suggested
I(Bac)/I(H$\beta$)=1.6$\sim$15, and other theoretical work also
confirmed a large range in flux ratios (\citealt{Puetter82a};
\citealt{Kwan84}; \citealt{Hubbard85}).  The observed ranges in
I(Bac)/I(H$\beta$) are also large. \citet{Canfield81} showed an
observed range of 0.5$\sim$3 for I(Bac)/I(H$\alpha$). \citet{Wills85}
observed 9 intermediate redshift QSOs whose I(Bac)/I(H$\beta$) ranges
from 4.65$\sim$9.5. Thus we were unable to constrain the intensity of
the whole Balmer continuum by using a standard flux ratio fixed to
the other Balmer emission lines. As a result, we must rely on the shape of the
observed Balmer bump, and then adopt the model's best fit parameters.

However, this limitation in defining the Balmer bump introduces
uncertainties in modeling the underlying continuum, because
over-subtraction of the Balmer bump will depress the slope of the remaining
underlying continuum, and vice-versa.
In the course of the broadband SED fitting described in section~\ref{SED-modelling}, 
we found that the temperature of accretion disc (determined by black hole mass) 
is sensitive to the slope of optical continuum, 
unless the continuum slope is in the opposite sense to that of the accretion disc 
model and thus can not be fitted, or
there are OM points providing stronger constraints.
We also found that a flatter optical 
continuum  may lead to a lower best-fit black hole mass, although this 
also depends on other factors. Therefore, the subtraction of the Balmer 
continuum can have an impact on the modeling of broadband SED and the best-fit
black hole mass. The influence of this depends on the relative importance of 
other SED restrictions. This is the reason why the Balmer continuum must be
carefully modeled and subtracted.

\subsection{The Intrinsic Underlying Continuum}
\label{SDSS-continuum}
Our basic assumption is that the residual optical spectrum, after
subtraction of the Balmer continuum, FeII emission and other
emission lines mentioned previously, arises mainly from the accretion disc
emission.  As a reasonable approximation over a limited wavelength range we
use a powerlaw of the following form to fit the underlying continuum,
\begin{equation}
\label{powerlaw}
F(\lambda) = C_{1}{\cdot}({\lambda}/5100\AA)^{-C_{2}}
\end{equation}
The powerlaw approximation for the optical underlying disc continuum is also
widely adopted in previous and recent AGN optical spectral studies.
(e.g. \citealt{Grandi82}; \citealt{Tsuzuki06}; \citealt{Zhou06}; \citealt{Landt11}).

We model the dust reddening using the \citet{Seaton79}'s 1100{\AA} to
10000{\AA} reddening curve, and we apply this to the overall model,
i.e. emission lines, Balmer continuum and the disc continuum. There
are also other reddening curves available such as \citet{Fitzpatrick86} for the
Large Magellanic Cloud, \citet{Prevot84} and \citet{Bouchet85} for the
Small Magellanic Cloud and \citet{Calzetti00} for starburst galaxies,
but over the wavelength range of 2500{\AA} to 10000{\AA}, the
difference between these reddening curves is small, except for
\citet{Calzetti00}'s curve which is appropriate for starburst galaxies,
and is thus not applicable for our AGN sample. 

\subsection{The Host Galaxy Contribution}
\label{HG:Contamination}
Many previous studies on AGN's optical/infrared spectra have 
adopted a powerlaw as a reasonable
approximation for the accretion disc continuum blueward of 1$\mu$m
(e.g. \citealt{Mei09}; \citealt{Bian10}), but these studies also
needed to include additional contributions from the host galaxy and emission 
from the dusty torus to account for the extra continuum emission
at long wavelengths of the optical spectrum
(e.g. \citealt{Kinney96}; \citealt{Mannucci01}; \citealt{Landt11}).
In our work we have also identified an inconsistency 
between the 3000{\AA}$\sim$8000{\AA} spectral
shape and a single powerlaw shape (i.e. the flat optical spectrum
problem discussed in Section~\ref{FOC-problem}). The blue end of the optical
spectrum, presumed to arise from a standard accretion disc, often
shows a steeper spectral slope than the red end.

However, in our sample we found evidences suggesting only a weak if any, contribution 
from the host galaxy. For example, the optical spectra of our sample do not show the 
strong curvature characteristic of the presence of a stellar component in a host galaxy.
Furthermore, the good quality optical spectra do not exhibit  
stellar absorption features (see Section~\ref{FOC-problem} and
Figure~\ref{hostgalaxy-example}). In fact the 3$\arcsec$ diameter
fibre used to obtain the SDSS spectra also helps to reduce the
contribution of stellar emission from a host galaxy, particularly
for nearby sources in our sample such as KUG 1031+398. These evidences
argues against the possibility that the red optical continuum
is primarily dominated by host galaxy emission. In fact, it is possible that the 
observed additional component arises due to emission from the outer regions of
a standard accretion disc (e.g. \citealt{Soria02}; \citealt{Collin04};
\citealt{Hao10}). The existence of such an additional red optical continuum
component reduces the consistency of a powerlaw fit to the optical spectra.

\subsection{The Optical Spectrum Fitting}
Our optical spectral fitting is performed only for data blueward
of 7000{\AA}. The choice to truncate the model at 7000{\AA} is made for several 
reasons. We wish to include H$\alpha$ line in the spectral
fitting range, and the broad wing of H$\alpha$ profile sometimes
extend to $\sim$7000{\AA} (e.g. PG 1352+183, RBS 1423, Mrk 926). There
are some objects whose SDSS spectra extend only to $\sim$6700{\AA}
(e.g. 2XMM J080608.0+244421, HS 0810+5157, 2XMM J100025.2+015852). The
choice of 7000\AA, rather than a longer wavelength, is to maintain 
consistency of optical spectral fitting for the whole sample. The
final reason concerns an aspect of the powerlaw fitting. We found 
that in some objects (e.g. PG 1115+407, LBQS 1228+1116, PG1352+183), 
a flat slope power-law under predicts the observed emission at $\sim$7000\AA.
Therefore, if we include longer wavelengths than
7000\AA, our powerlaw fitting for the standard accretion disc
continuum towards the blue optical spectra would be biased 
by other continuum emission at these longer
wavelengths, and so affect the broadband SED
fitting. Consequently, we chose to truncate our optical spectral
fitting at 7000\AA.

However, we still cannot be sure that the underlying continuum is
totally free from other non-disc continuum components. So after
completing the fitting procedure, we then checked the 
spectral fitting status within two narrow wavebands,
i.e. 4400{\AA} - 4800{\AA} and 5100{\AA} - 5600{\AA}.
Emission features if present in these two wavebands are mainly from
FeII emission, and the underlying continua of these two wavebands
should be totally dominated by the accretion disc emission.
Assuming that the FeII emission lines within these two
wavebands have similar relative intensity ratios as in the
FeII template described in Section~\ref{FeII:problem},
the best-fit underlying powerlaw plus FeII emission model should 
have good fitting status in both of these two wavebands.
In general, the best-fit model derived from the full
optical spectrum fit also gives reasonably good
fitting status in both of these two narrow wavebands. However, in
some cases the model over-predicted the flux in 5100{\AA} - 5600{\AA}
but under-predicted the flux in 4400{\AA} - 4800{\AA}, 
so that we should slightly increase the slope of powerlaw to 
produce better spectral fitting in these two wavebands.
We adopted these parameter values in preference to those 
directly from the full spectrum fit, as they should be more immune 
to problems such as host galaxy or hot dust contamination.

\begin{table*}
 \centering
  \begin{minipage}{165mm}
   \caption{Broadband SED Fitting Parameters, and Model Outputs (L$_{bol}$, {\it f}$_{d}$, {\it f}$_{c}$, {\it f}$_{p}$).
ID: object number, the same as Table~\ref{object list}; N$_{H,gal}$ and N$_{H,int}$: the fixed galactic and free intrinsic neutral hydrogen column densities in 10$^{20}$ $cm^{-2}$; $\Gamma_{pow}$: the powerlaw component's slope in the SED fitting, (*) denotes the objects whose powerlaw slopes hit the uplimit of 2.2 and were fixed there; Fpl: the fraction of powerlaw component in the total reprocessed disc emission; R$_{cor}$: corona (truncation) radius in unit of Gravitational radii (r$_{g}$) within which all disc emission is reprocessed into the Comptonisation and powerlaw components; T$_{e}$: temperature of the Compton up-scattering electron population; Tau: optical depth of the Comptonisation component; log(M$_{BH}$): the best-fit black hole mass; log($\dot{M}$): total mass accretion rate; L$_{bol}$: bolometric luminosity integrated from 0.001 keV to 100 keV; {\it f}$_{d}$, {\it f}$_{c}$, {\it f}$_{p}$: luminosity fractions of disc emission, soft Comptonisation and hard X-ray Compotonisation components in the bolometric luminosity; $\chi^{2}$: the reduced $\chi^{2}$ of the broadband SED fitting.}
   \label{SED-fitting-parameters}
     \begin{tabular}{@{}ccccccccccccccc@{}}
\hline\hline
ID & N$_{H,gal}$ & N$_{H,int}$ & $\Gamma_{pow}$ & Fpl & R$_{cor}$ & T$_{e}$ & Tau & {\it log}(M$_{BH}$) & {\it log}($\dot{M}$) & L$_{bol}$ &{\it f\/}$_{d}$ & {\it f\/}$_{c}$ & {\it f\/}$_{p}$ & $\chi^{2}$\\

 & $\times{\it 10}^{\it 20}$ & $\times{\it 10}^{\it 20}$ & & & $r_{g}$ & {\it keV} & & {\it M}$_{\sun}$ & $g$ $s^{-1}$ & ${\it 10}^{\it 44}$ & & & & {\it reduced} \\

\hline
1 & 1.79 & 0.00 & 1.71 & 0.69 & 100. & 0.262 & 17.2 & 8.61 & 26.06 & 58.9 & 0.19 & 0.25 & 0.56 & 1.00 \\
2 & 2.43 & 1.06 & 1.77 & 0.39 & 100. & 0.226 & 15.7 & 7.85 & 25.21 & 8.28 & 0.19 & 0.49 & 0.32 & 0.97 \\
3 & 6.31 & 9.88 & 1.91 & 0.25 & 11.9 & 0.108 & 20.0 & 7.41 & 25.92 & 42.9 & 0.87 & 0.10 & 0.03 & 1.57 \\
4 & 3.49 & 2.81 & 1.66 & 0.50 & 100. & 0.312 & 15.4 & 8.78 & 25.41 & 13.3 & 0.19 & 0.41 & 0.40 & 1.15 \\
5 & 3.53 & 4.03 & 2.12 & 0.36 & 54.9 & 0.205 & 14.9 & 7.87 & 26.28 & 98.4 & 0.32 & 0.44 & 0.24 & 1.10 \\
6 & 4.24 & 0.00 & 1.93 & 0.46 & 23.9 & 0.347 & 12.6 & 8.50 & 26.33 & 111 & 0.59 & 0.22 & 0.19 & 1.02 \\
7 & 1.33 & 3.74 & 2.20$^{*}$ & 0.29 & 8.37 & 0.137 & 40.3 & 7.00 & 26.53 & 175 & 0.26 & 0.53 & 0.21 & 1.20 \\
8 & 3.12 & 7.35 & 1.82 & 0.15 & 24.1 & 1.380 & 3.44 & 7.09 & 25.85 & 36.6 & 0.58 & 0.35 & 0.06 & 1.39 \\
9 & 1.30 & 1.36 & 1.71 & 0.71 & 12.9 & 0.360 & 11.1 & 6.96 & 25.94 & 45.0 & 0.84 & 0.05 & 0.11 & 17.2 \\
10 & 1.74 & 0.00 & 1.91 & 0.32 & 100. & 0.295 & 13.8 & 8.47 & 26.20 & 81.5 & 0.19 & 0.55 & 0.26 & 1.72 \\
11 & 1.72 & 2.00 & 1.71 & 0.49 & 20.2 & 0.449 & 9.23 & 7.80 & 26.02 & 53.8 & 0.65 & 0.18 & 0.17 & 1.01 \\
12 & 1.20 & 1.08 & 1.68 & 0.48 & 20.6 & 0.402 & 11.4 & 7.79 & 25.27 & 9.46 & 0.65 & 0.18 & 0.17 & 1.20 \\
13 & 3.56 & 0.00 & 1.37 & 0.87 & 10.9 & 0.146 & 17.9 & 9.20 & 26.52 & 170 & 0.90 & 0.01 & 0.09 & 3.12 \\
14 & 1.76 & 0.00 & 1.72 & 0.71 & 100. & 0.294 & 16.0 & 8.24 & 25.82 & 33.6 & 0.19 & 0.23 & 0.58 & 1.07 \\
15 & 1.31 & 2.43 & 2.20$^{*}$ & 0.09 & 14.2 & 0.214 & 12.3 & 6.23 & 25.31 & 10.4 & 0.80 & 0.18 & 0.02 & 2.27 \\
16 & 1.70 & 0.65 & 1.72 & 0.31 & 100. & 0.327 & 13.0 & 8.33 & 25.85 & 36.2 & 0.19 & 0.56 & 0.25 & 1.44 \\
17 & 0.65 & 0.85 & 1.74 & 0.14 & 48.7 & 0.326 & 11.4 & 7.97 & 25.85 & 36.5 & 0.35 & 0.56 & 0.09 & 1.08 \\
18 & 1.45 & 0.19 & 2.20$^{*}$ & 0.24 & 29.5 & 0.254 & 13.6 & 8.17 & 26.18 & 76.9 & 0.51 & 0.37 & 0.12 & 1.37 \\
19 & 3.70 & 1.41 & 1.98 & 0.19 & 45.8 & 0.142 & 21.5 & 7.71 & 24.85 & 3.61 & 0.37 & 0.52 & 0.12 & 1.10 \\
20 & 1.91 & 4.77 & 2.20$^{*}$ & 0.36 & 9.63 & 0.210 & 16.8 & 6.46 & 25.36 & 11.9 & 0.94 & 0.04 & 0.02 & 1.39 \\
21 & 1.77 & 0.00 & 1.79 & 0.75 & 22.7 & 0.206 & 19.6 & 7.98 & 26.09 & 63.4 & 0.61 & 0.10 & 0.29 & 3.59 \\
22 & 2.75 & 8.84 & 1.86 & 0.21 & 50.5 & 0.108 & 25.1 & 7.84 & 25.42 & 13.5 & 0.34 & 0.52 & 0.14 & 1.09 \\
23 & 1.59 & 0.00 & 1.41 & 0.45 & 86.9 & 0.626 & 9.59 & 7.99 & 25.02 & 5.40 & 0.22 & 0.43 & 0.35 & 0.99 \\
24 & 1.63 & 0.00 & 1.82 & 0.94 & 32.2 & 0.182 & 32.2 & 8.26 & 25.96 & 46.5 & 0.48 & 0.03 & 0.49 & 2.13 \\
25 & 2.34 & 0.00 & 1.79 & 0.40 & 25.7 & 0.351 & 12.9 & 8.49 & 26.26 & 94.2 & 0.56 & 0.27 & 0.17 & 1.83 \\
26 & 2.31 & 7.25 & 2.10 & 0.03 & 33.8 & 0.310 & 9.69 & 7.37 & 26.23 & 87.7 & 0.46 & 0.52 & 0.02 & 1.14 \\
27 & 2.75 & 0.00 & 1.85 & 0.22 & 37.6 & 0.554 & 8.29 & 7.50 & 25.44 & 14.2 & 0.43 & 0.45 & 0.12 & 1.12 \\
28 & 1.45 & 0.00 & 1.69 & 0.60 & 71.3 & 0.353 & 13.7 & 8.24 & 25.81 & 33.0 & 0.26 & 0.30 & 0.45 & 1.26 \\
29 & 1.18 & 1.36 & 2.00 & 0.12 & 30.9 & 0.389 & 8.85 & 7.76 & 26.03 & 55.1 & 0.49 & 0.45 & 0.06 & 1.24 \\
30 & 1.87 & 2.64 & 2.20$^{*}$ & 0.36 & 9.67 & 0.234 & 16.9 & 6.79 & 25.77 & 30.3 & 0.94 & 0.04 & 0.02 & 1.03 \\
31 & 0.84 & 0.00 & 1.68 & 0.54 & 100. & 0.404 & 12.9 & 8.70 & 25.84 & 35.9 & 0.19 & 0.37 & 0.43 & 0.99 \\
32 & 0.90 & 0.14 & 1.80 & 0.44 & 100. & 0.388 & 12.2 & 7.69 & 25.15 & 7.30 & 0.19 & 0.46 & 0.35 & 1.66 \\
33 & 1.07 & 0.82 & 2.18 & 0.57 & 15.0 & 0.226 & 15.6 & 7.78 & 25.96 & 47.3 & 0.78 & 0.10 & 0.13 & 1.15 \\
34 & 1.83 & 0.93 & 1.90 & 0.33 & 100. & 0.252 & 14.8 & 8.71 & 26.03 & 55.1 & 0.19 & 0.54 & 0.26 & 1.11 \\
35 & 1.76 & 0.90 & 1.80 & 0.83 & 100. & 0.202 & 20.4 & 7.67 & 26.13 & 69.8 & 0.19 & 0.14 & 0.67 & 1.05 \\
36 & 1.18 & 3.94 & 2.18 & 0.22 & 16.2 & 2.000 & 2.71 & 6.52 & 25.13 & 6.90 & 0.75 & 0.20 & 0.05 & 1.81 \\
37 & 1.82 & 0.00 & 2.04 & 0.38 & 100. & 0.219 & 17.2 & 8.23 & 25.88 & 39.3 & 0.19 & 0.50 & 0.31 & 1.33 \\
38 & 1.42 & 0.37 & 1.58 & 0.97 & 100. & 0.251 & 25.0 & 7.69 & 24.54 & 1.80 & 0.19 & 0.02 & 0.79 & 1.28 \\
39 & 1.36 & 4.77 & 2.10 & 0.11 & 40.6 & 0.281 & 11.4 & 7.01 & 25.17 & 7.57 & 0.40 & 0.53 & 0.07 & 1.90 \\
40 & 0.77 & 5.21 & 2.05 & 0.06 & 24.0 & 0.930 & 4.28 & 7.41 & 26.26 & 93.6 & 0.59 & 0.39 & 0.02 & 2.27 \\
41 & 1.81 & 0.00 & 1.90 & 0.39 & 28.9 & 0.298 & 14.0 & 8.39 & 26.10 & 65.2 & 0.52 & 0.30 & 0.19 & 1.63 \\
42 & 2.86 & 3.29 & 1.84 & 0.41 & 100. & 0.083 & 31.3 & 7.74 & 24.68 & 2.45 & 0.19 & 0.47 & 0.33 & 1.01 \\
43 & 2.69 & 0.00 & 1.71 & 0.58 & 55.8 & 0.406 & 11.9 & 8.07 & 26.10 & 64.7 & 0.31 & 0.29 & 0.40 & 1.29 \\
44 & 2.78 & 5.90 & 2.17 & 0.04 & 27.6 & 0.501 & 6.71 & 7.26 & 26.13 & 68.6 & 0.53 & 0.45 & 0.02 & 2.33 \\
45 & 1.46 & 0.00 & 1.82 & 0.49 & 41.0 & 0.286 & 14.1 & 8.62 & 26.75 & 290 & 0.40 & 0.30 & 0.30 & 2.42 \\
46 & 4.02 & 0.55 & 1.81 & 0.81 & 100. & 0.207 & 20.3 & 8.56 & 25.58 & 19.4 & 0.19 & 0.15 & 0.66 & 1.12 \\
47 & 3.78 & 16.69 & 1.82 & 0.25 & 100. & 0.115 & 29.8 & 7.96 & 25.62 & 21.5 & 0.19 & 0.61 & 0.20 & 0.99 \\
48 & 2.11 & 0.87 & 1.85 & 0.19 & 18.1 & 0.525 & 8.61 & 7.19 & 25.16 & 7.40 & 0.70 & 0.24 & 0.06 & 1.19 \\
49 & 4.90 & 0.36 & 2.20$^{*}$ & 0.33 & 72.5 & 0.211 & 19.6 & 7.73 & 25.15 & 7.33 & 0.25 & 0.50 & 0.25 & 1.15 \\
50 & 4.51 & 0.00 & 2.20$^{*}$ & 0.80 & 7.88 & 0.131 & 48.5 & 7.86 & 27.42 & 1350 & 0.98 & 0.00 & 0.01 & 1.39 \\
51 & 2.91 & 1.53 & 1.79 & 0.95 & 100. & 0.112 & 45.2 & 7.65 & 25.32 & 10.8 & 0.19 & 0.04 & 0.77 & 1.38 \\
\hline\hline
   \end{tabular}
  \\
 \end{minipage}
\end{table*}

\begin{table*}
 \centering
  \begin{minipage}{160mm}
   \caption{Broadband SED Key Parameters.
ID: object number, the same as Table~\ref{object list}; $\Gamma_{2-10keV}$: the slope of the single powerlaw fitted to 2-10 keV spectrum. L$_{2-10keV}$: 2-10 keV luminosity (in 10$^{44}$ erg s$^{-1}$); $\kappa_{2-10keV}$: the 2-10keV bolometric correction coefficient; $\lambda$L$_{2500\AA}$: the monochromatic luminosity at 2500{\AA} (in 10$^{43}$ erg s$^{-1}$); $\nu$L$_{2keV}$: the monochromatic luminosity at 2keV (in 10$^{43}$ erg s$^{-1}$); $\alpha_{ox}$: the optical X-ray spectral index; $\lambda$L$_{5100}$: the monochromatic luminosity at 5100{\AA} (in 10$^{44}$ erg s$^{-1}$); $\kappa_{5100}$: the 5100{\AA} bolometric correction coefficient; FWHM$_{H\beta}$: the narrow component subtracted H$\beta$ FWHM; L$_{bol}$/L$_{Edd}$: the Eddington Ratio.}
   \label{SED-key-parameters}
     \begin{tabular}{ccccccccccc}
\hline\hline
ID & $\Gamma_{2-10keV}$ & L$_{2-10keV}$ & $\kappa_{2-10keV}$ & $\lambda$L$_{2500\AA}$ & $\nu$L$_{2keV}$ & $\alpha_{ox}$ & $\lambda$L$_{5100}$ & $\kappa_{5100}$ & FWHM$_{H\beta}$ & L$_{bol}$/L$_{Edd}$ \\

 & & $\times{\it 10}^{\it 44}$ & & $\times{\it 10}^{\it 43}$ & $\times{\it 10}^{\it 43}$ & & $\times{\it 10}^{\it 44}$ & & {\it km s}$^{\it -1}$ & \\

\hline
1 & 1.69$\pm$0.06 & 4.941 & 11.9 & 81.3 & 25.6 & 1.19 & 8.15 & 7.24 & 13000 & 0.11 \\
2 & 1.67$\pm$0.10 & 0.469 & 17.7 & 18.4 & 2.47 & 1.33 & 0.791 & 10.5 & 6220 & 0.089 \\
3 & 1.77$\pm$0.07 & 0.289 & 149 & 41.0 & 1.91 & 1.51 & 1.35 & 31.7 & 2310 & 1.3 \\
4 & 1.80$\pm$0.11 & 0.567 & 23.6 & 12.8 & 3.15 & 1.23 & 1.91 & 6.98 & 10800 & 0.017 \\
5 & 2.10$\pm$0.22 & 2.284 & 43.2 & 134 & 12.9 & 1.39 & 5.48 & 18.0 & 2720 & 1.0 \\
6 & 1.93$\pm$0.18 & 4.855 & 22.9 & 290 & 27.6 & 1.39 & 14.8 & 7.52 & 5430 & 0.27 \\
7 & 2.39$\pm$0.22 & 0.267 & 657 & 61.3 & 2.43 & 1.54 & 1.95 & 89.6 & 1980 & 13 \\
8 & 1.84$\pm$0.04 & 0.418 & 87.7 & 23.5 & 2.89 & 1.35 & 0.539 & 68.1 & 2840 & 2.3 \\
9 & 1.76$\pm$0.01 & 0.839 & 53.8 & 22.7 & 5.35 & 1.24 & 0.113 & 399 & 3030 & 3.8 \\
10 & 1.92$\pm$0.05 & 3.532 & 23.1 & 205 & 23.1 & 1.36 & 7.59 & 10.8 & 4810 & 0.21 \\
11 & 1.71$\pm$0.11 & 1.811 & 29.8 & 78.9 & 9.03 & 1.36 & 3.75 & 14.4 & 5640 & 0.66 \\
12 & 1.68$\pm$0.23 & 0.502 & 18.9 & 21.2 & 1.57 & 1.43 & 1.04 & 9.12 & 4390 & 0.12 \\
13 & 1.37$\pm$0.12 & 0.751 & 227 & 790 & 2.99 & 1.93 & 42.6 & 4.00 & 10800 & 0.082 \\
14 & 1.69$\pm$0.04 & 3.189 & 10.6 & 50.2 & 17.0 & 1.18 & 3.91 & 8.60 & 7060 & 0.15 \\
15 & 2.35$\pm$0.12 & 0.042 & 251 & 2.89 & 0.353 & 1.35 & 0.204 & 51.1 & 988 & 4.7 \\
16 & 1.78$\pm$0.07 & 1.502 & 24.2 & 90.8 & 8.24 & 1.40 & 4.26 & 8.53 & 3560 & 0.13 \\
17 & 1.80$\pm$0.20 & 0.779 & 46.9 & 71.7 & 3.62 & 1.50 & 3.31 & 11.1 & 2250 & 0.30 \\
18 & 2.23$\pm$0.08 & 1.254 & 61.5 & 157 & 9.67 & 1.46 & 6.11 & 12.6 & 2310 & 0.40 \\
19 & 1.98$\pm$0.18 & 0.084 & 43.1 & 8.59 & 0.497 & 1.47 & 0.443 & 8.19 & 2000 & 0.054 \\
20 & 2.34$\pm$0.12 & 0.053 & 224 & 4.44 & 0.476 & 1.37 & 0.215 & 55.4 & 774 & 3.1 \\
21 & 1.70$\pm$0.04 & 3.856 & 16.5 & 109 & 20.5 & 1.28 & 2.22 & 28.6 & 6090 & 0.51 \\
22 & 1.70$\pm$0.09 & 0.396 & 34.1 & 27.3 & 2.17 & 1.42 & 0.983 & 13.8 & 7050 & 0.15 \\
23 & 1.80$\pm$0.19 & 0.145 & 37.5 & 11.5 & 0.907 & 1.42 & 0.708 & 7.66 & 1980 & 0.043 \\
24 & 1.83$\pm$0.18 & 4.735 & 9.84 & 106 & 25.1 & 1.24 & 6.64 & 7.01 & 13900 & 0.20 \\
25 & 1.88$\pm$0.03 & 3.054 & 30.9 & 249 & 20.0 & 1.42 & 8.44 & 11.2 & 4980 & 0.24 \\
26 & 2.09$\pm$0.25 & 0.362 & 243 & 63.3 & 2.60 & 1.53 & 2.04 & 43.2 & 1720 & 2.9 \\
27 & 1.94$\pm$0.04 & 0.277 & 51.5 & 20.3 & 2.51 & 1.35 & 0.988 & 14.4 & 4310 & 0.34 \\
28 & 1.71$\pm$0.14 & 2.951 & 11.2 & 63.6 & 13.2 & 1.26 & 4.80 & 6.91 & 4240 & 0.15 \\
29 & 2.00$\pm$0.12 & 0.726 & 76.0 & 76.3 & 4.75 & 1.46 & 3.25 & 17.0 & 3560 & 0.73 \\
30 & 2.46$\pm$0.09 & 0.146 & 207 & 13.4 & 1.28 & 1.39 & 0.452 & 67.2 & 954 & 3.8 \\
31 & 1.69$\pm$0.14 & 2.420 & 14.9 & 53.6 & 11.9 & 1.25 & 6.49 & 5.54 & 6810 & 0.055 \\
32 & 1.88$\pm$0.03 & 0.464 & 15.8 & 13.7 & 2.97 & 1.26 & 0.512 & 14.3 & 3100 & 0.12 \\
33 & 2.14$\pm$0.21 & 1.157 & 41.0 & 69.7 & 7.55 & 1.37 & 4.03 & 11.8 & 5690 & 0.60 \\
34 & 1.90$\pm$0.14 & 2.489 & 22.2 & 140 & 13.5 & 1.39 & 10.8 & 5.13 & 3310 & 0.082 \\
35 & 1.76$\pm$0.07 & 3.918 & 17.9 & 67.5 & 51.5 & 1.04 & 3.59 & 19.5 & 2790 & 1.2 \\
36 & 2.20$\pm$0.08 & 0.091 & 76.3 & 3.31 & 0.651 & 1.27 & 0.244 & 28.4 & 1890 & 1.6 \\
37 & 1.95$\pm$0.08 & 1.768 & 22.3 & 88.8 & 12.3 & 1.33 & 5.39 & 7.30 & 3960 & 0.18 \\
38 & 1.55$\pm$0.09 & 0.175 & 10.3 & 1.89 & 0.768 & 1.15 & 0.197 & 9.16 & 6630 & 0.028 \\
39 & 2.17$\pm$0.20 & 0.079 & 96.5 & 6.89 & 0.737 & 1.37 & 0.233 & 32.6 & 991 & 0.56 \\
40 & 2.02$\pm$0.06 & 0.468 & 200 & 70.0 & 3.54 & 1.50 & 2.05 & 45.7 & 2790 & 2.8 \\
41 & 1.94$\pm$0.05 & 2.444 & 26.7 & 167 & 15.8 & 1.39 & 6.26 & 10.4 & 2610 & 0.20 \\
42 & 1.76$\pm$0.11 & 0.158 & 15.5 & 4.92 & 0.804 & 1.30 & 0.265 & 9.28 & 4920 & 0.034 \\
43 & 1.74$\pm$0.07 & 4.524 & 14.3 & 109 & 25.7 & 1.24 & 4.36 & 14.9 & 4550 & 0.43 \\
44 & 2.25$\pm$0.05 & 0.236 & 292 & 45.5 & 2.13 & 1.51 & 2.36 & 29.2 & 1070 & 2.9 \\
45 & 1.82$\pm$0.06 & 17.502 & 16.6 & 645 & 98.4 & 1.31 & 30.4 & 9.58 & 10900 & 0.53 \\
46 & 1.81$\pm$0.12 & 2.175 & 8.93 & 19.0 & 10.4 & 1.10 & 2.97 & 6.55 & 9930 & 0.041 \\
47 & 1.45$\pm$0.25 & 0.868 & 24.9 & 36.6 & 4.39 & 1.35 & 0.931 & 23.2 & 4100 & 0.18 \\
48 & 2.03$\pm$0.11 & 0.101 & 73.2 & 8.71 & 0.734 & 1.41 & 0.278 & 26.7 & 1190 & 0.37 \\
49 & 2.40$\pm$0.22 & 0.200 & 36.8 & 14.5 & 1.69 & 1.36 & 0.719 & 10.2 & 1340 & 0.11 \\
50 & 2.41$\pm$0.18 & 3.299 & 411 & 860 & 27.3 & 1.57 & 29.5 & 46.0 & 2200 & 14 \\
51 & 1.67$\pm$0.03 & 1.659 & 6.50 & 12.5 & 8.30 & 1.07 & 0.624 & 17.3 & 11100 & 0.19 \\
\hline\hline
   \end{tabular}
  \\
 \end{minipage}
\end{table*}


\section{The Broadband SED Modeling}
\label{SED-modelling}

\subsection{Data Preparation}
For each object we extracted the original data files (ODFs) 
and the pipeline products (PPS) from XMM-Newton Science Archive (XSA)
\footnote{http://xmm.esac.esa.int/external/xmm\_data\_acc/xsa/index.\\shtml}. 
In the following data reduction process, tasks from XMM-Newton Science
Analysis System (SAS) v7.1.0 were used. 
First, {\tt EPCHAIN/EMCHAIN} tasks were used to extract events 
unless the events files had already been extracted
for each exposure by PPS. Then {\tt ESPFILT} task was used to define
background Good Time Intervals (GTIs) that are free from flares. In each
available EPIC image, a 45\arcsec radius circle was used to extract the source
region, and an annulus centered on the source with inner and outer
radii of 60\arcsec and 120\arcsec was used to define the background region.
For other sources listed in the region files of PPS that are included in 
these regions, these were subtracted using the default radii generated by PPS, 
which scaled with the source brightness. Then the GIT filter, source and
background region filters were applied to the corresponding events files to
produce a set of source and background events files. We only accepted photons with
quality flag =0 and pattern 0$\sim$4. The {\tt EPATPLOT} task was then used to
check for pile-up effects. When pile-up was detected, an annulus with inner and
outer radii of 12\arcsec and 45\arcsec was used instead of the previous
45\arcsec radius circle to define the source region. Then source events files were
reproduced using the new source region filter. Source and background spectra
were extracted from these events files for each available EPIC exposure.
Tasks {\tt RMFGEN/ARFGEN} were used to produce response matrices and
auxiliary files for the source spectra. These final spectra
were grouped with a minimum of 25 counts per bin using the {\tt GRPPHA v3.0.1}
tool for spectral fitting in {\sc Xspec v11.3.2}. To prepare the OM data, 
the {\it om\_filter\_default.pi} file and all response files for the V,B,U, UVW1,
UVM2, UVW2 filters were downloaded from the OM response file directory in
HEASARC Archive\footnote{http://heasarc.gsfc.nasa.gov/FTP/xmm/data/responses/om/}.
We then checked the OM source list file for each object to see if there were
any available OM count rates. Each count rate and its
associated error were entered into the {\it om\_filter\_default.pi} file and then
combined with the response file of the corresponding OM filter, again by 
using the {\tt GRPPHA} tool to produce OM data that could be used in 
{\sc Xspec}.

Finally, the XMM-Newton EPIC spectra are combined 
with the aperture corrected OM photometric points, and the optical 
continuum points produced from the optical underlying continuum (obtained
from the full optical spectrum fitting) using {\tt FLX2XSP} tool.
From these data we constructed a broadband nuclear SED of each AGN. 
There is a ubiquitous data gap in the far UV region which is due to
photo-electric absorption by Galactic gas. Unfortunately, in most
cases of low-redshift AGN, their intrinsic SED also peaks in this very
UV region, and so this unobservable energy band often conceals a large
portion of the bolometric luminosity. In order to account for this,
and to estimate the bolometric luminosity, we fit the X-ray and
UV/optical continua all together using a new broadband SED model
(\citealt{Done11}, {\sc Xspec} model: {\it optxagn\/}).
We then
calculate the bolometric luminosity by summing up the integrated
emission using the best-fit parameters obtained for each continuum
component.

\subsection{The Broadband SED Model}
\label{broadband-sed-model}

A standard interpretation of the broadband SED is that the
emission is dominated by a multi-temperature accretion disc component which
peaks in the UV (e.g. \citealt{Gierlinski99}, {\sc Xspec} model: {\it
diskpn}). This produces the seed photons for Compton up-scattering by a
hot, optically thin electron population within a corona situated above the disc,
resulting in a power law component above 2~keV (e.g. \citealt{Haardt91};
\citealt{Zdziarski00}, {\sc Xspec} model: {\it bknpl}).  However, the
X-ray data clearly show that there is yet another component which
rises below 1~keV in almost all high mass accretion rate AGNs. The ubiquity of
this component can be seen, for example, in the compilation of AGN SEDs presented 
in \citet{Middleton07}, and one of the strongest cases is the  
NLS1 RE J1034+396 (\citealt{Casebeer06};
\citealt{Middleton09}) and RX J0136.9-3510 (\citealt{Jin09}). The
origin of this so-called soft X-ray excess is still unclear 
(e.g. \citealt{Gierlinski04}; \citealt{Crummy06}; \citealt{Turner07};
\citealt{Miller08}), and so some previous
broadband SED modeling studies have explicitly excluded data below 1~keV.
An obvious consequence is that in such studies a soft excess   
component cannot influence the models, so making it
possible to fit the data using just a disc and (broken) power law
continuum (VF07; \citealt{Vasudevan09}). However, in our current 
study we include all of the data, and so we require a self-consistent model which 
incorporates this soft component.

Whatever the true origin of the soft X-ray excess, the simplest model which can 
phenomenologically fit its shape is the optically thick, low temperature
thermal Comptonisation model ({\tt compTT}). But the observed data are used to 
constrain the three separate components, {\tt discpn + compTT + bknpl}, 
which is generally problematic given the 
gap in spectral coverage between the UV and soft X-ray regions caused
by interstellar absorption. So instead, we combine these three components together
using a local model in {\sc xspec}, assuming that they are all 
ultimately powered by gravitational energy released in accretion.
A complete description of this model can be found in the {\sc Xspec} website$^{5}$ and is also given in
\citet{Done11}. It is in essence a faster version of the models 
recently applied to black hole binary spectra observed close
to their Eddington limit (\citealt{Done06}) and to the (possibly
super Eddington) Ultra-Luminous X-ray sources (\citealt{Gladstone09}; 
\citealt{Middleton10}), thus this model is more appropriate for fitting a medium sized
sample of objects. A comprehensive comparison with the model of \citet{Done06} is given in \citet{Done11}.
To make this paper self contained we give a brief synopsis
of the model. We assume that the gravitational energy released in the disc 
at each radius is emitted as a blackbody only down to a given radius, $R_{corona}$. 
Below this radius, we further assume that the energy can no longer be completely thermalised,
and is distributed between the soft excess component and the high energy tail.
Thus the model includes all three components which are known to
contribute to AGN SED in a self consistent way. As such
it represents an improvement on the fits in VF07 in several respects, 
by including the soft excess and by requiring energy conservation, and it improves on
\citet{Done06} by including the power law tail.

In our SED fitting, the optical/UV data constrains the mass accretion rate through the 
outer disc, provided we have an estimate of the black hole mass.
We constrain this by our analysis of the $H\beta$ emission line
profile. The main difference from previous studies based on non-reverberation samples
is that we do not directly use the FWHM of the H$\beta$ profile to derive the black hole mass.
Rather, we use the FWHM of the intermediate and broad line 
component determined from the emission line fitting results presented in
Section~\ref{emission line fitting}. These are then used in Equation~\ref{BH-mass}
(\citealt{Woo02} and references therein) to derive the black hole mass limits required for
the SED fitting:
\begin{equation}
\label{BH-mass}
M_{BH}=4.817{\times}[{{{\lambda}L_{\lambda}(5100{\AA})}\over{10^{44}erg s^{-1}}}]^{0.7}{FWHM}^{2}
\end{equation}
where $L_{\lambda}$(5100{\AA}) is measured directly from the SDSS spectra.
The rms difference between the black hole masses from this equation and from the reverberation mapping 
study is $\sim$0.5 dex. Thus we also adopted any best-fit values that fell below the original lower limit 
(which was set by FWHM of the intermediate component) by less than 0.5 dex. 
With this method, the best-fit black hole mass found by SED fitting is always consistent 
with the prediction from the H$\beta$ profile.
Section~\ref{BH mass compare} discusses the differences between the best-fit black hole
masses and those estimated using other methods.

Once the black hole mass is constrained, the optical data then sets the mass
accretion rate $\dot{M}$, and hence the total energy available is
determined by the accretion efficiency. We assume a stress-free
(Novikov-Thorne) emissivity for a Schwarzschild black hole, i.e. an
overall efficiency of 0.057 for $R_{in}=6R_g$. Thus the total
luminosity of the soft excess and power law is
$0.057\dot{M}c^2 (1-R_{in}/R_{corona})$. This constrains the model in
the unobservable EUV region, with the input free parameter $R_{corona}$
setting the model output of the luminosity ratio between the standard disc emission and
Comptonisation components. The upper limit of $R_{corona}$ is set to be 100 $R_{g}$,
which corresponds to 81\% released accretion disc energy. This upper limit is based
on the requirement that the seed photons should be up-scattered (\citealt{Done11}).
We assume that both the Comptonisation components 
scatter seed photons from the accretion disc with temperature corresponding to
$R_{corona}$. The other model input parameters are; the temperature ($kT_e$) and 
optical depth ($\tau$) of the soft Comptonisation component which are determined by the 
shape of the soft X-ray excess, the spectral index ($\Gamma$) of the 
hard X-ray Comptonisation that produces the 2-10~keV power law, 
with electron temperature fixed at 100~keV. The model output $f_{pl}$ represents 
the fraction of the non-thermalised accretion energy (i.e. given by the 
luminosity originating from the region of $R_{corona}$ to $R_{in}$), 
which is emitted in the hard X-ray Comptonisation.

We also included two sets of corrections for attenuation ({\it reddening}-{\it wabs}),
to account for the line of sight Galactic absorption and for the 
absorption intrinsic to each source, the latter is redshifted ({\it zred} and {\it zwabs} in
{\sc Xspec}). The Galactic HI column density is fixed at the value
taken from \citet{Kalberla05}, but the intrinsic HI column density is left as
a free parameter.
The standard dust to gas conversion formula of E(B-V)=$1.7\times 10^{-22} N_H$ (\citealt{Bessell91}) is used 
for both Galactic and intrinsic reddening. We
set the initial value of the powerlaw photon index to be that of the
photon index in 2-10 keV energy band, but it can vary during the fitting
process. However, we set an upper limit of 2.2 for the powerlaw photon
index, not only because the photon index is $<$ 2.2 
for the majority of Type 1 AGNs (\citealt{Middleton07}),
but also because otherwise the much higher signal-to-noise 
in the soft excess in some observed spectra can artificially steepen
the hard X-ray powerlaw and result in unphyiscal best-fit models.

All free parameters used in the broadband SED fitting are listed in 
Table~\ref{SED-fitting-parameters}. For completeness,
we also explicitly calculate the fraction of the total luminosity carried by each
component of the model 
(i.e. disc: $f_{d}$; soft Comptonization: $f_{c}$; hard X-ray 
Comptonization: $f_p$) from the model fit parameters R$_{cor}$ and
F$_{pl}$ (see Table~\ref{SED-fitting-parameters})
\footnote{a full description of the model parameters can be found on the {\sc Xspec\/}
web page: \\http://heasarc.nasa.gov/xanadu/xspec/models/optxagn.html}.
Table~\ref{SED-key-parameters} lists
the important characteristic parameters. The main uncertainty
in these parameters, especially the black hole mass,  
is dominated by other systematic uncertainties introduced by 
the observational data, model assumptions (e.g. the assumption of 
a non-spinning black hole and the inclination dependence of the disc emission) 
and the analysis methods involved. Therefore the parameter fitting uncertainties
which are often less than 10\%, are not significant in comparison, and thus are not listed. 
The statistical properties of these parameters are discussed in 
section~\ref{stat-property}.

\begin{figure*}
\centering
\includegraphics[bb=40 460 595 842,scale=0.95,clip=, angle=0]{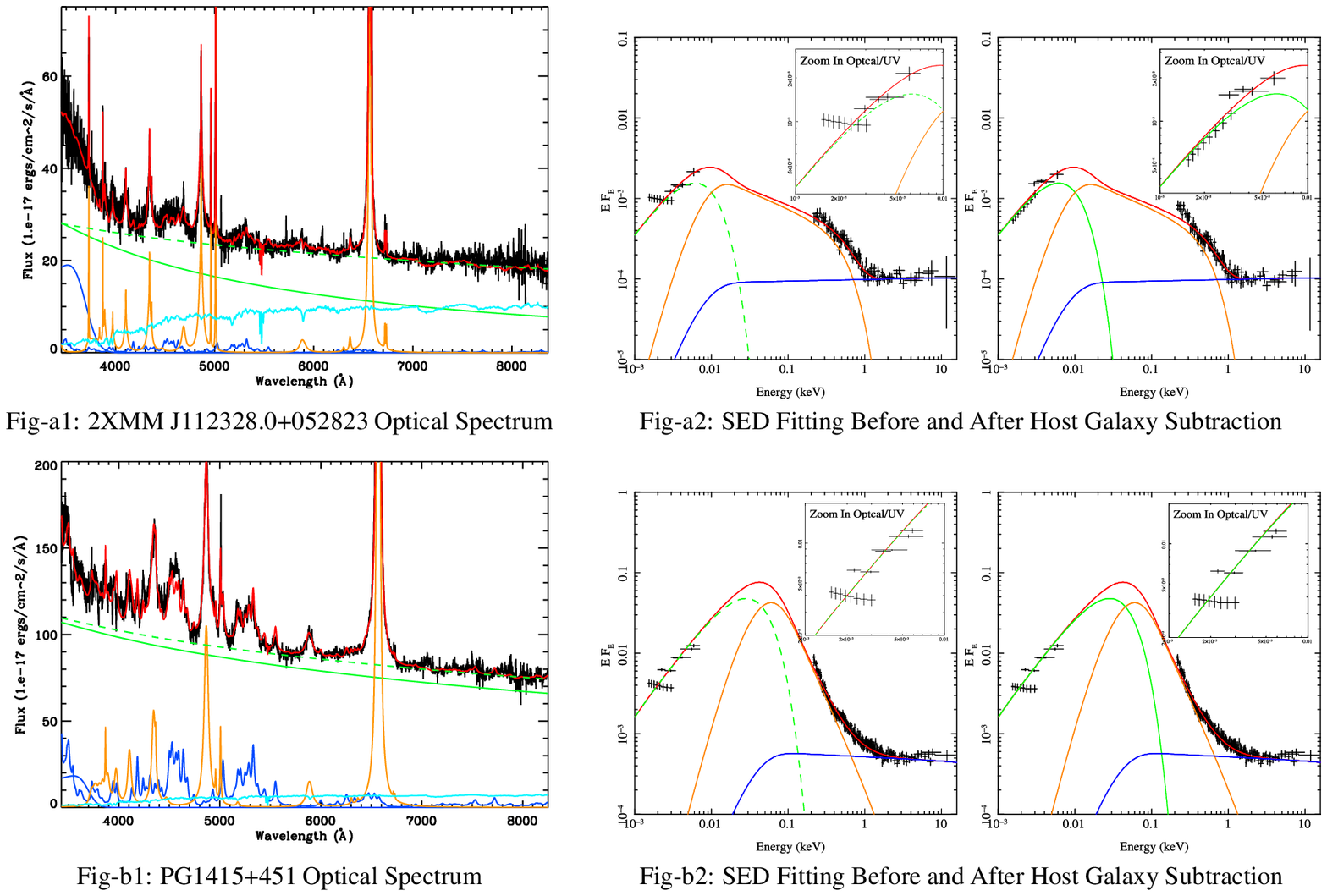}
\caption{A comparison between the results of two subtractions of host galaxy contribution.
2XMM J112328+052823 (Fig-a1 and Fig-a2) shows an underlying continuum 
that more closely resembles a disc continuum (solid green line in Fig-a1)
after modelling and subtracting the
host galaxy contribution (light blue spectrum in Fig-a1).
The left panel of Fig-a2 shows the original broadband SED fitting without subtracting the host
galaxy contribution. The dashed green line shows the modelled accretion disc emission in the best-fit SED.
The inserted panel shows a magnification of the fit in the optical/UV region, where a big discrepancy
exists between the SDSS data and best-fit SED model.
The right panel of Fig-a2 is the new SED fit using the new
underlying disc continuum (shown as solid green line in Fig-a1)
after subtracting the host galaxy contribution. The new fit is improved in the optical
region compared with the previous results in the left panel of Fig-a2.
In contrast to the above example, PG 1415+451 (Fig-b1 and Fig-b2)
has little host galaxy contribution in the SDSS optical spectrum (see the light blue component
in Fig-b1), and its broadband SED fitting in the optical region remains poor regardless of the
amount of host galaxy subtraction applied (see the two panels in Fig-b2).
The spectral template for Elliptical galaxies in \citet{Kinney96} was used in both cases since their
host galaxies both have elliptical morphologies in SDSS image.}
\label{hostgalaxy-example}
\end{figure*}
\subsection{Problems in The SED Fitting}

We further discuss two problems we encountered during the fitting procedure in
the following subsections. The first problem is the discrepancy between
the OM and SDSS continuum points (mentioned in
Section~\ref{OM:correction}). The second problem is that of the 
observed flat optical continuum, whose shape cannot be accounted 
for in our SED model (mentioned in Section~\ref{HG:Contamination}).

\subsubsection{The discrepancy between the OM photometry and the SDSS continuum}
\label{discrepancy}
There remains a significant discrepancy between many of the OM and
SDSS continuum points, even after applying the aperture correction
discussed in Section~\ref{aperture correction} (see
Figure~\ref{SED}). The OM points often appear above (brighter) the 
extrapolation of the SDSS continuum to the OM wavelengths.  We identify
three possible reasons for this discrepancy:

(1) Remaining aperture
effects: There is an aperture difference between the SDSS fibres
(3{\arcsec}diameter) and the OM apertures we used
(6{\arcsec}diameter). Clearly the OM points will still include more
host galaxy starlight than the SDSS points, and so will appear above
the SDSS spectrum.

(2) Contamination from emission lines: The wavelength
ranges for each OM filter (over which the effective transmission is
greater than 10\% of the peak effective transmission) are as follows:
UVW2 1805-2454\AA, UVM2 1970-2675\AA, UVW1 2410-3565\AA, U
3030-3890\AA, B 3815-4910\AA, V 5020-5870\AA. We exclude the 
contribution from strong optical emission lines within 
the OM U, B, V bandpass (and also the Balmer continuum contribution in U band)
by using the best-fit optical underlying continuum which excludes such 
features from the SDSS spectral fitting. In fact, this was an important initial motivation
of the study, i.e. to obtain more accurate estimates
of the true underlying continuum rather than simply to use the SDSS `ugriz'
photometric data. Inclusion of strong emission lines within
these photometric data would result in over-estimation of the optical 
continuum, and so compromise our aim to study the shape of the optical underlying continuum.
This is an important spectral characteristic used to constrain the
accretion disc component in the SED fitting (see also  
the discussion in section 5.1.2). There are some strong
emission lines within the UV bandpasses such as Ly$\alpha$, CIV 1549,
CIII 1909 and MgII 2798, whose fluxes are not available from SDSS spectrum.
Accurate subtraction of these line fluxes for each object would  
require new UV spectroscopy. We conclude that inclusion of emission line 
flux within the OM photometric points may account for some of the 
observed discrepancy.

(3) Intrinsic source variability: AGN are well known to be variable 
across their SEDs. In general there is a significant 
time difference between acquistion of the SDSS and OM-UV data, so 
intrinsic variation may contribute to any observed
discrepancy. Mrk 110 is the most extreme example of this phenomena in our sample, as its SDSS
spectrum has a very large discrepancy compared with the OM data. The recent paper by \citet{Landt11}
gives another set of optical spectra for Mrk 110, which  
is more consistent with our best-fit model.
It shows that the inclusion of OM data is useful to help identify cases such as this.
As an additional test for variability, we assembled all available
GALEX data for our sample. We find that 43 objects in our sample have
GALEX data. Using a GALEX aperture of 12\arcsec, which is limited by the PSF and which is also similar to 
the UV OM apertures, we compare these values with the SED model. The ratio of the GALEX
data and our SED model
within the same bandpass differ by less than a factor of 2 for the
majority of our sample, and significantly the flux ratio distribution is almost symmetric
and is centered close to unity. This suggests that the non-simultaneous OM and
SDSS data is not likely to be a major impediment to our modeling.

In effect, these three factors will merge together to produce the observed
discrepancy between the SDSS and OM data. Since the combined effects
of Point (1) and (2) which will add flux and generally be greater than that caused by optical/UV variability
as shown by previous long term reverberation mapping studies
(\citealt{Giveon99}; \citealt{Kaspi00}), we should 
treat the OM points included in our SED modeling as upper limits when
interpreting the results of our modeling. Indeed, the 90\%
confidence uncertainties in the BH masses derived directly from the {\sc Xspec\/}
fitting are almost certainly small compared with the systematic
errors introduced by the above uncertainties.

\subsubsection{The observed flat optical continuum}
\label{FOC-problem}
A related problem in our fitting is about the SDSS continuum
shape. For some AGNs, their SDSS continuum data points exhibit a very 
different spectral slope from that of the SED model. This cannot
be reconciled by adjusting the parameters of the accretion disc model, and thus
implies the presence of an additional component at longer optical wavelengths,  
which flattens compared with that predicted by the accretion disc models. One obvious explanation
for this flux excess is the contribution from the host galaxy. In 
late type host galaxies such as elliptical and S0 galaxies,
emission from their old stellar populations peaks at near
infrared wavelengths. \citet{Kinney96} combined spectra of quiescent
galaxies and constructed an average spectral template for each
morphological type, including bulge, elliptical, S0, Sa, Sb, Sc and
starburst galaxies. For some objects in our sample with high S/N SDSS spectra which show
at least marginal stellar absorption features, we have added the
corresponding type of host galaxy spectral template taken from
\citet{Kinney96}, into the overall SDSS spectral fitting. This revised
underlying continuum in the optical, and was then used in the
broadband SED fitting. We are then able to compare it with the original
fit, to see how the subtraction of a stellar population template
effects the overall SED fitting.

Figure~\ref{hostgalaxy-example} shows two examples. The first
is 2XMM J112328.0+052823, in which
after subtracting the host galaxy component, the observed optical
continuum is closer to the slope of the SED model. 
However, the results for PG1415+451 in Figure~\ref{hostgalaxy-example}
lower panel imply that its host galaxy cannot be the origin of the
flat optical spectrum. The reason is that its optical spectrum does not show
any strong stellar absorption features. This means that the maximum
amount of host galaxy contribution is small, and so and there remains
a substantial inconsistency in the slope versus the SED model. In addition to
2XMM J112328.0+052823 above, only Mrk1018 and
2XMM J125553.0+272405 show clear stellar absorption
features. Also the 3\arcsec diameter fibre excludes much of the host
galaxy component at these redshifts. Therefore, on these general grounds we
conclude that host galaxy contamination is small for most sources in our sample,  
and consequently cannot fully account for the observed flat optical continuum.
Additional support for this view comes from good correlations between the X-ray components
and the red optical continuum, suggesting that this extra optical flux is
likely related to the intrinsic activity 
(e.g. \citealt{Soria02}; \citealt{Collin04}; \citealt{Hao10}; \citealt{Landt11}).

\begin{figure*}
\centering
\includegraphics[scale=0.75,clip=1]{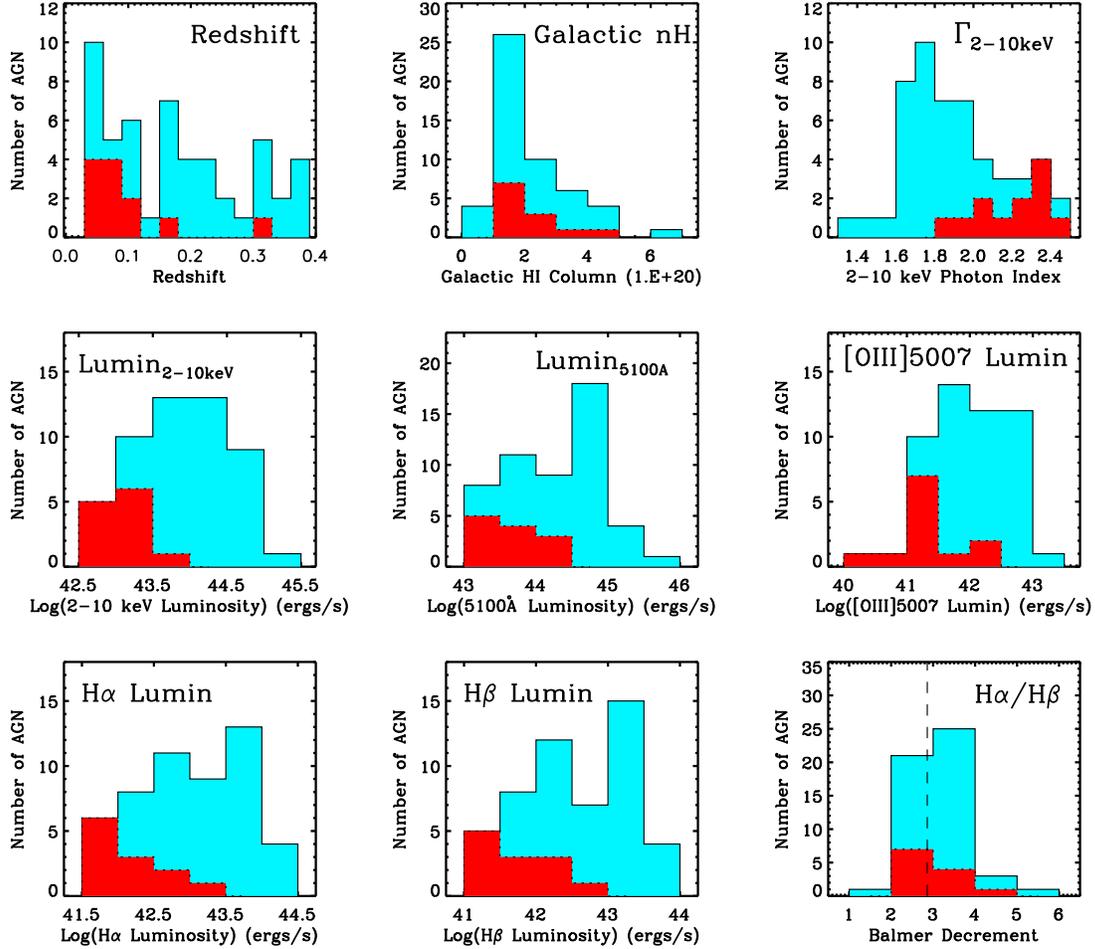}
\caption{Distributions of our sample for different properties. In each panel the blue
areas show the distribution for the whole sample, while the red areas
show the distribution for the 12 NLS1s in our sample.
 We note that the H$\alpha$, H$\beta$ and [OIII] $\lambda$5007 luminosities are based on 
results of line profile fitting, after subtracting the blends from other nearby emission lines
(see Section~\ref{emission line fitting}).
For comparison we also indicate the Balmer decrement value of 2.86, found under case B recombination,
as shown by dashed line in the same panel.}
\label{all-histo}
\end{figure*}
\begin{figure*}
\centering
\includegraphics[scale=0.75,clip=1]{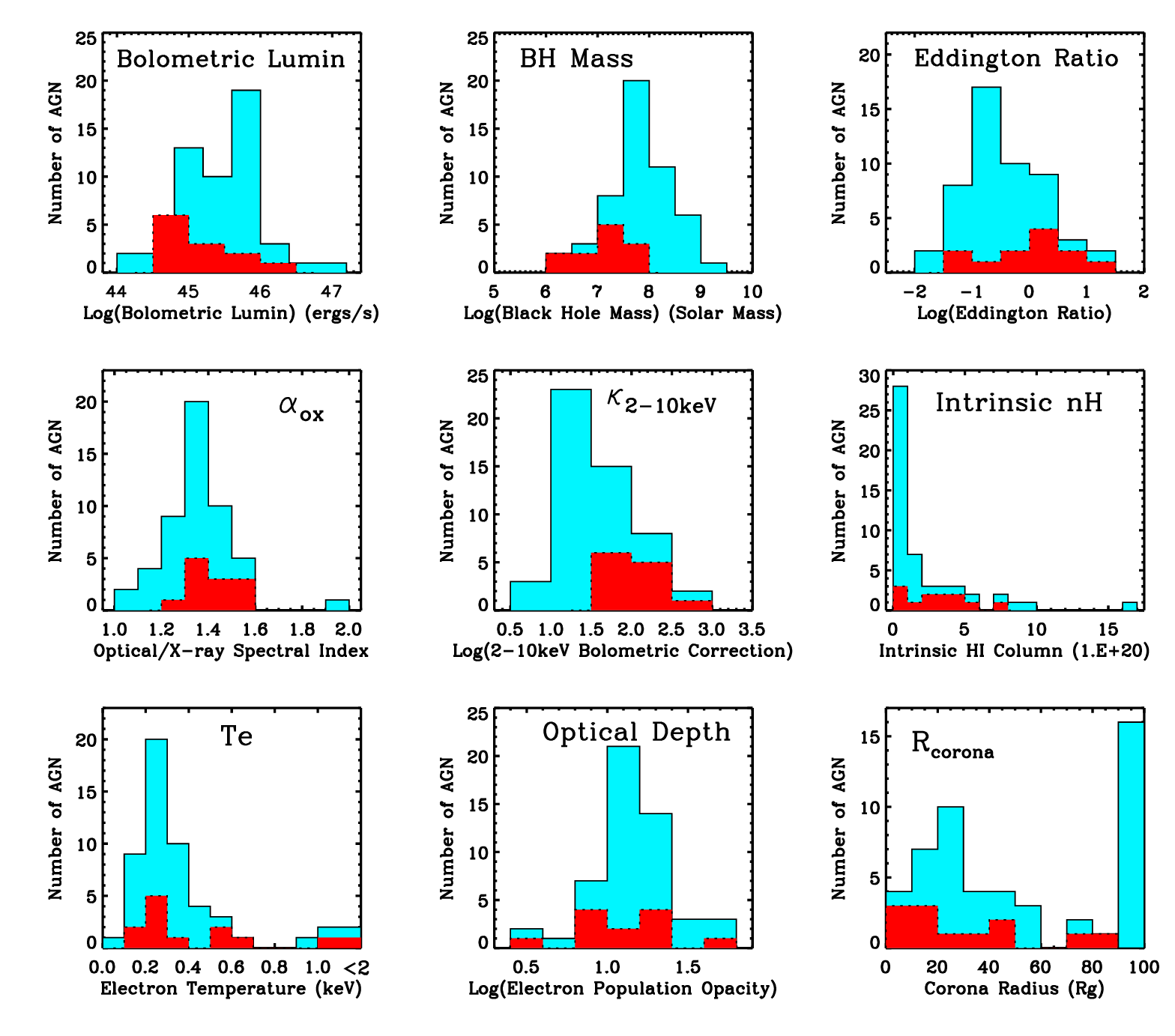}
\caption{The distribution of model dependent parameters using the same
colour coding as in Figure~\ref{all-histo}. Comments on each distribution
are given in Section~\ref{model-para1}.}
\label{model-histo}
\end{figure*}
\section{Statistical properties of the sample}
\label{stat-property}
Histograms of data on our sample are shown in
Figure~\ref{all-histo}, Figure~\ref{balmer-histo} and
Figure~\ref{bolometric-histo}, including redshift, HI column density,
optical and X-ray modeling parameters etc. The red region in
the histograms show the distributions for the 12 NLS1s in our sample. It is
clear that NLS1s are distinct among the whole sample in several respects.

\subsection{General Properties}

Figure~\ref{all-histo} shows some basic properties of our sample which
are not model dependent:\\
(1). Redshift: the sample's redshift ranges from 0.031 (Mrk 493) to
0.377 (HS 0810+5157). The NLS1s are found mainly at lower redshifts,
with $<z>_{n} = 0.12$ compared to the $<z>_{n} = 0.19$ for the 
BLS1s. For comparison we see that the sample of VF07 has a similar redshift
range, but it has a lower average redshift of 0.10.\\ 
(2). The Galactic nH: the average Galactic nH is $2.25{\times}10^{20}$.\\ 
(3). The photon indexes obtained from simple power law fits to the restricted energy range
of 2-10~keV. The NLS1s cluster on the higher photon index side, with an
average of $2.21{\pm}0.20$, which differs from the sample
average of $1.92{\pm}0.25$ and the BLS1s' average of $1.83{\pm}0.18$.
This means that NLS1s tend to have softer
X-ray spectra, which is further confirmed in the following section 
on the mean SEDs.\\ 
(4). The X-ray continuum and 2-10 keV luminosity: this
distribution shows that NLS1s have lower 2-10 keV luminosities in spite of
their steeper slopes. We note that the VF07 sample has a similar
distribution, except for their inclusion of three extremely low X-ray luminosity
AGN (i.e. NGC4395, NGC3227 and NGC6814), these objects were not included
in our sample due to our selection criteria and/or a lack of
SDSS spectra.\\ 
(5). The optical continuum luminosity at 5100 \AA. On average the NLS1 have lower optical
luminosities than BLS1. \\
(6-8). The [OIII] $\lambda$5007, H$\alpha$ and  H$\beta$ emission line luminosities.
Again the NLS1s have on average lower luminosities than BLS1s. \\
(9). The Balmer decrement. The average value for the whole sample is 3.14$\pm$0.62,
and for NLS1s is 3.05$\pm$0.38. This difference is not statistically 
significant, but we return to the issue in our next paper (paper II), 
where we consider the separate components as well as the overall profile.

\subsection{Results from The Broadband SED Modeling}
\label{model-para1}
\begin{figure*}
\centering
\includegraphics[scale=0.6,clip=1,angle=90]{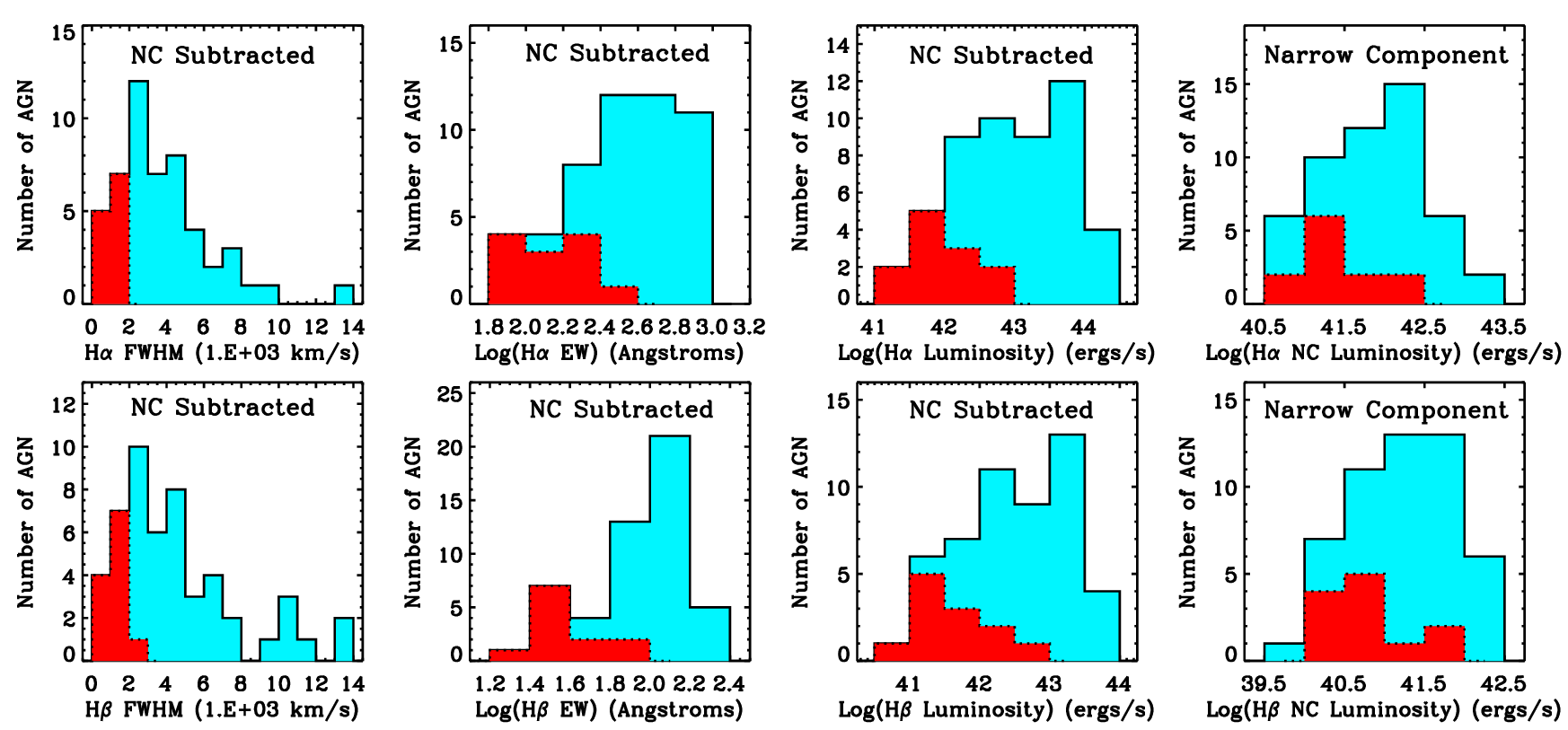}
\caption{The Balmer line parameter distributions. The first row is for H$\alpha$ and
the second is for H$\beta$. We combine the intermediate and broad components in each
Balmer line profile to form the total broad line properties, giving values of the FWHM, EW and luminosity.
The final panel shows the luminosity distribution of the narrow component for comparison.
The distributions for the 12 NLS1s are indicated by the
red regions, as in Figure~\ref{all-histo}.}
\label{balmer-histo}
\end{figure*}
Figure~\ref{model-histo} shows properties derived from the SED fits:\\
(1). The bolometric luminosity: the distribution range is between
$1.8{\times}10^{44} ergs$ $s^{-1}$ (Mrk 464) and $1.4{\times}10^{47} ergs$
$s^{-1}$ (PG 2233+134).  There is no clear difference in the
distribution of the complete sample and the sub-set of NLS1s.
The average luminosity is Log(L$_{bol}$)=45.49$\pm$0.55, which is consistent with 
the value of 45.19$\pm$1.01 found in VF07 sample, except for the three extremely 
nearby and low luminosity AGNs in VF07.
\\ 
(2). The black hole mass: using the best-fit black hole masses, 
the whole sample peaks between $10^{7}M_{\odot}$ and
$10^{8}M_{\odot}$. Equation~\ref{BH-mass} suggests that the
black hole mass should depend on both H$\beta$ FWHM and L$_{5100}$, and the results from
our SED fitting suggest that NLS1s with smaller Balmer line FWHM do indeed harbour lower mass
black holes. KUG 1034+396 has the lowest black hole mass in our
sample. The value of $1.7{\times}10^{6}M_{\odot}$ is consistent
with the estimate based on the first firmly detected AGN QPO (quasi periodic
oscillation) found in this source (\citealt{Gierlinski08}). 
Again we can compare our results with those of VF07 sample. We find that their average black 
hole mass is 7.89$\pm$0.82, calculated using the M(L$_{5100}$, FWHM$_{H\beta}$) relation. 
Adopting this same method for our sample, we find a very similar average of 7.99$\pm$0.93. 
Our best-fit masses have a slightly lower average value of 7.83$\pm$0.64 (also
see Section~\ref{BH mass compare} for a comparison of different estimates of black hole 
masses).\\
(3). The Eddington ratio: the
average values are 3.21$\pm$3.07 for NLS1 which display a wide dispersion, and
0.57$\pm$0.50 for BLS1 and 0.93$\pm$0.85 for
the whole sample. Of the eight objects whose Eddington ratios are above 1, six are NLS1
galaxies, and the highest value is 14.2 (PG 2233+134). Clearly, NLS1s tend to
have larger Eddington ratios.
Our Eddington ratio distribution is also similar to that found in the sample of VF07
whose average value is 0.47$\pm$0.44, except that their distribution has a 
more pronounced peak at $\sim$0.1.\\
(4). The $\alpha_{ox}$ index, is defined between restframe continuum points at 2500 {\AA}
and 2 keV (see \citealt{Lusso10} and references therein).
The distribution for NLS1 is peaked at marginally higher values than for BLS1.\\
(5). The $\kappa_{2-10}$ bolometric correction, is defined as $L_{bol}/L_{2-10}$
(see VF07 and references therein).
We find that NLS1s have a significantly higher fraction of their bolometric luminosity
emitted as hard X-rays than the BLS1s.
Compared with the VF07 sample, both distributions peak at 
$\kappa_{2-10}$=10$\sim$30, but our sample shows
a smoother distribution decreasing as $\kappa_{2-10}$ increases after $\sim$30,
and so results in a slightly higher average value of $\kappa_{2-10}$.\\
(6). The intrinsic nH: This distribution shows that the intrinsic equivalent neutral hydrogen column
densities are low for our sample, which is a natural consequence of our initial
sample selection criteria. The NLS1s have slightly higher intrinsic
absorption than BLS1s, which may imply a slightly higher dust
reddening. However the distribution of Balmer decrements shows no significant
difference between these two types of AGNs.\\
(7). The temperature of the Comptonisation component used to describe
the soft X-ray excess. This is close to 0.2~keV in all objects,
confirming the trend seen in previous studies for this component to
exhibit a narrow range of peak energy (\citealt{Czerny03}; \citealt{Gierlinski04}).
The distribution peak at this energy is more marked for
the BLS1 than for NLS1, although the small number statistics means that this difference cannot be considered 
as definitive for our sample.\\  
(8). The optical depth of the soft excess Comptonised
component.  It is clear that this component is always optically thick,
with most objects having $\tau{\sim}10-30$.
There is no significant difference in temperature or optical depth
between the broad and narrow line objects. \\ 
(9). There is a difference in the coronal radii distribution between the BLS1s and NLS1s.
Corona radius controls the relative amount of power emerging from the accretion disc
and the soft X-ray excess/hard tail. There are two peaks in the distribution 
for the broad line objects, one between 10 and 20 R$_{g}$ 
(where $R_g=GM/c^2$), and the other at 100 R$_{g}$
(which is set as the upper limit of this parameter in our
broadband SED model). By contrast these radii in NLS1 are consistent with just
the first peak. 
At first sight this is surprising, since NLS1 are
expected to be those with the strongest soft X-ray excess. However,
their similar soft excess temperatures around 0.2~keV suggests that
atomic processes may be significant (reflection and/or absorption from
partially ionized material), and this may influence our fits.
The average coronal radii are 32$\pm$26 R$_{g}$ for NLS1, 59$\pm$37 R$_{g}$ for BLS1 and 
53$\pm$36 R$_{g}$ for the whole sample. This supports the conclusion of VF07 that
high Eddington ratio AGN have lower coronal fractions compared to those with low Eddington ratios.

\subsection{Balmer Line Parameter Distribution}

Figure~\ref{balmer-histo} shows further details of the modeled
profiles of $H\alpha$ (first row), and $H\beta$ (second row). \\
(1). The FWHM of the broad emission profile. This is calculated from 
co-adding the two best fit Gaussian profiles for the 
broad and intermediate line components, and then using the resultant
profile to determine the FWHM. This is
equivalent to subtracting the narrow line core from the observed profile
and measuring the resultant FWHM. \\
Note, the NLS1s by definition have H${\alpha} < 2000$ km $s^{-1}$.\\
(2-3). The equivalent widths and line luminosities are  
again measured using the total broad emission line profile as above. The NLS1s have  
both lower equivalent widths and line luminosities.\\
(4). By contrast, there is no pronounced difference between NLS1s and BLS1s in
their Balmer narrow line component. This suggests that the narrow line region
is less influenced by whatever difference in properties is responsible for 
the defining difference between NLS1s and BLS1s in the broad line region.

\subsection{The Bolometric Luminosities}
\begin{figure}
\centering
\includegraphics[width=3.4in,height=3.15in,clip=1]{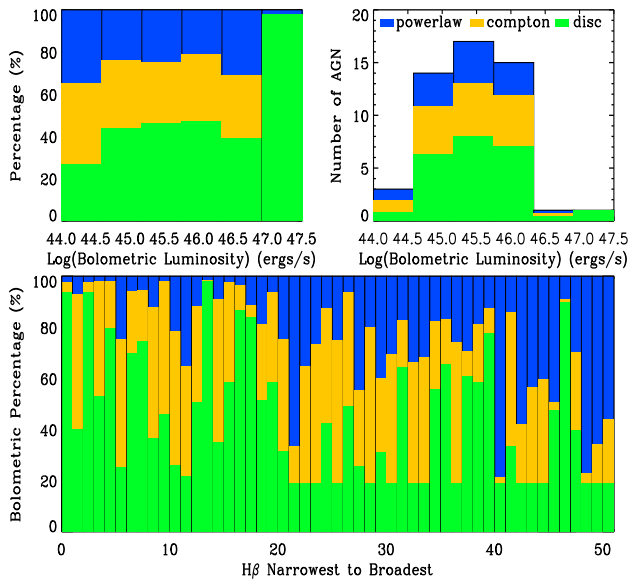}
\caption{The bolometric luminosity distribution for the
different continuum components of the SED, i.e. accretion disc (green),
Comptonisation (orange) and hard X-ray Comptonisation (blue).
The upper left panel shows the percentage within each luminosity bin
for each of these three SED components.
The Upper right panel shows the luminosity distribution of the whole sample,
with each bin divided into three regions according to the fractional
contribution from the different components in that luminosity bin.
The lower panel shows how the contribution from each component changes 
as a function of rank order in H$\beta$ FWHM, after the narrow line component
has been removed.}
\label{bolometric-histo}
\end{figure}

The fraction of the total luminosity contained in each 
component of the SED model 
is shown in Figure~\ref{bolometric-histo}. The upper left panels show
these fractions as a function of the bolometric luminosity.
It seems that as the bolometric luminosity increases, 
the disc component slightly increases in importance.  
However, the total numbers of objects at high luminosities 
is small, as seen in the upper right panels, where the fraction 
is multiplied by the number of objects in the bin, 
so we should be cautious about this finding. 

The lower panel shows this fraction for each of the objects ranking from the smallest to biggest $H\beta$ FWHM. 
Thus low rank objects have the narrowest $H\beta$ (and hence are
by definition NLS1s). These also have the lowest black hole masses 
and highest Eddington ratios. They are more likely
to have a smaller fraction of their total luminosity emitted in the soft X-ray
excess component, than the BLS1s. This relates to the issue of the coronal radii,
see Point (9) of Section~\ref{model-para1}.
There are also some BLS1s which have an apparently high 
fraction of power in their soft X-ray excesses, but they may also have alternative
spectral fits including reflection and/or absorption. 

We note that in all these plots the lower limit to the disc fraction of $0.19$ results from
setting an upper limit of 100 R$_{g}$ for the coronal radius parameter,
as mentioned in Section~\ref{broadband-sed-model}

\begin{table}
 \centering
   \caption{The average black hole masses, as shown in Figure~\ref{bhmass_dist}.}
   \label{BHmass-averages}
     \begin{tabular}{cccc}
\hline\hline
& NLS1 & BLS1 &ALL\\
\hline
$<$M$_{BH, IC}>$ & 6.58$\pm$0.49 & 8.09$\pm$0.56 & 7.73$\pm$0.84\\
$<$M$_{BH, BC}>$ & 7.72$\pm$0.49 & 9.05$\pm$0.55 & 8.74$\pm$0.78\\
$<$M$_{BH, IC+BC}>$ & 6.75$\pm$0.49 & 8.37$\pm$0.65 & 7.99$\pm$0.93\\
$<$M$_{BH, \sigma}>$ & 6.57$\pm$0.46 & 7.89$\pm$0.47 & 7.58$\pm$0.73\\
$<$M$_{BH, FIT}>$ & 7.11$\pm$0.54 & 8.05$\pm$0.48 & 7.83$\pm$0.64\\
$<$M$_{BH, RP}>$ & 7.42$\pm$0.39 & 8.44$\pm$0.53 & 8.20$\pm$0.66\\
\hline\hline
   \end{tabular}
\end{table}

\subsection{The Black Hole Mass}
\label{BH mass compare}
The black hole mass is one of the key parameters used in
our SED fitting, and it largely determines the continuum shape in
the optical/UV region. The masses derived from reverberation mapping are
considered to be the most accurate, but the total number of objects
which have been studied using this technique is still relatively small
(e.g. \citealt{Peterson04}; \citealt{Denney10}; \citealt{Bentz10}).
In the absence of reverberation mapping, the empirical relation between
M$_{BH}$ and H$\beta$ linewidth and L$_{5100}$ is often used as a proxy to estimate the black hole mass
(\citealt{Peterson04}). A serious limitation of this method is that it is still not clear
which specific measure of the H$\beta$ profile provides the closest
association with the velocity dispersion of the gas in the broad line region.

There are various alternative measures of the velocity width used for determining
the black hole mass, including the FWHMs of the intermediate 
component (IC) and the broad component (BC) (e.g. \citealt{Zhu09}). 
One could also use the model independent second momentum
(e.g. \citealt{Peterson04}; \citealt{Bian08}), or more simply the FWHM of 
the H$\beta$ line after subtracting the narrow component (NC) 
(e.g. \citealt{Peterson04}).
The NC subtracted FWHM and the second momentum estimates often lie within 
the range of values covered by the IC and BC FWHMs, except 
for some peculiar objects such as those with broad double-peaked profiles, for example UM 269.
Given all these uncertainties we decided to adopt the 
the best-fit black hole mass obtained from the SED model, rather than simply fixing it at a value
determined from a specific linewidth measurement. Moreover, it is now suggested that
radiation pressure may be important in modifying the black hole
mass derived using the relation between M$_{BH}$ and L$_{5100}$ and H$\beta$ FWHM, 
especially for objects with high Eddington ratios such as most NLS1s
(e.g. \citealt{Marconi08}).

\begin{figure}
\centering
\includegraphics[bb=85 360 414 855, scale=0.75,clip=1]{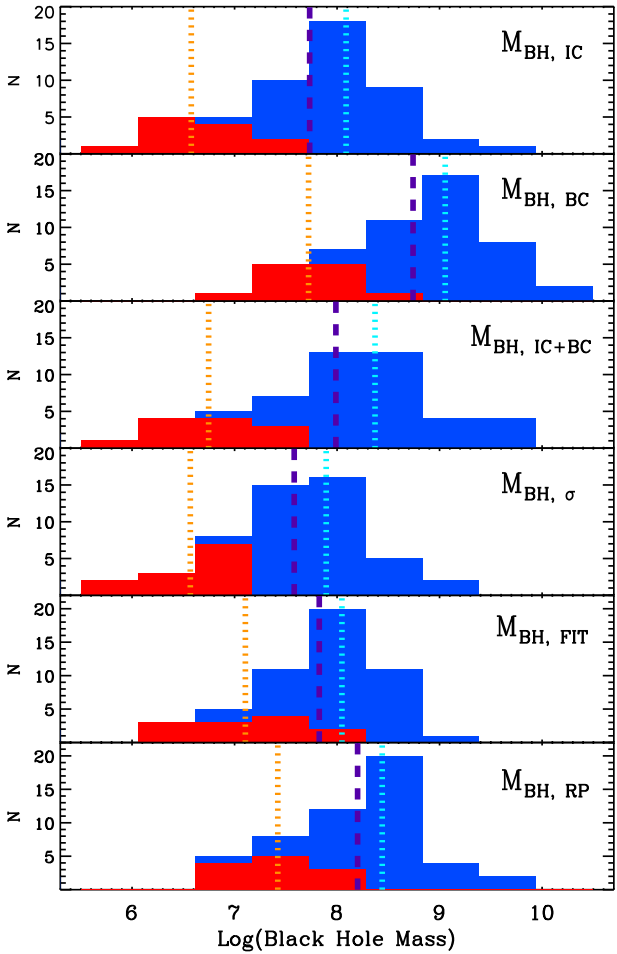}
\caption{A comparison of various methods used to derive black hole mass. 
The total distributions are shown with the 12 NLS1s show by the red regions.
 The purple dashed line indicate the average black hole mass for the whole sample.
 The orange and cyan dotted lines indicate the average masses of NLS1s and BLS1s, respectively.
 The average values are listed in Table~\ref{BHmass-averages}. Values for individual objects 
are listed in Table~\ref{black hole mass sum}.}
\label{bhmass_dist}
\end{figure}

In order to compare our results with those from other studies,
we have made various estimates of black hole masses for every source in our sample as follows:\\
(1) M$_{BH, IC}$, M$_{BH, BC}$ and M$_{BH, IC+BC}$ are derived using 
Equation~\ref{BH-mass} with different H$\beta$ FWHMs obtained from
our Balmer line fitting procedure.\\
(2) M$_{BH,\sigma}$ is the black hole mass calculated from the second momentum of the total
H$\beta$ line profile (see \citet{Peterson04} for details of the definition of `second momentum'), 
by using R$_{BLR}\propto$L$_{5100}^{0.518}$ and a geometry factor of $f=3.85$. 
These assumptions are considered to be appropriate when
using second momentum as a measure of the velocity dispersion in BLR (\citealt{Bentz06};
\citealt{Collin06}; \citealt{Bian08}). \\
(3) M$_{BH, RP}$ is the black hole mass corrected for radiation pressure, using equation (9) in
\citet{Marconi08} with $f=3.1$, $log(g)=7.6$.\\
We compare the black hole mass distributions obtained from these different methods in 
Figure~\ref{bhmass_dist}. The mean values are listed in Table~\ref{BHmass-averages}.

The M$_{BH,IC}$ and M$_{BH,BC}$
represent the two extreme estimates of black hole masses. The M$_{BH,
IC}$ could still be influenced by contamination from a NLR component, especially for NLS1s where
deconvolution of the narrow and broad components is very difficult. If there
is a residual narrow line component, it will introduce a bias that underestimates 
black hole masses. Conversely, using M$_{BH, BC}$ is more likely
to bias towards higher black hole masses, due to the presence of low contrast 
very broad wings often seen in H$\beta$ profiles. We found
FWHM$_{IC+BC}$/$\sigma_{H\beta}$=1.30$\pm$0.39 for our sample, which
is consistent with 1.33$\pm$0.36 found by \citet{Bian08}. This leads
to slightly lower values of M$_{BH,\sigma}$ than M$_{BH, IC+BC}$, but
these two methods both give black hole masses between M$_{BH, IC}$ and
M$_{BH,BC}$, with M$_{BH, IC+BC}$ spanning a broader mass range.

\begin{figure*}
\centering
\includegraphics[bb= 120 0 540 828, scale=0.55, clip=1, angle=90]{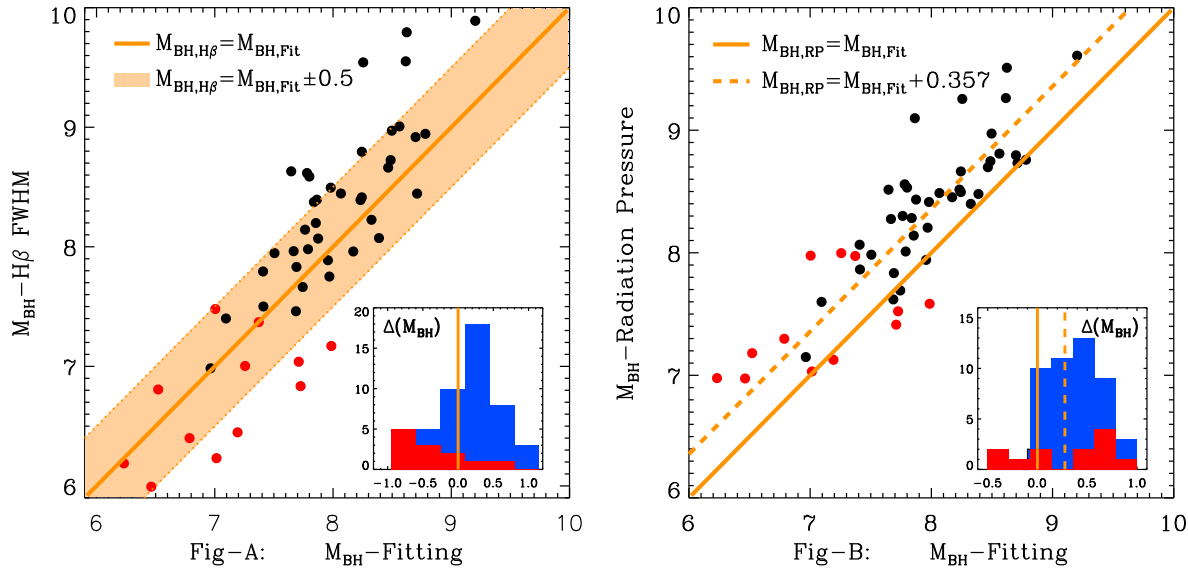}
\caption{Correlations of best-fit black hole mass (`M$_{BH}$-Fitting' or `M$_{BH,FIT}$') vs. H$\beta$ FWHM determined black hole mass (`M$_{BH}$-H$\beta$ FWHM' or `M$_{BH,IC+BC}$') and vs. radiation pressure corrected black hole mass (`M$_{BH}$-Radiation Pressure' or `M$_{BH,RP}$'). Red points represent the 12 NLS1s. The inserted panel in panel-A shows the distribution of the mass difference between M$_{BH,IC+BC}$ and M$_{BH,FIT}$, while the inserted panel in panel-B shows the distribution of the mass difference between M$_{BH,RP}$ and M$_{BH,FIT}$. Red regions highlight the distribution of NLS1s.}
\label{bhmass-directcompare}
\end{figure*}

Our best-fit SED black hole masses (M$_{BH,FIT}$) are also distributed between M$_{BH, IC}$
and M$_{BH,BC}$, with similar average masses as M$_{BH,IC+BC}$ (a comparison is
shown in Figure~\ref{bhmass-directcompare} Panel-A).
Note that M$_{BH,FIT}$ is a free parameter in the SED fitting unless it hits
the lower or upper limits set by M$_{BH,IC}$ and M$_{BH,BC}$,
which occasionally happened (see Tabel~\ref{black hole mass sum}).
It is clearly shown in Figure~\ref{bhmass-directcompare} that the
black hole masses from the SED fitting are not consistent with estimates 
based on either extremely narrow or extremely broad lines. 
So for NLS1s, the mean M$_{BH, FIT}$ is 0.36 dex higher than M$_{BH,IC+BC}$;
while for BLS1s, the mean M$_{BH, FIT}$ is 0.22 dex lower than M$_{BH,IC+BC}$. 
Interestingly, this also implies
that the M$_{BH,FIT}$ of NLS1s may have less deviation from the established M-Sigma
relation than that using the M(L$_{5100}$, FWHM$_{H\beta}$) relation as shown in several
previous studies (e.g. \citealt{Wang01}; \citealt{Bian04}; \citealt{Zhou06}).

The situation may be further complicated as \citet{Marconi08} showed that
NLS1s could be consistent with the M-$\sigma_{*}$ relation if a correction 
for radiation pressure is applied to black hole masses derived from M(L$_{5100}$, FWHM$_{H\beta}$).
In our sample, correction for radiation pressure adds to the average
M$_{BH,IC+BC}$ by 0.67 dex for NLS1, 0.07 dex for BLS1 and 0.21 dex
for the whole sample. We also found a very similar mass distribution
between M$_{BH,RP}$ and M$_{BH,FIT}$, except for an average of 0.36 dex higher
in M$_{BH,RP}$. The differences between the average mass of
NLS1s and BLS1s are 0.78 dex and 0.72 dex in M$_{BH,RP}$ and M$_{BH,FIT}$, separately 
(see Figure~\ref{bhmass-directcompare} Panel-B).
Therefore, if M$_{BH,RP}$ can provide a good match to the M-Sigma relation
even down to low mass NLS1s as proposed by \citet{Marconi08}, then our SED determined
M$_{BH,FIT}$ may also give similar results. This implies that the
suggested deviation from the M-$\sigma_{*}$ relation for NLS1s may not be an intrinsic
property, but rather a consequence of using black hole estimates based on 
M(L$_{5100}$, FWHM$_{H\beta}$) relation, which may not be appropriate
for NLS1s (e.g. \citealt{Grupe04}; \citealt{Komossa08}).

\subsection{The Average Spectral Energy Distributions}
\citet{Elvis94} constructed SED templates for both radio-loud and
radio-quiet AGN, based on a sample of 47 quasars between redshift
0.025 and 0.94. VF07 modeled optical-to-X-ray SED for a
sample of 54 AGNs with redshifts between 0.001 and 0.371, and showed
that the SED was related to Eddington ratio. They also suggested
that $\kappa_{2-10keV}$ is well correlated with Eddington ratio.
In a later study (\citealt{Vasudevan09}) based on SED modeling of 29
local AGNs from \citet{Peterson04}, the SED dependence on Eddington ratio was reinforced.
Recently, \citet{Lusso10} studied 545 X-ray selected type 1 AGN over the
redshift range of 0.04 to 4.25. They computed SEDs at different
redshifts, and investigated $\alpha_{ox}$ correlations with other
parameters such as redshift, $\kappa_{2-10keV}$, $\lambda_{Edd}$ etc.

We present a mean SED for our sample which is sub-divided according to their $H\beta$
FWHM. This gave three sub-samples, those with the narrowest lines, those with
moderately broad lines, and those with very broad lines. 
All objects were de-redshifted to their local frame. First, each of the best-fit SEDs was
divided into 450 energy bins between 1 eV and 100 keV. For each energy bin we
calculated the monochromatic luminosity for the sub-sample with 12 NLS1s, using
their individual SED models. Then an average value and standard deviation
in each energy bin were calculated in logarithm space.  Thus a mean
SED for the 12 NLS1s was constructed.  Using the same method for the
12 moderate and 12 broadest line objects, their mean SEDs were produced.
The total SED energy range is 1 eV to 100 keV, but we note that only
spectral ranges from 1.5-6 eV and 0.3-10 keV are actually covered by the 
observational data, and all other ranges are based on model extrapolations.

Obviously, limitations of our mean SEDs include the relatively small 
sample sizes composing the SEDs, and the redshift restriction z $<$ 0.4. 
On the other hand, we have assembled high quality data sets of optical, UV 
and X-ray observations. The exclusion of objects with high intrinsic absorption  
in the optical/UV helps to simplify the modeling assumptions. 
Our exclusion of warm-absorber objects may have introduced 
unknown selection
effects, but again this simplified the SED modeling. Our model of the
accretion flow also includes more detailed physical
assumptions on the optical-to-X-ray spectrum than in previous broadband
SED studies. These advantages make our broadband SED fitting more
physically plausible. Thus our mean SEDs too should be more reliable, especially in the 
unobservable far UV region, where often the peak of the energy is emitted. 

\begin{figure*}
\centering \includegraphics[angle=90,scale=0.6,clip=]{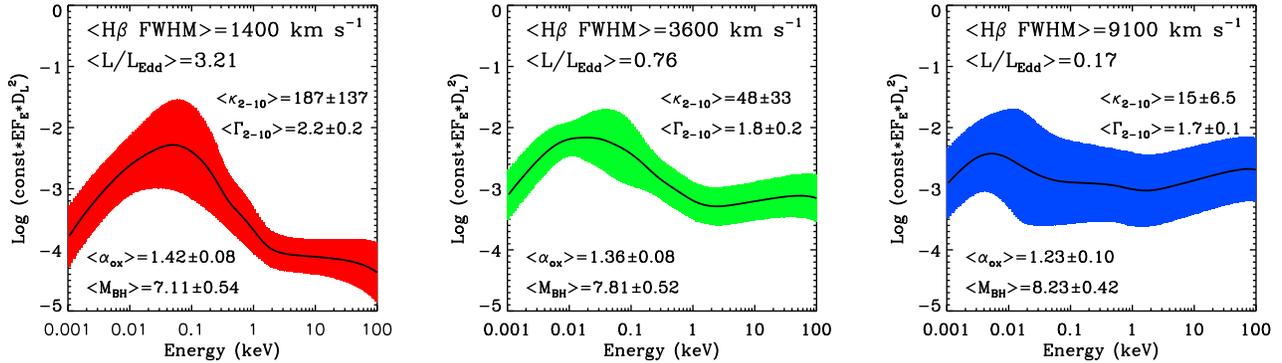}
\caption{The average SED of our sample.  The panel on the left shows the averaged 
SED for the 12
NLS1s (including two marginal NLS1s, 2XMM 112328.0+052823 and 1E
1346+26.7). The average H$\beta$ FWHM is 1400 $\pm$ 500 km s$^{-1}$.
The red area indicates a one standard deviation region on either side of the
average spectrum. The central panel is for 12 objects with moderate
line width. The average FWHM is 3700 $\pm$ 600 km s$^{-1}$. The green
region indicates one standard deviation. The panel on the right is the mean
SED for the 12 broadest line objects in our sample, including the one 
double-peak source. The average FWHM is 9800 $\pm$ 2900 km s$^{-1}$. We
also show the average value of the 2-10 keV powerlaw photon index, the 2-10 keV
bolometric correction, and the $\alpha_{ox}$ value with a one sigma
error. D$_{L}$ on the Y-axis title is the luminosity distance.  The unit of Y-axis 
is `keV (ergs s$^{-1}$ keV$^{-1}$)' in logarithm. The same arbitrary constant of 
1.31$\times$10$^{-46}$ is used for rescaling each plot.}
\label{averaged-spectra}
\end{figure*}

Figure~\ref{averaged-spectra} shows the mean SEDs for the three subsets
of our sample.  We caution that there is still substantial
spectral diversity within each subsample, and echo 
\citet{Elvis94}'s warning that if AGN SEDs are simply averaged without
considering their detailed intrinsic properties, then the dispersion in
the resultant mean SED will be large, so the mean SED may lose some
useful information about AGN properties. Nevertheless, there appears to be a 
clear SED connection with H$\beta$ FWHM. As the line width
increases, so the big blue bump (BBB) in the UV region becomes weaker
relative to the hard X-rays, and its peak shifts towards lower energy.
Also the spectral slope at high energies becomes harder.

This evolution in spectral shape is similar to that found by 
VF07 and \citet{Vasudevan09}, in which two mean SEDs of
different mean Eddington ratio were compared. This relation might be 
expected since the FWHM and Eddington ratio are also strongly
(anti)correlated in our sample. VF07 interpreted the
spectral diversity as a scaled up version of the different accretion
states of Galactic black hole binaries. The low Eddington ratio AGN could 
be analogous to the low/hard state in black hole binaries in having a weak disc giving a 
strong high energy tail, and the high Eddington ratio sources are analogous
to the high/soft state, in which the disc emission dominates. Our SED templates 
do not extend down to such low Eddington ratios as in
VF07, but we still see a similar behaviour.

\section{Summary and Conclusions}

In this paper we presented a spectral study of 51 unobscured Type 1 AGNs,
including 12 NLS1s. We assembled X-ray data from the EPIC monitor on board
the XMM-Newton satellite, and optical data from the SDSS DR7. In addition we
added optical/UV data from the XMM-Newton OM monitor when available. Our
results confirm some previously known correlations. For example, NLS1s often have
softer powerlaw fits from 2-10 keV, and have lower 2-10 keV
luminosities. Their H$\alpha$, H$\beta$ and [OIII]$\lambda$5007 lines
are also less luminous on average than found in BLS1s.

We use detailed models to fit the $H\alpha$ and $H\beta$ line profiles,
with multi-components to deblend the narrow, intermediate and broad
components by means of simultaneous modeling of the FeII continuum and other
blended lines.  We then use results from the $H\beta$ line fitting to
constrain the black hole mass. The FWHM of the intermediate and
broad components give a lower and upper limit for the mass,
respectively. This supports previous studies which find that NLS1s
tend to have lower black hole masses and higher Eddington ratios,
although their bolometric luminosities are not significantly different
from those of BLS1s.

We include the Balmer continuum and permitted iron emission, and extend the 
modeling across the entire SDSS spectrum in order to isolate the intrinsic 
optical underlying continuum.
However, this pure optical continuum is often (in 32/51 objects) flatter than
is predicted by the standard accretion disc model. This could indicate some contamination from
the host galaxy, but the lack of stellar absorption features in most of the SDSS
spectra suggests that this cannot be a general explanation. Instead it
seems more likely that there is an additional component in the optical region
related to the AGN, which is as yet not well understood.

We also show that the Balmer continuum is not well modeled if the edge
wavelength is fixed at its laboratory value of 3646{\AA}.  It is
shifted redwards, and smoothed by more than predicted by the FWHM of
the Balmer emission lines. These effects could both be produced by
density broadening. Potentially more detailed models of the optical 
emission could employ this as a new diagnostic tool for studying the physical
conditions e.g. electron density and temperature, in the innermost
Balmer emitting regions.

The optical, UV and X-ray data were fitted using a new broadband SED model,
which assumes that the gravitational potential
energy is emitted as optically thick blackbody emission at each radius
down to some specific coronal radius. Below this radius the remaining energy down
to the last stable orbit is divided between a soft X-ray excess
component and a hard X-ray tail. This energetically constrains the
model fits in the unobservable EUV region. We construct the resulting SEDs
for each of the sources.

A multi-component decomposition of the broadband SED shows that
relative contributions
to the bolometric luminosity from the accretion disc, Compotonisation and
powerlaw components vary among sources with different luminosity
and H$\beta$ linewidth. We find a slight increase in contribution from
the accretion disc as the luminosity increases, but a larger sample with
more sources at both low and high luminosities is needed to
confirm this.

Our study also supports the distinctiveness of the NLS1s among the whole
sample. We find that NLS1s tend to have a softer 2-10 keV spectrum,
lower 2-10 keV luminosity, lower black hole mass, higher Eddington
ratio and higher $\alpha_{ox}$ index.  However NLS1s do not stand out from
the whole sample in terms of their bolometric luminosity distribution.
We estimate the corona radii for every AGN in our sample from the SED
fitting. This shows that on average NLS1s have smaller corona radii,
and correspondingly a smaller coronal component contribution. 

We compare the best-fit black hole masses with those corrected for
radiation pressure, and other estimates of black hole mass
based on the R$_{BLR}$-L$_{5100}$ relation, including 
numerous options for measuring the velocity width of the H$\beta$ emission line.
These results show that the black holes masses derived from 
SED fitting have a similar distribution to that derived from profiles 
corrected for radiation pressure effects, except for
an offset of 0.3 dex lower in both the NLS1 and BLS1 subsamples.
The black hole mass difference between NLS1s and BLS1s from these two methods
(i.e. SED fitting and radiation pressure corrected profiles)
are both smaller than inferred from other mass measurements.
This implies that compared with black hole mass estimates based 
only on the H$\beta$ FWHM, NLS1s may lie closer to the 
established M-$\sigma_{*}$ relation at the low mass end, when their 
black hole masses are corrected for radiation pressure, and when we 
use masses derived from our SED fitting.

Finally, we form three broadband SED templates by co-adding SEDs in three subsamples
(consist of 12 objects in each) to examine how the broadband SED depends on $H\beta$ FWHM velocity
width, and by extension the Eddington ratio. The results show that
there is a change in the SED shape as the FWHM increases,
with NLS1s having the largest big blue bump in the extreme UV region.
Other important parameters such as $\Gamma_{2-10keV}$, $\kappa_{2-10keV}$ and
$\alpha_{ox}$, also change as the H$\beta$ FWHM increases. The
implications of correlations among these parameters will be discussed
in our next paper.

\section*{Acknowledgements}
C. Jin acknowledges financial support through the award of Durham
Doctoral Fellowship. This work is partially based on the data from
SDSS, whose funding is provided by the Alfred P. Sloan Foundation, the
Participating Institutions, the National Science Foundation, the
U.S. Department of Energy, the National Aeronautics and Space
Administration, the Japanese Monbukagakusho, the Max Planck Society,
and the Higher Education Funding Council for England.  This work is
also partially based on observations obtained with XMM-Newton, an ESA
science mission with instruments and contributions directly funded by
ESA Member States and the USA (NASA).  We also used GALEX data, which
is based on observations made with the NASA Galaxy Evolution Explorer.
GALEX is operated for NASA by the California Institute of Technology
under NASA contract NAS5-98034.

\appendix
\onecolumn
\centering

\section{The Spectral Modelling Results}

\begin{figure}
\centering
\includegraphics[bb=0 280 595 850,scale=0.85,clip=]{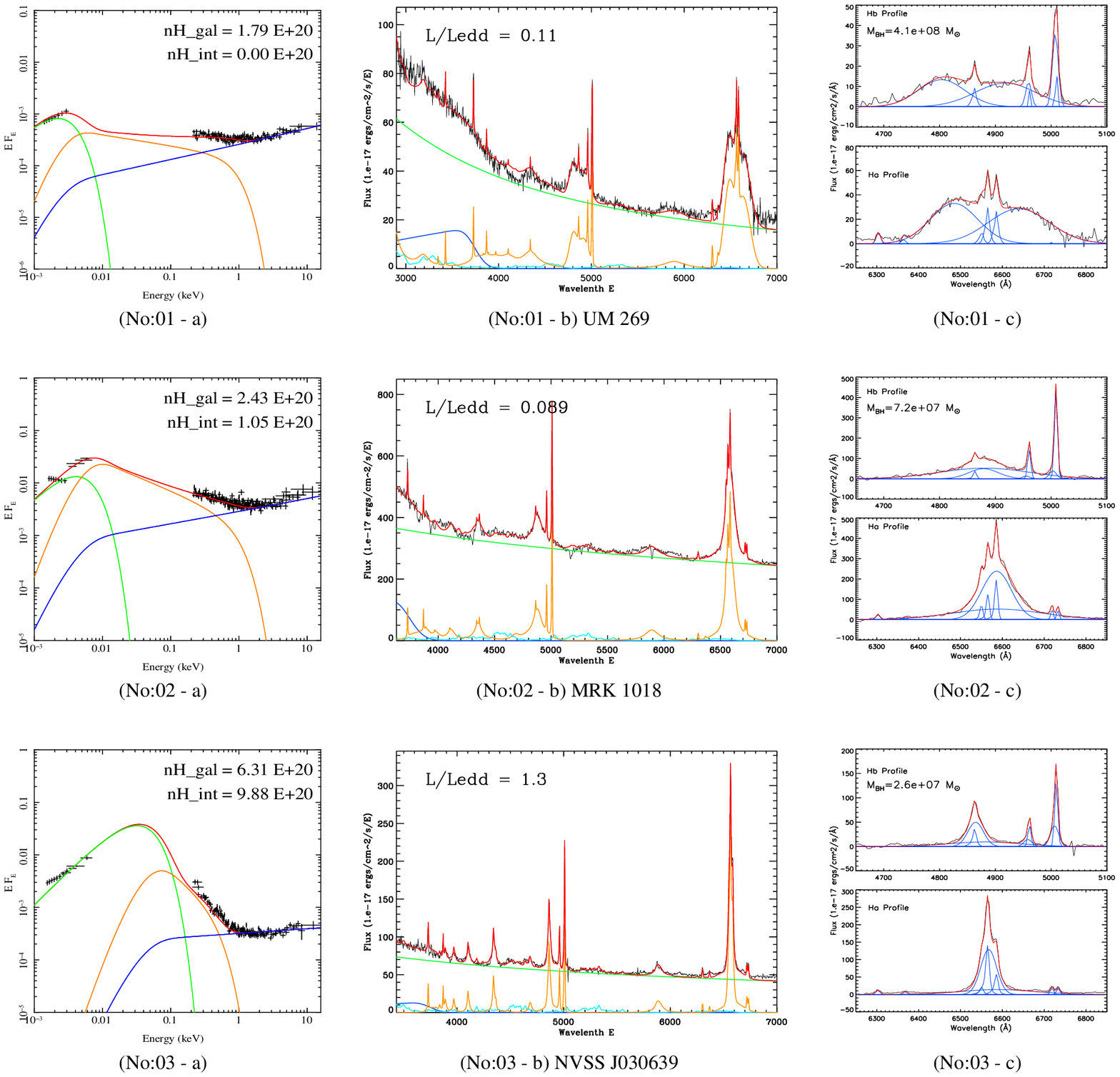}
\caption{The spectral fitting results.
Object order follows all other tables in this paper as increasing RA and DEC.
1. Broadband SED fitting plot (panel-a): X-ray data has been rebinned for each object.
Green solid line is the pure accretion disc component peaking at optical/UV region,
orange line is Comptonisation component producing soft X-ray excess below 2 keV,
blue line is the hard X-ray Comptonisation component dominating 2-10 keV spectrum,
and red is the total broadband SED model.
2. SDSS spectrum fitting plot (panel-b): only the fitted spectrum below 7000{\AA}
is plotted. Green solid line is the best-fit underlying continuum from accretion disc.
Orange line shows all best-fit emission lines,
including the results from detailed Balmer line fitting in panel-c.
FeII emission is plotted as light blue, while Balmer continuum being dark blue.
The total best-fit model with reddening is drawn in red solid line.
3. Balmer emission line fitting plot(panel-c): spectral ranges containing
H$\alpha$ and H$\beta$ profiles are plotted separately, with blue lines showing individual
line components and red line showing the whole best-fit model.
These are also the corresponding zoom-in plots of nearby
regions of H$\alpha$ and H$\beta$ in panel-b.
The given black hole mass is the broadband SED best-fit value,
see Section~\ref{SED-modelling} for detailed descriptions.}

\label{SED}
\end{figure}
\addtocounter{figure}{-1}
\begin{figure}
\centering
\includegraphics[bb=0 50 595 800,scale=0.85,clip=]{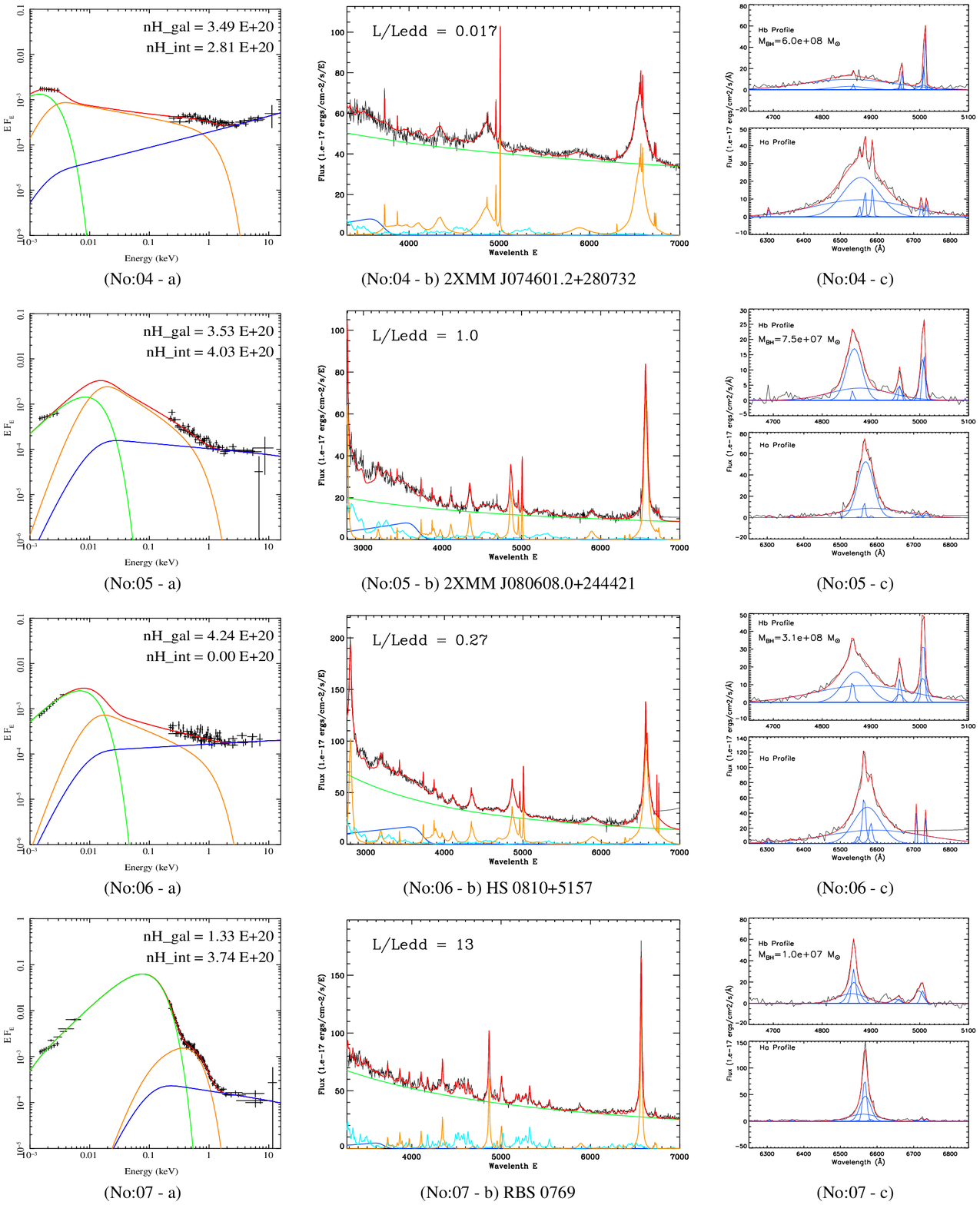}
\caption{\it continued}
\label{SED2}
\end{figure}
\addtocounter{figure}{-1}
\begin{figure}
\centering
\includegraphics[bb=0 50 595 800,scale=0.85,clip=]{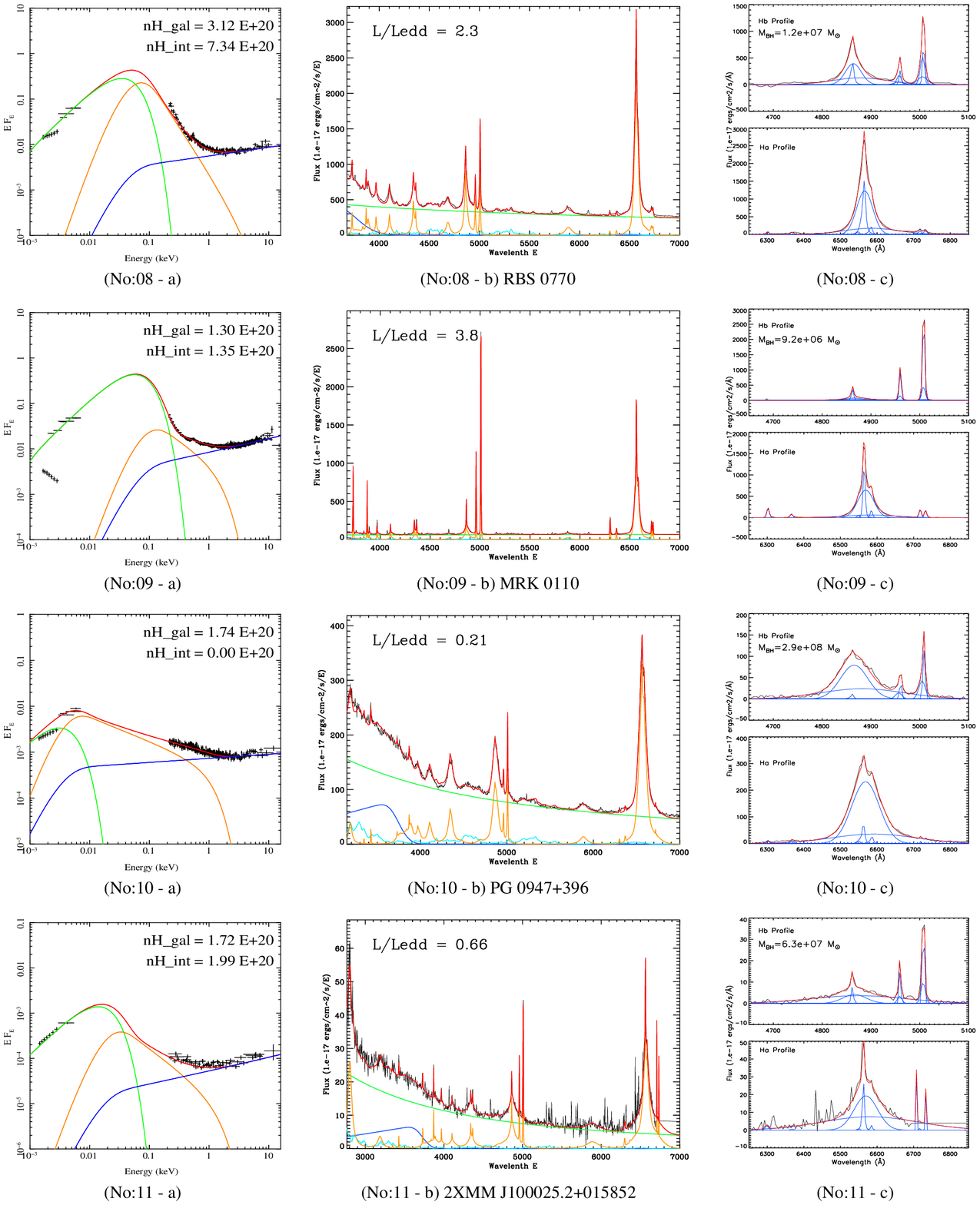}
\caption{\it continued}
\label{SED3}
\end{figure}
\addtocounter{figure}{-1}
\begin{figure}
\centering
\includegraphics[bb=0 50 595 800,scale=0.85,clip=]{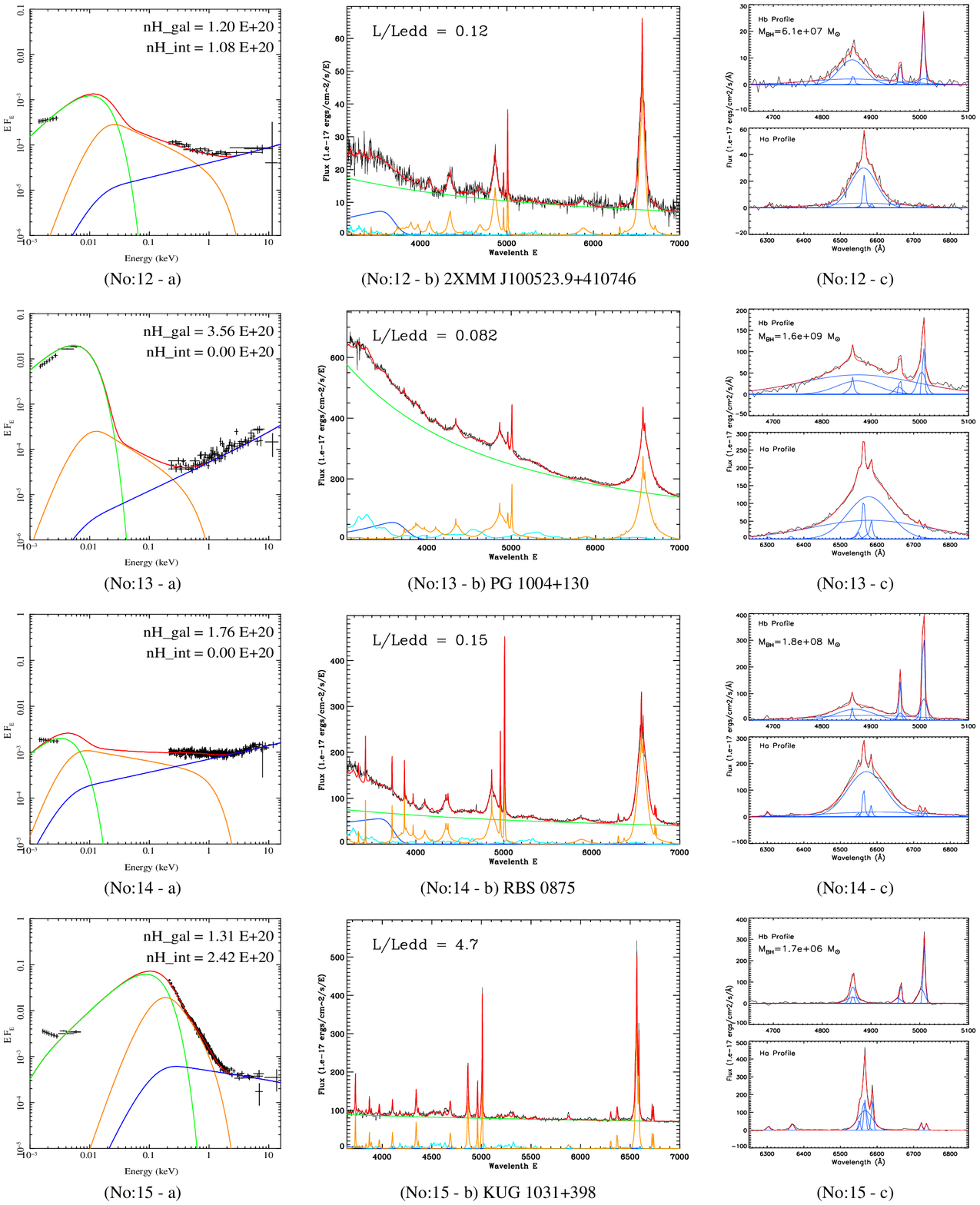}
\caption{\it continued}
\label{SED4}
\end{figure}
\addtocounter{figure}{-1}
\begin{figure}
\centering
\includegraphics[bb=0 50 595 800,scale=0.85,clip=]{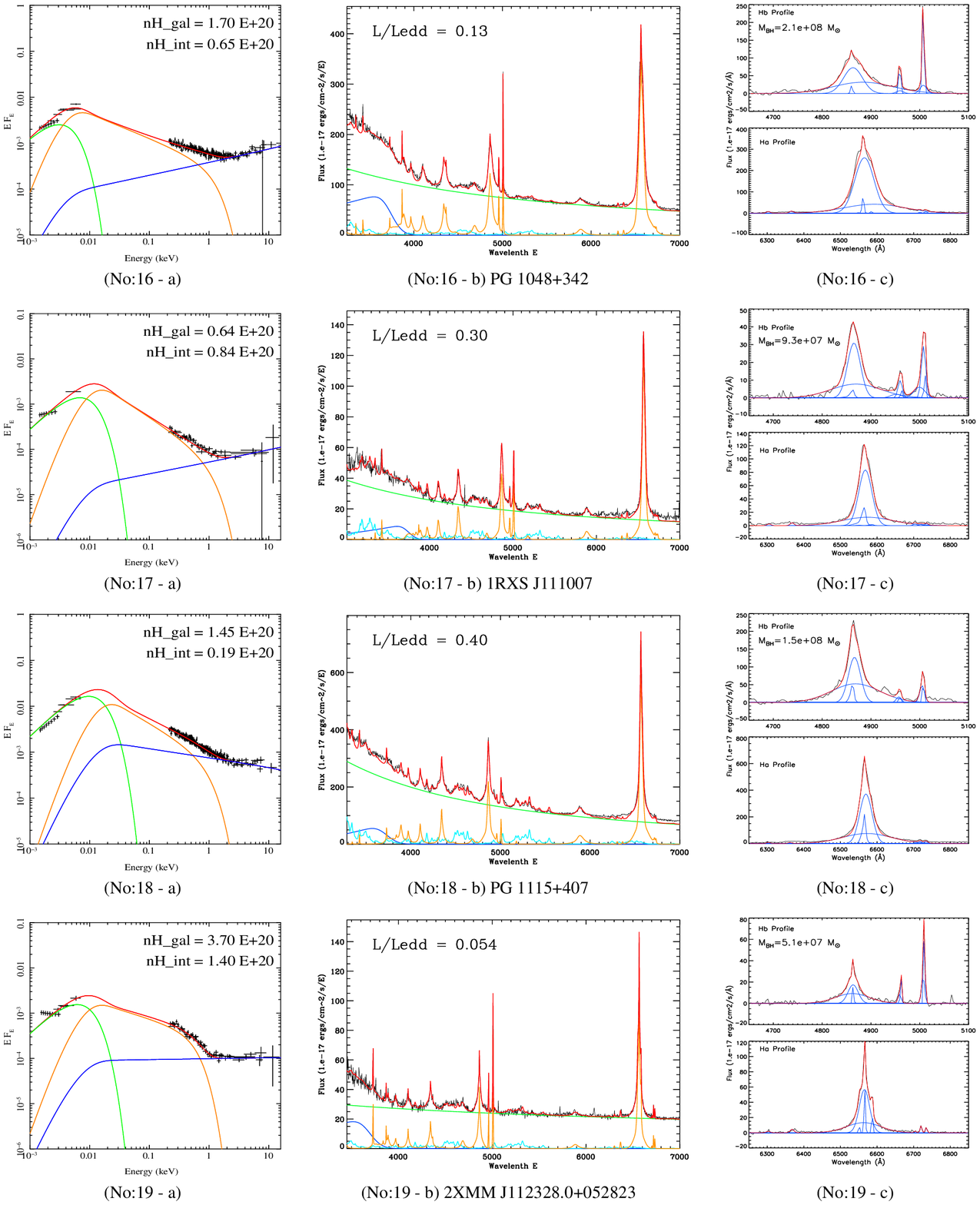}
\caption{\it continued}
\label{SED5}
\end{figure}
\addtocounter{figure}{-1}
\begin{figure}
\centering
\includegraphics[bb=0 50 595 800,scale=0.85,clip=]{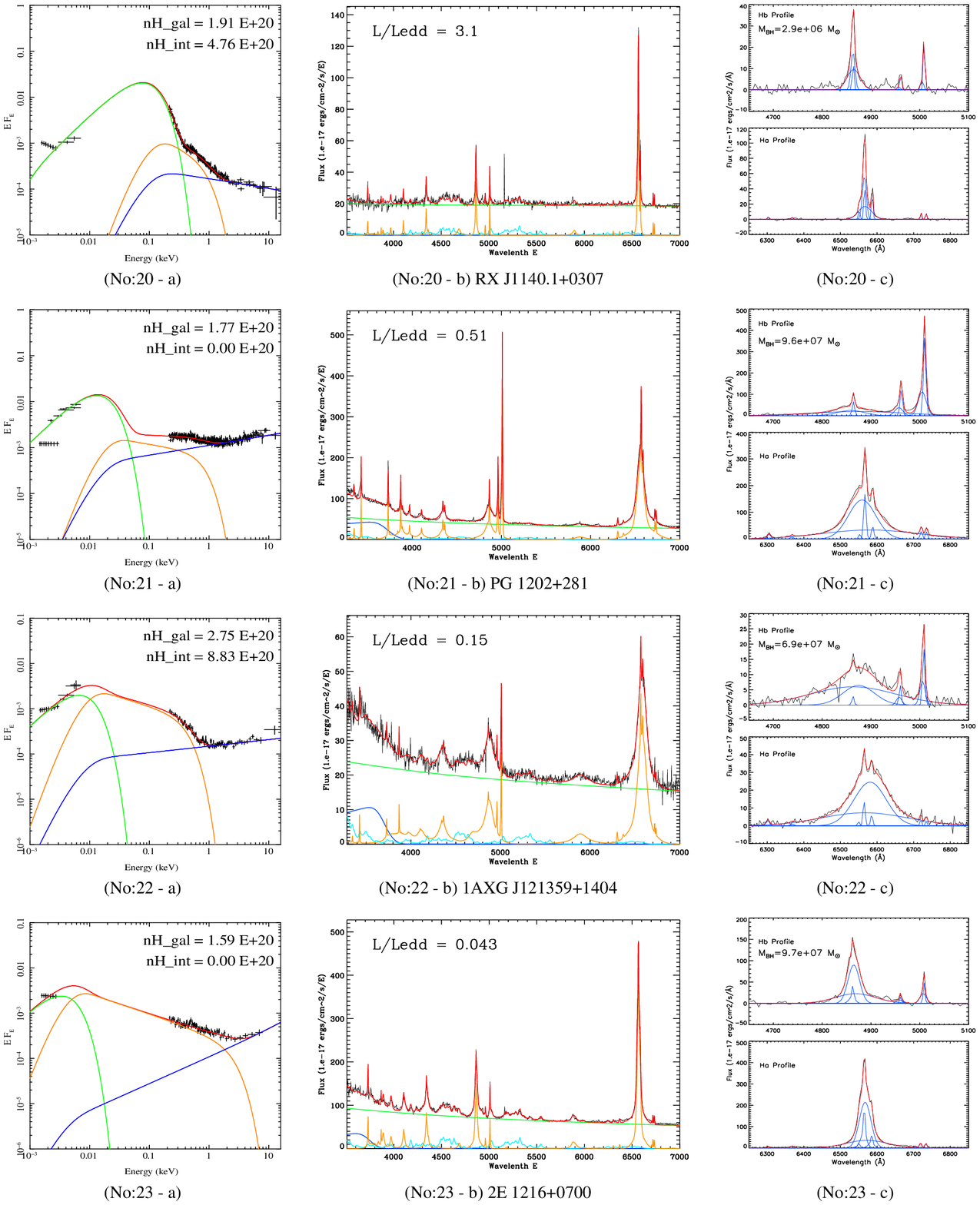}
\caption{\it continued}
\label{SED6}
\end{figure}
\addtocounter{figure}{-1}
\begin{figure}
\centering
\includegraphics[bb=0 50 595 800,scale=0.85,clip=]{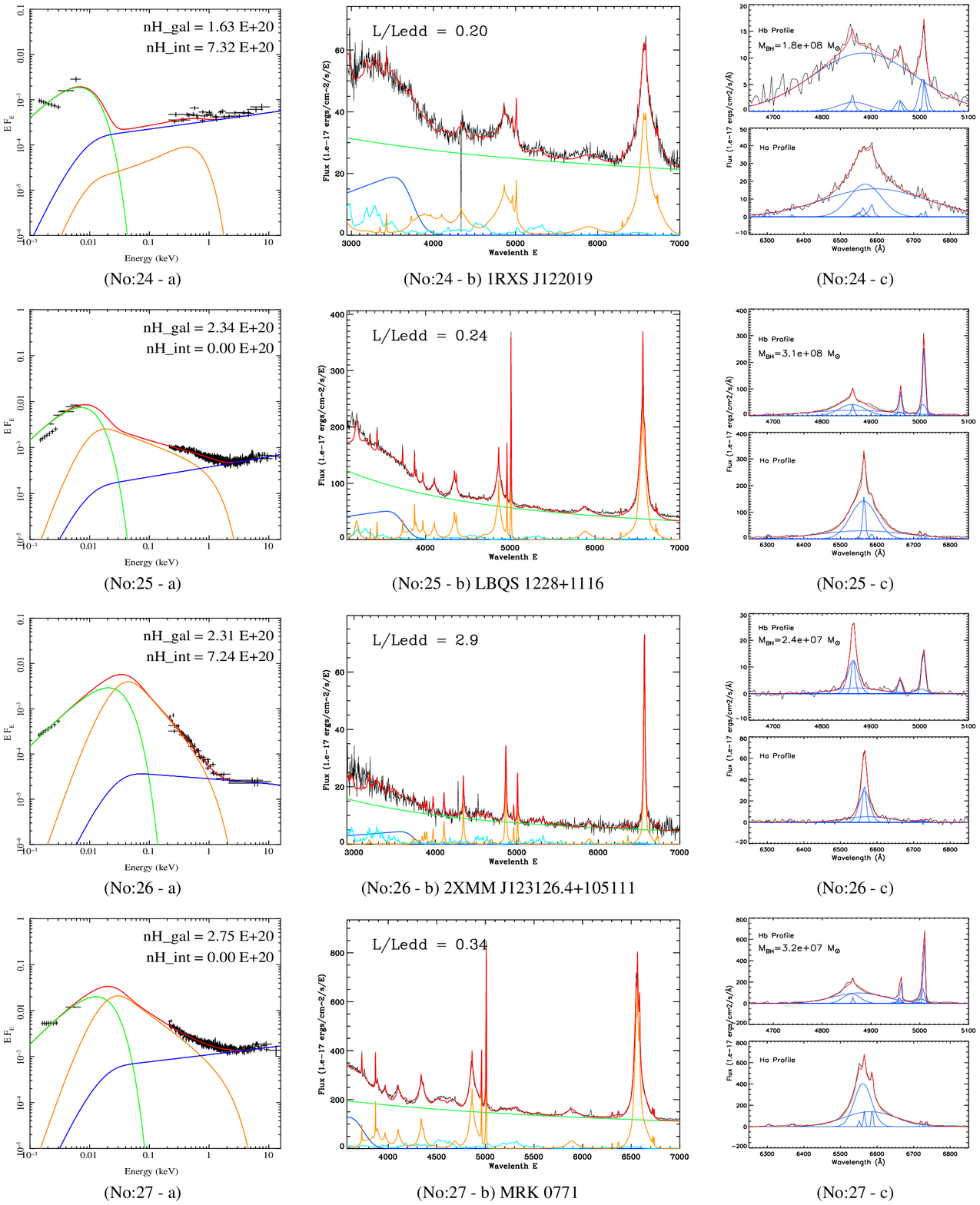}
\caption{\it continued}
\label{SED7}
\end{figure}
\addtocounter{figure}{-1}
\begin{figure}
\centering
\includegraphics[bb=0 50 595 800,scale=0.85,clip=]{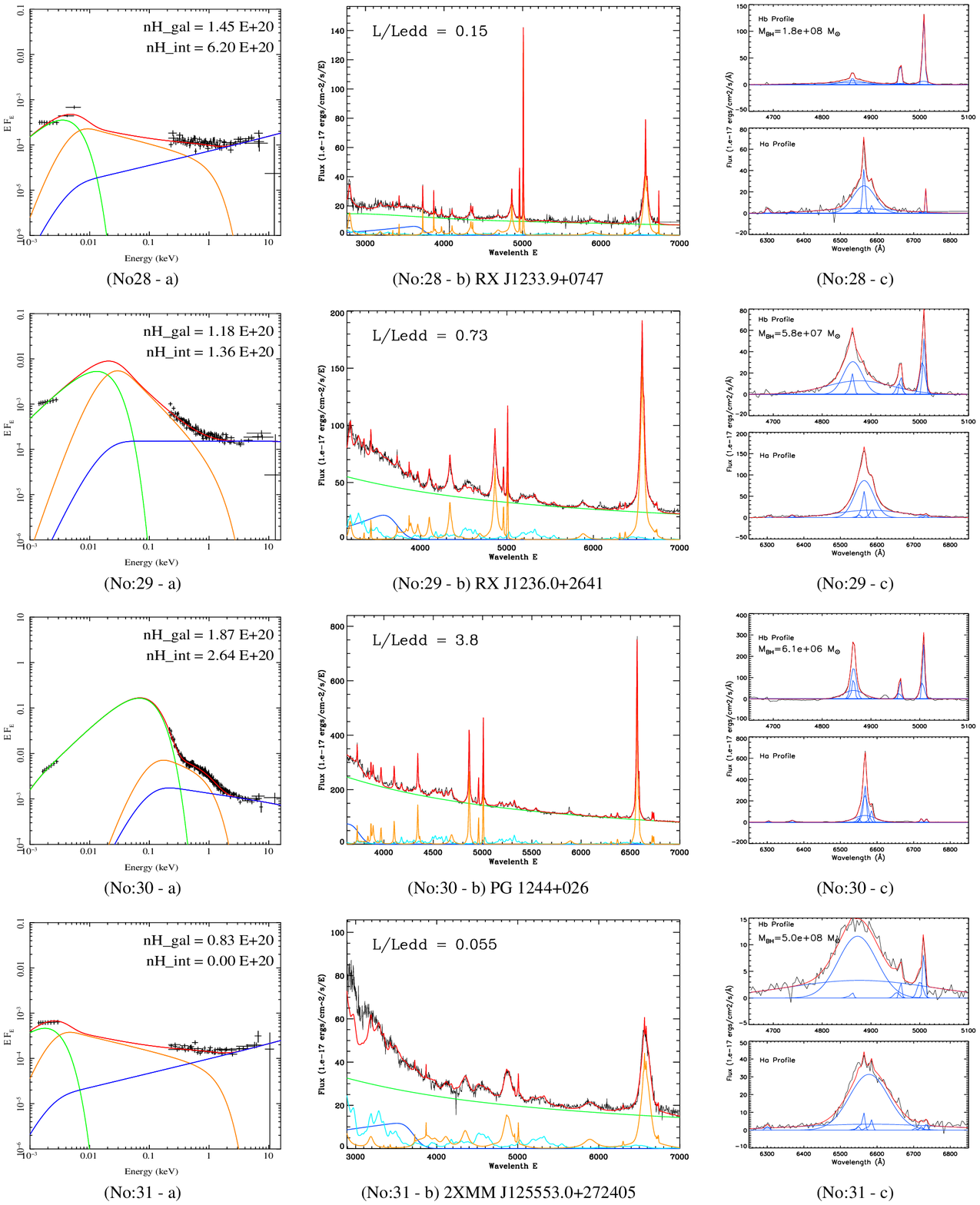}
\caption{\it continued}
\label{SED8}
\end{figure}
\addtocounter{figure}{-1}
\begin{figure}
\centering
\includegraphics[bb=0 50 595 800,scale=0.85,clip=]{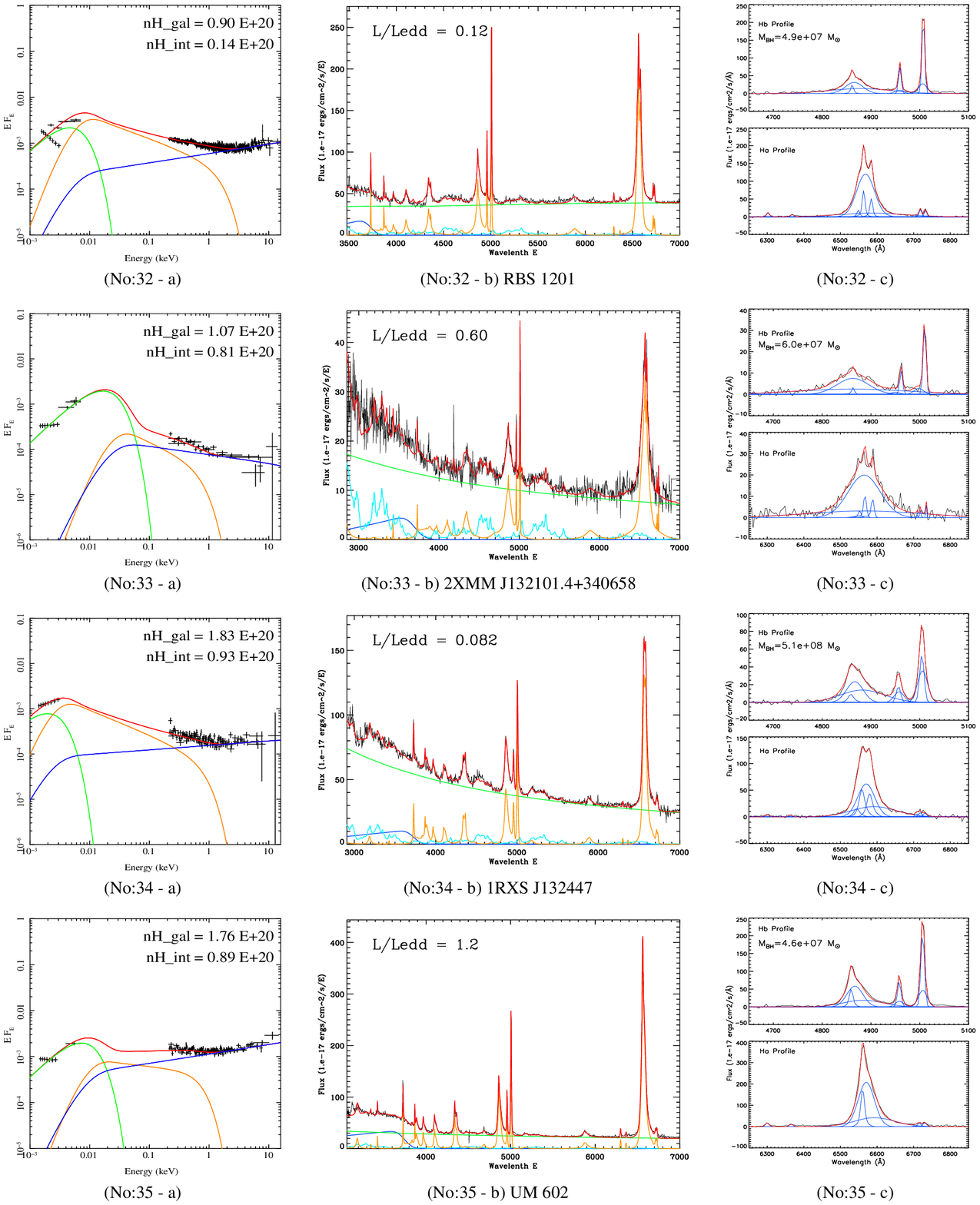}
\caption{\it continued}
\label{SED9}
\end{figure}
\addtocounter{figure}{-1}
\begin{figure}
\centering
\includegraphics[bb=0 50 595 800,scale=0.85,clip=]{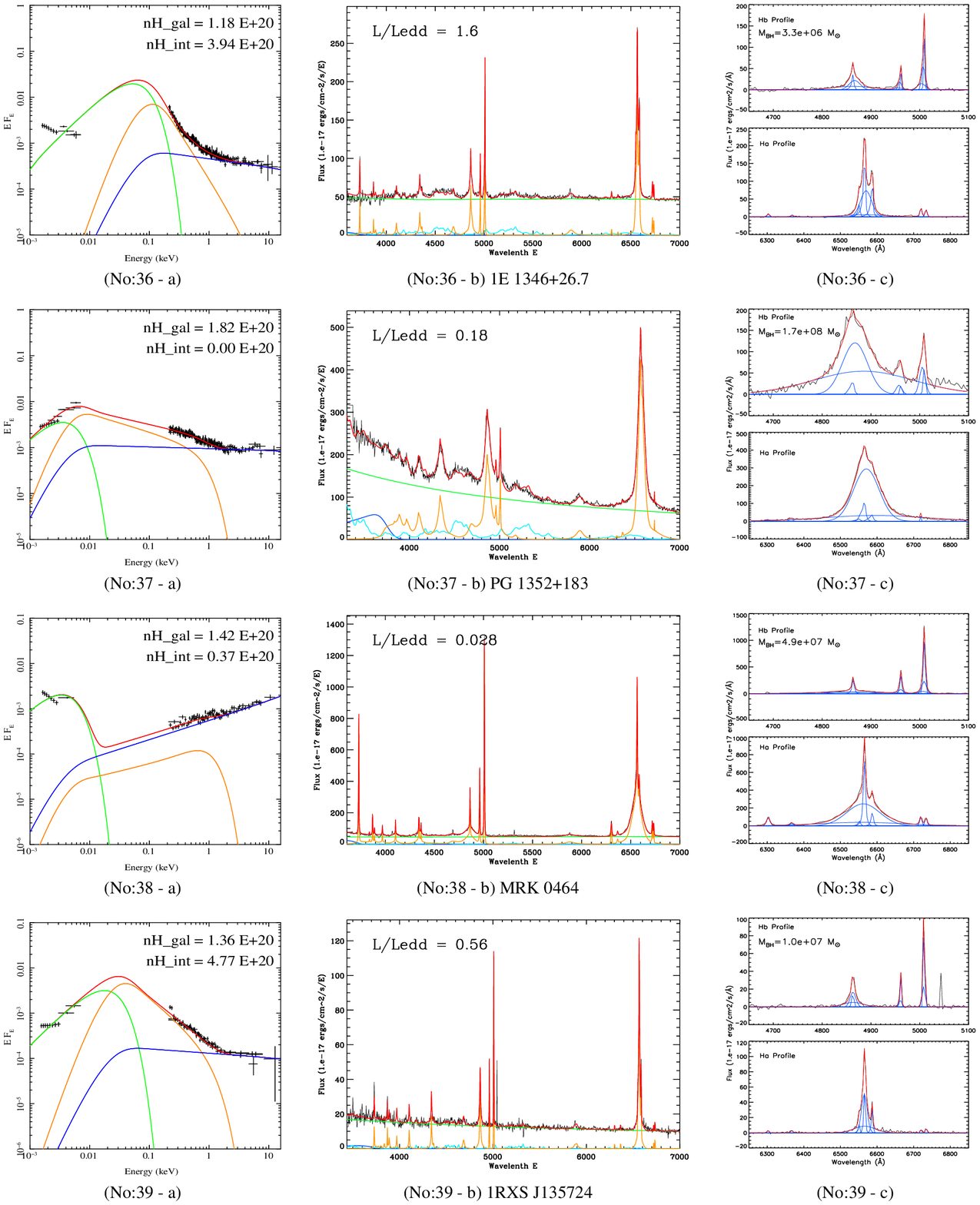}
\caption{\it continued}
\label{SED10}
\end{figure}
\addtocounter{figure}{-1}
\begin{figure}
\centering
\includegraphics[bb=0 50 595 800,scale=0.85,clip=]{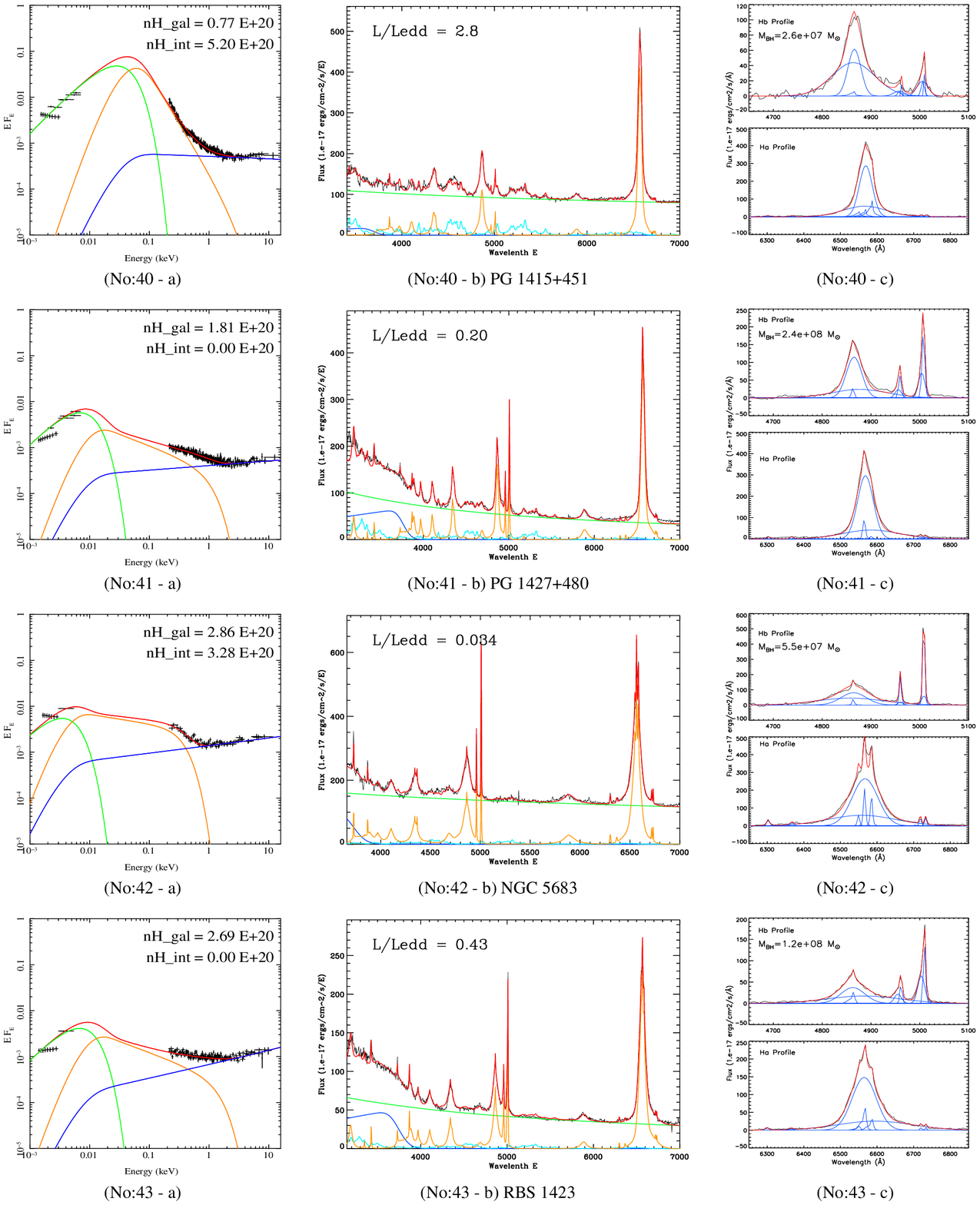}
\caption{\it continued}
\label{SED11}
\end{figure}
\addtocounter{figure}{-1}
\begin{figure}
\centering
\includegraphics[bb=0 50 595 800,scale=0.85,clip=]{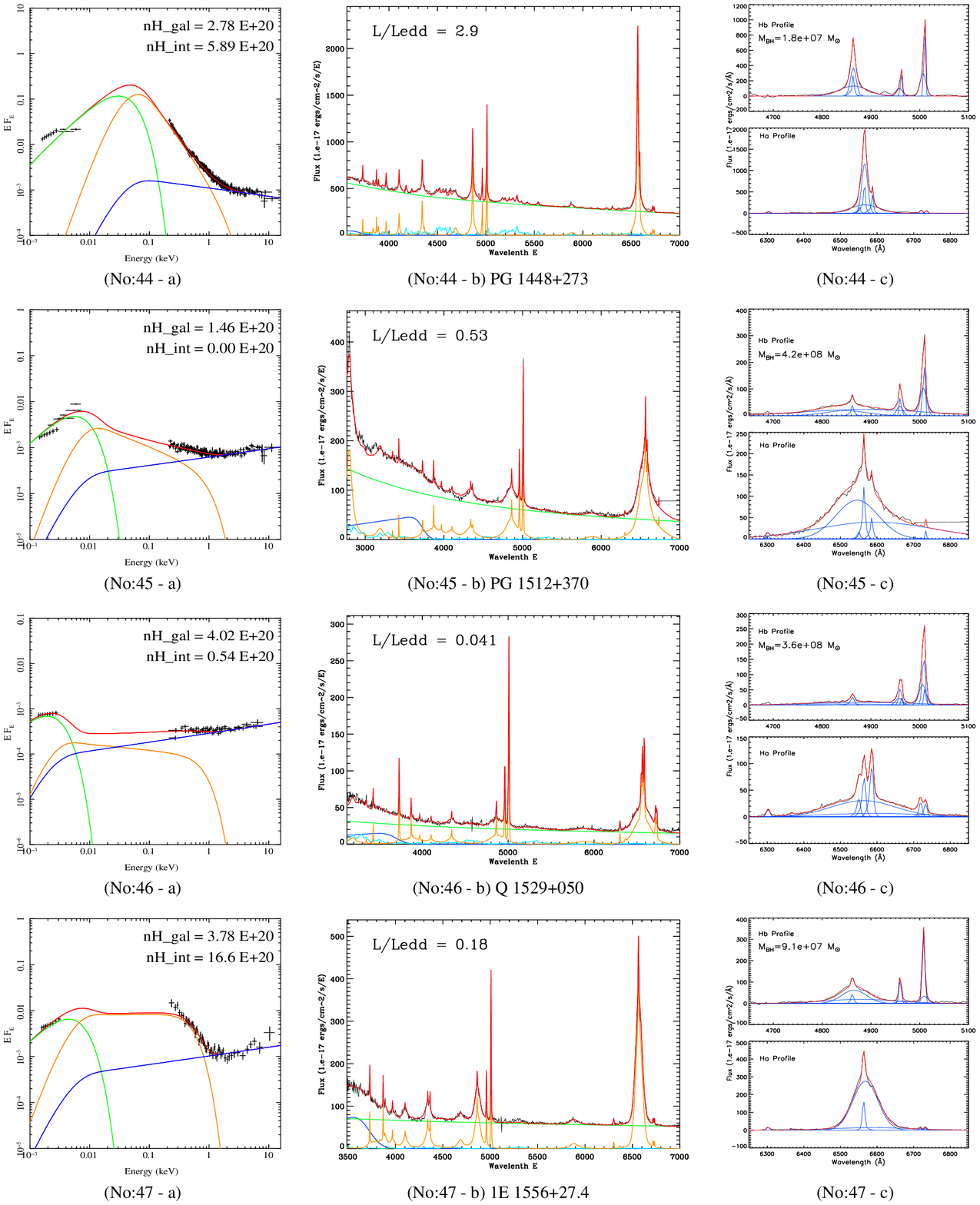}
\caption{\it continued}
\label{SED12}
\end{figure}
\addtocounter{figure}{-1}
\begin{figure}
\centering
\includegraphics[bb=0 50 595 800,scale=0.85,clip=]{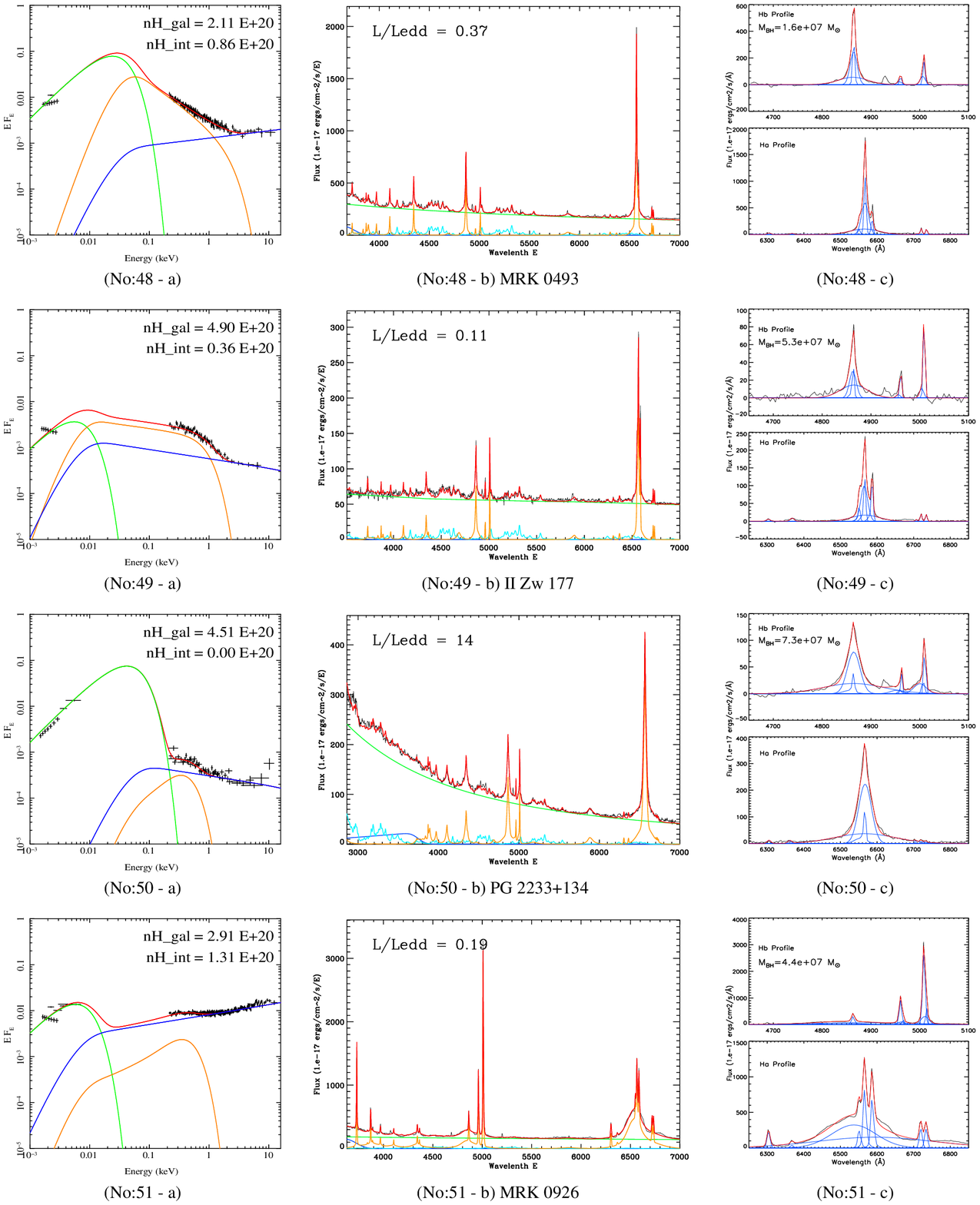}
\caption{\it continued}
\label{SED13}
\end{figure}

\clearpage
\section{XMM-Newton and SDSS DR7 Source Position And Seperation of Our Sample}
\begin{table*}
 \centering
  \begin{minipage}{160mm}
   \caption{XMM-Newton and SDSS DR7 source position and separation of our sample.
ID: object number, the same as Table~\ref{object list}; XMM\_Ra and XMM\_Dec: source's
right ascension and declination in the corresponding XMM-Newton observation; XMM\_PosErr:
X-ray position uncertainty from XMM-Newton; SDSS\_Ra and SDSS\_Dec: source's
right ascension and declination measured by SDSS; Separation: the angular separation between
source's XMM-Newton and SDSS coordinates; Sep./XMM\_PosErr: the ratio between coordinates
separation and X-ray position uncertainty, showing the significance of coordinate separation.}
   \label{source-match}
     \begin{tabular}{@{}cccccccc@{}}
\hline\hline
ID& XMM\_Ra&XMM\_Dec&XMM\_PosErr&SDSS\_Ra&SDSS\_Dec&Separation&Sep./XMM\_PosErr\\
  &\it degree\/&\it degree\/ & \it arcsec\/ &\it degree\/&\it degree\/& \it arcsec\/&\\
\hline
1 &10.83216 &0.85443 &0.35 &10.83227 &0.85425 &0.75 &2.10\\
2 &31.56642 &-0.29178 &1.03 &31.56664 &-0.29144 &1.44 &1.40\\
3 &46.66479 &0.06204 &0.35 &46.66487 &0.06200 &0.33 &0.93\\
4 &116.50527 &28.12559 &0.36 &116.50530 &28.12559 &0.09 &0.25\\
5 &121.53373 &24.73937 &0.40 &121.53390 &24.73919 &0.86 &2.14\\
6 &123.59218 &51.81109 &0.38 &123.59217 &51.81095 &0.50 &1.32\\
7 &140.69583 &51.34385 &0.35 &140.69595 &51.34390 &0.33 &0.92\\
8 &140.92903 &22.90931 &0.35 &140.92918 &22.90907 &1.00 &2.86\\
9 &141.30347 &52.28644 &0.35 &141.30355 &52.28621 &0.85 &2.44\\
10 &147.70155 &39.44735 &0.35 &147.70161 &39.44737 &0.19 &0.54\\
11 &150.10520 &1.98110 &0.17 &150.10519 &1.98115 &0.18 &1.04\\
12 &151.34968 &41.12950 &0.38 &151.34980 &41.12941 &0.44 &1.14\\
13 &151.85868 &12.81567 &0.38 &151.85876 &12.81562 &0.34 &0.89\\
14 &157.74620 &31.04878 &0.35 &157.74623 &31.04884 &0.21 &0.61\\
15 &158.66084 &39.64129 &0.35 &158.66082 &39.64119 &0.36 &1.03\\
16 &162.93283 &33.99096 &0.35 &162.93290 &33.99075 &0.76 &2.17\\
17 &167.52841 &61.42283 &0.37 &167.52898 &61.42262 &1.23 &3.36\\
18 &169.62621 &40.43171 &0.35 &169.62619 &40.43167 &0.17 &0.47\\
19 &170.86692 &5.47319 &0.36 &170.86718 &5.47311 &0.98 &2.71\\
20 &175.03644 &3.11972 &0.35 &175.03633 &3.11984 &0.58 &1.63\\
21 &181.17565 &27.90348 &0.35 &181.17545 &27.90328 &0.95 &2.69\\
22 &183.48412 &14.07530 &0.26 &183.48415 &14.07537 &0.27 &1.05\\
23 &184.87880 &6.72630 &0.28 &184.87863 &6.72623 &0.66 &2.38\\
24 &185.07680 &6.68898 &0.37 &185.07683 &6.68878 &0.71 &1.94\\
25 &187.72544 &11.00311 &0.20 &187.72550 &11.00310 &0.23 &1.16\\
26 &187.86003 &10.85327 &0.22 &187.86020 &10.85314 &0.75 &3.46\\
27 &188.01513 &20.15831 &0.35 &188.01511 &20.15821 &0.38 &1.07\\
28 &188.48376 &7.79869 &0.37 &188.48381 &7.79888 &0.71 &1.92\\
29 &189.01670 &26.69323 &0.36 &189.01677 &26.69335 &0.46 &1.28\\
30 &191.64732 &2.36918 &1.00 &191.64687 &2.36910 &1.64 &1.63\\
31 &193.97112 &27.40152 &0.27 &193.97104 &27.40146 &0.33 &1.23\\
32 &195.09236 &28.40082 &0.13 &195.09234 &28.40073 &0.32 &2.45\\
33 &200.25592 &34.11620 &0.38 &200.25590 &34.11609 &0.38 &1.00\\
34 &201.19851 &3.40888 &0.37 &201.19856 &3.40908 &0.71 &1.92\\
35 &205.30811 &-0.88743 &0.35 &205.30807 &-0.88755 &0.45 &1.26\\
36 &207.14581 &26.51932 &0.25 &207.14562 &26.51943 &0.72 &2.90\\
37 &208.64863 &18.08835 &0.35 &208.64872 &18.08820 &0.64 &1.80\\
38 &208.97286 &38.57458 &0.36 &208.97302 &38.57464 &0.49 &1.37\\
39 &209.35241 &65.41847 &0.25 &209.35220 &65.41831 &0.67 &2.69\\
40 &214.25318 &44.93513 &0.35 &214.25341 &44.93510 &0.60 &1.69\\
41 &217.42952 &47.79076 &0.35 &217.42947 &47.79061 &0.54 &1.52\\
42 &218.71847 &48.66196 &0.25 &218.71857 &48.66188 &0.38 &1.50\\
43 &221.06099 &6.55192 &0.35 &221.06111 &6.55188 &0.47 &1.32\\
44 &222.78667 &27.15737 &0.35 &222.78651 &27.15748 &0.64 &1.82\\
45 &228.67944 &36.84746 &0.35 &228.67946 &36.84734 &0.46 &1.29\\
46 &233.12007 &4.89952 &0.37 &233.11998 &4.89956 &0.34 &0.93\\
47 &239.62260 &27.28773 &0.37 &239.62235 &27.28729 &1.80 &4.86\\
48 &239.79023 &35.02983 &0.35 &239.79012 &35.02986 &0.35 &1.00\\
49 &334.82724 &12.13148 &0.35 &334.82721 &12.13144 &0.18 &0.52\\
50 &339.03200 &13.73203 &0.37 &339.03201 &13.73205 &0.11 &0.29\\
51 &346.18072 &-8.68642 &1.00 &346.18116 &-8.68573 &2.95 &2.95\\
\hline\hline
   \end{tabular}
  \\
  \\
 \end{minipage}
\end{table*}

\clearpage
\section{Black Hole Masses from Different Methods}
\begin{table*}
 \centering
  \begin{minipage}{160mm}
   \caption{Black hole masses from different methods. 
M$_{BH,IC}$: black hole mass calculated from the FWHM of H$\beta$ intermediate component in logarithm and solar mass;
M$_{BH,BC}$: black hole mass calculated from the FWHM of H$\beta$ broad component;
M$_{BH,IC+BC}$: black hole masses calculated from the FWHM of superposing H$\beta$ intermediate component (IC) and broad component (BC) (i.e. narrow component subtracted), using Equation~\ref{BH-mass};
M$_{BH,\sigma}$: black hole mass calculated from the second momentum of the whole H$\beta$ line profile, see subsection~\ref{BH mass compare} for details;
M$_{BH,Fit}$: the best-fit black hole masses in logarithm, which is constrained by M$_{BH,IC}$ and M$_{BH,BC}$, but values within 0.5 lower than log(M$_{BH,IC}$) were also adopted in the fitting, see subsection~\ref{BH mass compare};
log(M$_{BH,RP}$): the radiation pressure corrected black hole mass using Equation 9 in \citet{Marconi08} with {\it f}=3.1 and log{\it g}=7.6;
(*): note that M$_{BH,IC+BC}$ is always within the range of M$_{BH,IC}$ and M$_{BH,BC}$, except for UM269 whose H$\beta$ shows double-peak profile.}
   \label{black hole mass sum}
     \begin{tabular}{clcccccc}
\hline\hline
ID & Common Name & M$_{BH, IC}$ & M$_{BH, BC}$ & M$_{BH, IC+BC}$ & M$_{BH,~\sigma}$ & M$_{BH, Fit}$ & M$_{BH, RP}$ \\
 & & {\it log}, {\it M}$_{\sun}$ &{\it log}, {\it M}$_{\sun}$ &{\it log}, {\it M}$_{\sun}$ &{\it log}, {\it M}$_{\sun}$ &{\it log}, {\it M}$_{\sun}$ &{\it log}, {\it M}$_{\sun}$\\
\hline
1 & UM269 & 8.89 & 9.11 & 9.55$^{*}$ & 8.26 & 8.61 & 9.26 \\
2 & MRK1018 & 7.77 & 8.75 & 8.20 & 7.79 & 7.85 & 8.14 \\
3 & NVSSJ030639 & 7.40 & 8.72 & 7.50 & 7.47 & 7.41 & 7.86 \\
4 & 2XMMi/DR7 & 8.32 & 9.15 & 8.94 & 8.15 & 8.78 & 8.76 \\
5 & 2XMMi/DR7 & 7.94 & 9.11 & 8.07 & 7.71 & 7.87 & 8.43 \\
6 & HS0810+5157 & 8.70 & 9.82 & 8.97 & 8.45 & 8.50 & 8.97 \\
7 & RBS0769 & 7.28 & 8.15 & 7.48 & 6.89 & 7.00 & 7.98 \\
8 & RBS0770 & 7.24 & 8.31 & 7.40 & 7.22 & 7.09 & 7.60 \\
9 & MRK0110 & 6.77 & 7.74 & 6.98 & 6.76 & 6.96 & 7.15 \\
10 & PG0947+396 & 8.52 & 9.48 & 8.66 & 8.13 & 8.47 & 8.70 \\
11 & 2XMMi/DR7 & 8.16 & 9.39 & 8.59 & 8.18 & 7.80 & 8.53 \\
12 & 2XMMi/DR7 & 7.86 & 8.90 & 7.98 & 7.64 & 7.79 & 8.01 \\
13 & PG1004+130 & 9.40 & 10.30 & 9.89 & 8.97 & 9.20 & 9.61 \\
14 & RBS0875 & 8.59 & 9.52 & 8.79 & 8.28 & 8.24 & 8.66 \\
15 & KUG1031+398 & 6.13 & 7.49 & 6.19 & 5.85 & 6.23 & 6.98 \\
16 & PG1048+342 & 8.02 & 9.02 & 8.23 & 7.80 & 8.33 & 8.40 \\
17 & 1RXSJ111007 & 7.62 & 8.79 & 7.75 & 7.46 & 7.97 & 8.20 \\
18 & PG1115+407 & 7.75 & 8.95 & 7.96 & 7.69 & 8.17 & 8.45 \\
19 & 2XMMi/DR7 & 6.79 & 7.83 & 7.04 & 6.82 & 7.71 & 7.41 \\
20 & RXJ1140.1+0307 & 5.74 & 6.80 & 5.99 & 5.83 & 6.46 & 6.97 \\
21 & PG1202+281 & 8.13 & 9.21 & 8.49 & 8.09 & 7.98 & 8.41 \\
22 & 1AXGJ121359+1404 & 8.02 & 8.88 & 8.37 & 7.85 & 7.84 & 8.28 \\
23 & 2E1216+0700 & 7.04 & 8.13 & 7.17 & 6.96 & 7.99 & 7.58 \\
24 & 1RXSJ122019 & 8.60 & 9.63 & 9.54 & 8.51 & 8.26 & 9.26 \\
25 & LBQS1228+1116 & 8.50 & 9.54 & 8.73 & 8.23 & 8.49 & 8.75 \\
26 & 2XMMi/DR7 & 7.27 & 8.56 & 7.37 & 7.13 & 7.37 & 7.97 \\
27 & MRK0771 & 7.48 & 8.46 & 7.95 & 7.49 & 7.50 & 7.98 \\
28 & RXJ1233.9+0747 & 8.19 & 9.24 & 8.41 & 7.90 & 8.24 & 8.50 \\
29 & RXJ1236.0+2641 & 7.94 & 9.02 & 8.14 & 7.78 & 7.76 & 8.30 \\
30 & PG1244+026 & 6.26 & 7.41 & 6.40 & 6.27 & 6.79 & 7.30 \\
31 & 2XMMi/DR7 & 8.77 & 9.92 & 8.92 & 8.54 & 8.70 & 8.80 \\
32 & RBS1201 & 7.29 & 8.29 & 7.46 & 7.38 & 7.69 & 7.62 \\
33 & 2XMMi/DR7 & 8.28 & 9.55 & 8.62 & 8.22 & 7.78 & 8.56 \\
34 & 1RXSJ132447 & 8.19 & 9.04 & 8.45 & 7.71 & 8.71 & 8.73 \\
35 & UM602 & 7.82 & 8.61 & 7.96 & 7.29 & 7.67 & 8.28 \\
36 & 1E1346+26.7 & 6.63 & 7.55 & 6.81 & 6.81 & 6.52 & 7.18 \\
37 & PG1352+183 & 8.27 & 9.20 & 8.39 & 8.33 & 8.23 & 8.52 \\
38 & MRK0464 & 7.56 & 8.36 & 7.83 & 7.39 & 7.69 & 7.83 \\
39 & 1RXSJ135724 & 6.08 & 7.20 & 6.23 & 6.10 & 7.01 & 7.03 \\
40 & PG1415+451 & 7.47 & 8.51 & 7.79 & 7.42 & 7.41 & 8.07 \\
41 & PG1427+480 & 7.96 & 9.08 & 8.07 & 7.68 & 8.39 & 8.48 \\
42 & NGC5683 & 7.43 & 8.27 & 7.66 & 7.33 & 7.74 & 7.69 \\
43 & RBS1423 & 8.23 & 9.20 & 8.45 & 7.96 & 8.07 & 8.49 \\
44 & PG1448+273 & 6.81 & 8.19 & 7.00 & 7.01 & 7.26 & 8.00 \\
45 & PG1512+370 & 9.12 & 10.19 & 9.79 & 8.84 & 8.62 & 9.51 \\
46 & Q1529+050 & 8.70 & 8.86 & 9.01 & 8.26 & 8.56 & 8.81 \\
47 & 1E1556+27.4 & 7.76 & 8.40 & 7.89 & 7.55 & 7.96 & 7.94 \\
48 & MRK0493 & 6.33 & 7.56 & 6.45 & 6.43 & 7.19 & 7.13 \\
49 & IIZw177 & 6.59 & 7.79 & 6.83 & 6.72 & 7.73 & 7.52 \\
50 & PG2233+134 & 8.26 & 9.62 & 8.39 & 8.10 & 7.86 & 9.10 \\
51 & MRK0926 & 8.15 & 9.01 & 8.63 & 8.06 & 7.65 & 8.51 \\
\hline\hline
   \end{tabular}
  \\
  \\
 \end{minipage}
\end{table*}
\clearpage
\section{Emission Line Fitting Parameters of The Whole Sample}
\begin{table*}
 \centering
  \begin{minipage}{175mm}
   \caption{Emission line parameters for the whole sample. Narrow component (NC), intermediate
component (IC), broad component (BC) and intermediate plus broad component (I+B) are shown
separately for H$\alpha$ and H$\beta$. IC and BC are both Gaussian, while NC
may have the same profile as the whole [OIII] $\lambda$5007 or only the narrowest Gaussian
component in [OIII] $\lambda$5007. [OIII] $\lambda$5007 has two (or sometimes three) Gaussian components.
In the case of [OIII] $\lambda$5007, 'I+B' raw shows the parameters for the whole
emission line rather than having narrow component subtracted
Only one Gaussian profile is used for HeII $\lambda$4686,
[FeVII] $\lambda$6087 and
[FeX] $\lambda$6374. Sometimes the S/N of our spectra is not high enough to resolve
all these lines, or they are too week to be resolved, thus they do not have their line
parameters measured.
'vel' means velocity of line center relative to the rest frame vacuum wavelength in $km s^{-1}$.
The velocity of 'NC' is small and may come from the redshift uncertainty in Sloan's final
redshift measurement, and thus should not be taken seriously.
FWHMs of 'NC', 'IC' and 'BC' are directly from the Gaussian profile parameters.
FWHM for 'I+B' is measured directly from the superposed model profile. The numbers are all
in $km s^{-1}$. 'lum' and 'ew' means luminosity in $Log_{10}(ergs s^{-1})$ and equivalent width in \AA.}
   \label{emission-line-parameter}
     \begin{tabular}{@{}ccccccccccccccccc@{}}
\hline\hline
  ID & &\multicolumn{4}{c}{H$\alpha$}& \multicolumn{4}{c}{H$\beta$}& \multicolumn{4}{c}{[OIII] 5007} &HeII&FeVII&FeX\\
  & & {\it vel\/} & {\it fwhm\/} & {\it lum\/} & {\it ew\/} & {\it vel\/} & {\it fwhm\/} & {\it lum\/} & {\it ew\/} & {\it vel\/} & {\it fwhm\/} & {\it lum\/} & {\it ew\/} & {\it lum\/} & {\it lum\/} & {\it lum\/} \\
\hline\hline
1$^{d}$&\it NC\/&--- & 456 & 42.09 & 21 &--- & 457 & 41.46 & 3.1 & --- & --- & --- & --- & --- & 42.3 & ---\\
(1)$^{f}$&\it IC\/ & -3700 & 6080 & 43.18 & 270 & -3700 & 6080 & 42.67 & 50 & 47 & 203 & 41.49 & 3.5 & --- & --- & ---\\
&\it BC\/& 3400& 7840 & 43.23 & 300 & 3400 & 7840 & 42.73 & 57 & -47 & 720 & 42.19 & 18 & --- & --- & ---\\
&\it I+B\/&--- & 13000 & 43.51 & 570 & --- & 13000 & 43.00 & 110 & --- & 462 & 42.27 & 21 & --- & --- & ---\\
\hline
2&\it NC\/&--- & 405 & 40.77 & 5.0 &--- & 401 & 40.13 & 0.94 & --- & --- & --- & --- & 41.1 & 41.0 & ---\\
(1)$^{f}$&\it IC\/ & 1000 & 3810 & 41.98 & 83 & 1000 & 3810 & 41.19 & 11 & 21 & 386 & 41.15 & 10.0 & --- & --- & ---\\
&\it BC\/& 1100 & 11700 & 41.81 & 55 & 1100 & 11700 & 41.70 & 34 & -300 & 892 & 40.46 & 2.0 & --- & --- & ---\\
&\it I+B\/&--- & 4330 & 42.21 & 140 & --- & 6220 & 41.82 & 45 & --- & 401 & 41.23 & 12 & --- & --- & ---\\
\hline
3&\it NC\/&--- & 470 & 41.84 & 42 &--- & 469 & 41.17 & 6.4 & --- & --- & --- & --- & 41.3 & 41.3 & 41.2\\
(1)$^{f}$&\it IC\/ & 120 & 2040 & 42.36 & 140 & 120 & 2040 & 41.86 & 32 & 52 & 396 & 41.59 & 17 & --- & --- & ---\\
&\it BC\/& 1100 & 9390 & 42.05 & 68 & 1100 & 9390 & 41.80 & 28 & -130 & 1030 & 41.51 & 15 & --- & --- & ---\\
&\it I+B\/&--- & 2190 & 42.53 & 210 & --- & 2310 & 42.14 & 59 & --- & 468 & 41.85 & 32 & --- & --- & ---\\
\hline
4$^{*}$&\it NC\/&--- & 251 & 40.89 & 3.5 &--- & 249 & 40.35 & 0.81 & -34 & 239 & 40.82 & 2.5 & 40.8 & --- & ---\\
(1)$^{f}$&\it IC\/ & -400 & 5260 & 42.23 & 77 & -400 & 5260 & 41.31 & 7.4 & 140 & 192 & 41.23 & 6.2 & --- & --- & ---\\
&\it BC\/& -450 & 13700 & 42.29 & 86 & -450 & 13700 & 42.20 & 57 & -110 & 1370 & 40.87 & 2.7 & --- & --- & ---\\
&\it I+B\/&--- & 6500 & 42.56 & 160 & --- & 10800 & 42.25 & 64 & --- & 248 & 41.49 & 11 & --- & --- & ---\\
\hline
5&\it NC\/&--- & 326 & 41.76 & 13 &--- & 325 & 41.01 & 1.6 & --- & --- & --- & --- & 41.9 & 41.9 & ---\\
(2)$^{f}$&\it IC\/ & 190 & 2360 & 43.15 & 320 & 190 & 2360 & 42.56 & 58 & -23 & 325 & 41.71 & 8.4 & --- & --- & ---\\
&\it BC\/& 900 & 9030 & 42.96 & 200 & 900 & 9030 & 42.52 & 53 & -150 & 794 & 42.01 & 17 & --- & --- & ---\\
&\it I+B\/&--- & 2610 & 43.37 & 520 & --- & 2720 & 42.84 & 110 & --- & 474 & 42.19 & 25 & --- & --- & ---\\
\hline
6&\it NC\/&--- & 461 & 42.66 & 51 &--- & 457 & 41.88 & 4.7 & --- & --- & --- & --- & 42.4 & --- & ---\\
(1)$^{f}$&\it IC\/ & 400 & 3980 & 43.40 & 280 & 400 & 3980 & 42.86 & 45 & -76 & 387 & 42.25 & 12 & --- & --- & ---\\
&\it BC\/& 1400 & 14400 & 43.53 & 370 & 1400 & 14400 & 43.16 & 90 & -120 & 1020 & 42.19 & 10. & --- & --- & ---\\
&\it I+B\/&--- & 4930 & 43.77 & 650 & --- & 5430 & 43.34 & 140 & --- & 462 & 42.52 & 22 & --- & --- & ---\\
\hline
7&\it NC\/&--- & 580 & 41.88 & 38 &--- & 575 & 41.38 & 8.0 & --- & --- & --- & --- & 41.5 & --- & ---\\
(2)$^{f}$&\it IC\/ & 160 & 1570 & 42.10 & 62 & 160 & 1570 & 41.60 & 13 & -160 & 580 & 40.96 & 3.2 & --- & --- & ---\\
&\it BC\/& -120 & 4300 & 41.99 & 48 & -120 & 4300 & 41.70 & 17 & -610 & 1200 & 41.27 & 6.5 & --- & --- & ---\\
&\it I+B\/&--- & 1820 & 42.35 & 110 & --- & 1980 & 41.95 & 30 & --- & 1030 & 41.45 & 9.7 & --- & --- & ---\\
\hline
8$^{*}$&\it NC\/&--- & 442 & 41.75 & 78 &--- & 445 & 41.08 & 12 & -27 & 300 & 41.10 & 13 & 41.6 & 41.0 & 41.0\\
(1)$^{f}$&\it IC\/ & 73 & 2360 & 42.28 & 260 & 73 & 2360 & 41.68 & 48 & -160 & 663 & 41.26 & 19 & --- & --- & ---\\
&\it BC\/& 920 & 8080 & 41.97 & 130 & 920 & 8080 & 41.72 & 53 & -170 & 1390 & 41.03 & 11 & --- & --- & ---\\
&\it I+B\/&--- & 2580 & 42.45 & 390 & --- & 2840 & 42.00 & 100 & --- & 444 & 41.62 & 43 & --- & --- & ---\\
\hline
9$^{*}$&\it NC\/&--- & 317 & 41.46 & 150 &--- & 319 & 40.81 & 34 & -10. & 297 & 41.66 & 240 & 40.6 & 40.4 & ---\\
(1)$^{f}$&\it IC\/ & 200 & 2360 & 42.02 & 530 & 200 & 2360 & 40.99 & 50 & 150 & 1450 & 40.44 & 14 & --- & --- & ---\\
&\it BC\/& 840 & 7230 & 41.48 & 150 & 840 & 7230 & 41.14 & 72 & -41 & 509 & 41.10 & 64 & --- & --- & ---\\
&\it I+B\/&--- & 2500 & 42.13 & 680 & --- & 3030 & 41.37 & 120 & --- & 316 & 41.79 & 320 & --- & --- & ---\\
\hline\hline
   \end{tabular}
 \end{minipage}
\end{table*}
\addtocounter{table}{-1}
\begin{table*}
 \centering
  \begin{minipage}{175mm}
   \caption{continued}
     \begin{tabular}{@{}ccccccccccccccccc@{}}
\hline\hline
  ID & &\multicolumn{4}{c}{H$\alpha$}& \multicolumn{4}{c}{H$\beta$}& \multicolumn{4}{c}{[OIII] 5007} &HeII&FeVII&FeX\\
  & & {\it vel\/} & {\it fwhm\/} & {\it lum\/} & {\it ew\/} & {\it vel\/} & {\it fwhm\/} & {\it lum\/} & {\it ew\/} & {\it vel\/} & {\it fwhm\/} & {\it lum\/} & {\it ew\/} & {\it lum\/} & {\it lum\/} & {\it lum\/} \\
\hline\hline
10&\it NC\/&--- & 405 & 42.06 & 18 &--- & 401 & 41.13 & 1.3 & --- & --- & --- & --- & 41.9 & 42.0 & 41.2\\
(1)$^{f}$&\it IC\/ & 180 & 4060 & 43.45 & 430 & 180 & 4060 & 42.87 & 70 & 28 & 330 & 41.96 & 8.9 & --- & --- & ---\\
&\it BC\/& 1200 & 12300 & 43.12 & 200 & 1200 & 12300 & 42.83 & 64 & -210 & 934 & 41.97 & 9.2 & --- & --- & ---\\
&\it I+B\/&--- & 4440 & 43.62 & 630 & --- & 4810 & 43.15 & 130 & --- & 407 & 42.27 & 18 & --- & --- & ---\\
\hline
11&\it NC\/&--- & 279 & 42.03 & 47 &--- & 281 & 41.38 & 5.8 & --- & --- & --- & --- & 41.7 & --- & ---\\
(1)$^{f}$&\it IC\/ & 230 & 3440 & 42.89 & 340 & 230 & 3440 & 42.11 & 31 & -43 & 249 & 42.06 & 29 & --- & --- & ---\\
&\it BC\/& 730 & 14100 & 43.10 & 550 & 730 & 14100 & 42.66 & 110 & -82 & 634 & 41.77 & 15 & --- & --- & ---\\
&\it I+B\/&--- & 4360 & 43.31 & 900 & --- & 5640 & 42.77 & 140 & --- & 279 & 42.24 & 44 & --- & --- & ---\\
\hline
12&\it NC\/&--- & 396 & 41.55 & 36 &--- & 394 & 40.60 & 2.8 & --- & --- & --- & --- & 41.6 & --- & ---\\
(1)$^{f}$&\it IC\/ & -76 & 3850 & 42.53 & 350 & -76 & 3850 & 41.90 & 57 & 36 & 1590 & 40.92 & 6.2 & --- & --- & ---\\
&\it BC\/& -140 & 12600 & 42.11 & 130 & -140 & 12600 & 41.77 & 43 & -23 & 372 & 41.33 & 16 & --- & --- & ---\\
&\it I+B\/&--- & 4130 & 42.67 & 480 & --- & 4390 & 42.14 & 100 & --- & 395 & 41.47 & 22 & --- & --- & ---\\
\hline
13&\it NC\/&--- & 396 & 42.50 & 10. &--- & 395 & 41.94 & 1.6 & --- & --- & --- & --- & 42.8 & 42.3 & ---\\
(1)$^{f}$&\it IC\/ & 580 & 6160 & 43.53 & 110 & 580 & 6160 & 42.85 & 13 & -44 & 304 & 42.16 & 2.8 & --- & --- & ---\\
&\it BC\/& 700 & 17200 & 43.62 & 140 & 700 & 17200 & 43.46 & 52 & -250 & 1190 & 42.37 & 4.5 & --- & --- & ---\\
&\it I+B\/&--- & 7730 & 43.88 & 240 & --- & 10800 & 43.55 & 65 & --- & 395 & 42.58 & 7.3 & --- & --- & ---\\
\hline
14&\it NC\/&--- & 298 & 41.97 & 23 &--- & 300 & 41.48 & 5.8 & --- & --- & --- & --- & 42.0 & 41.9 & ---\\
(1)$^{f}$&\it IC\/ & 300 & 5580 & 43.31 & 520 & 300 & 5580 & 42.59 & 74 & 6.0 & 893 & 42.10 & 25 & --- & --- & ---\\
&\it BC\/& 1900 & 16300 & 42.81 & 160 & 1900 & 16300 & 42.70 & 96 & -49 & 266 & 42.31 & 41 & --- & --- & ---\\
&\it I+B\/&--- & 5940 & 43.43 & 680 & --- & 7060 & 42.95 & 170 & --- & 297 & 42.52 & 65 & --- & --- & ---\\
\hline
15&\it NC\/&--- & 342 & 41.03 & 30 &--- & 345 & 40.43 & 7.4 & --- & --- & --- & --- & 40.7 & 40.0 & 40.4\\
(2)$^{f}$&\it IC\/ & 58 & 918 & 41.22 & 46 & 58 & 918 & 40.73 & 15 & 53 & 299 & 40.84 & 19 & --- & --- & ---\\
&\it BC\/& -20 & 4400 & 40.97 & 26 & -20 & 4400 & 40.48 & 8.3 & -320 & 1000 & 40.74 & 15 & --- & --- & ---\\
&\it I+B\/&--- & 994 & 41.42 & 73 & --- & 987 & 40.92 & 23 & --- & 340 & 41.09 & 34 & --- & --- & ---\\
\hline
16&\it NC\/&--- & 279 & 41.62 & 9.9 &--- & 281 & 40.98 & 1.5 & --- & --- & --- & --- & 42.1 & 41.4 & ---\\
(2)$^{f}$&\it IC\/ & 37 & 2810 & 43.14 & 330 & 37 & 2810 & 42.47 & 46 & -55 & 968 & 41.53 & 5.4 & --- & --- & ---\\
&\it BC\/& 1300 & 8860 & 42.85 & 170 & 1300 & 8860 & 42.61 & 63 & -76 & 280 & 42.00 & 16 & --- & --- & ---\\
&\it I+B\/&--- & 3080 & 43.32 & 500 & --- & 3560 & 42.84 & 110 & --- & 297 & 42.13 & 22 & --- & --- & ---\\
\hline
17$^{*}$&\it NC\/&--- & 608 & 42.00 & 37 &--- & 607 & 41.08 & 2.9 & -28 & 617 & 41.84 & 17 & 41.7 & 41.8 & 41.7\\
(3)$^{f}$&\it IC\/ & 160 & 1930 & 42.91 & 300 & 160 & 1930 & 42.35 & 53 & 140 & 235 & 41.14 & 3.4 & --- & --- & ---\\
&\it BC\/& 430 & 7450 & 42.67 & 170 & 430 & 7450 & 42.34 & 51 & -530 & 1600 & 41.57 & 9.2 & --- & --- & ---\\
&\it I+B\/&--- & 2120 & 43.10 & 470 & --- & 2250 & 42.64 & 100 & --- & 607 & 42.08 & 30 & --- & --- & ---\\
\hline
18&\it NC\/&--- & 400 & 42.23 & 32 &--- & 401 & 41.53 & 3.6 & --- & --- & --- & --- & 42.1 & --- & 41.9\\
(1)$^{f}$&\it IC\/ & 220 & 1810 & 43.02 & 200 & 220 & 1810 & 42.44 & 29 & -60 & 260 & 41.28 & 2.1 & --- & --- & ---\\
&\it BC\/& 360 & 7220 & 42.93 & 160 & 360 & 7220 & 42.66 & 48 & -210 & 701 & 41.61 & 4.6 & --- & --- & ---\\
&\it I+B\/&--- & 2060 & 43.28 & 360 & --- & 2310 & 42.86 & 76 & --- & 401 & 41.78 & 6.7 & --- & --- & ---\\
\hline
19&\it NC\/&--- & 188 & 40.90 & 13 &--- & 191 & 40.25 & 2.3 & --- & --- & --- & --- & 41.0 & 40.4 & 40.5\\
(2)$^{f}$&\it IC\/ & 46 & 1500 & 41.77 & 96 & 46 & 1500 & 41.16 & 19 & 75 & 186 & 40.93 & 11 & --- & --- & ---\\
&\it BC\/& -54 & 5010 & 41.66 & 74 & -54 & 5010 & 41.40 & 33 & -100 & 445 & 40.80 & 8.4 & --- & --- & ---\\
&\it I+B\/&--- & 1730 & 42.02 & 170 & --- & 2000 & 41.60 & 51 & --- & 223 & 41.17 & 20 & --- & --- & ---\\
\hline
20&\it NC\/&--- & 232 & 40.64 & 14 &--- & 230 & 40.03 & 3.1 & --- & --- & --- & --- & 41.2 & --- & 39.8\\
(2)$^{f}$&\it IC\/ & 29 & 578 & 41.12 & 40 & 29 & 578 & 40.51 & 9.5 & 43 & 234 & 40.29 & 5.8 & --- & --- & ---\\
&\it BC\/& 120 & 1970 & 41.12 & 41 & 120 & 1970 & 40.76 & 17 & -230 & 460 & 39.80 & 1.9 & --- & --- & ---\\
&\it I+B\/&--- & 686 & 41.42 & 82 & --- & 774 & 40.95 & 27 & --- & 254 & 40.41 & 7.6 & --- & --- & ---\\
\hline
21&\it NC\/&--- & 354 & 42.05 & 49 &--- & 357 & 41.53 & 11 & --- & --- & --- & --- & --- & 41.5 & ---\\
(2)$^{f}$&\it IC\/ & -250 & 3990 & 43.03 & 470 & -250 & 3990 & 42.14 & 44 & 110 & 357 & 42.29 & 64 & --- & --- & ---\\
&\it BC\/& 730 & 13900 & 42.93 & 380 & 730 & 13900 & 42.58 & 120 & -200 & 1160 & 42.27 & 63 & --- & --- & ---\\
&\it I+B\/&--- & 4570 & 43.29 & 840 & --- & 6090 & 42.71 & 170 & --- & 419 & 42.58 & 130 & --- & --- & ---\\
\hline
22&\it NC\/&--- & 317 & 40.86 & 6.5 &--- & 319 & 40.08 & 0.86 & --- & --- & --- & --- & 41.5 & --- & ---\\
(2)$^{f}$&\it IC\/ & 750 & 4700 & 42.28 & 170 & 750 & 4700 & 41.59 & 28 & -19 & 316 & 40.92 & 6.0 & --- & --- & ---\\
&\it BC\/& 59 & 12600 & 42.18 & 130 & 59 & 12600 & 41.98 & 68 & -150 & 764 & 40.90 & 5.8 & --- & --- & ---\\
&\it I+B\/&--- & 5510 & 42.53 & 300 & --- & 7050 & 42.13 & 96 & --- & 389 & 41.21 & 12 & --- & --- & ---\\
\hline\hline
   \end{tabular}
 \end{minipage}
\end{table*}

\addtocounter{table}{-1}
\begin{table*}
 \centering
  \begin{minipage}{175mm}
   \caption{continued}
     \begin{tabular}{@{}ccccccccccccccccc@{}}
\hline\hline
  ID & &\multicolumn{4}{c}{H$\alpha$}& \multicolumn{4}{c}{H$\beta$}& \multicolumn{4}{c}{[OIII] 5007} &HeII&FeVII&FeX\\
  & & {\it vel\/} & {\it fwhm\/} & {\it lum\/} & {\it ew\/} & {\it vel\/} & {\it fwhm\/} & {\it lum\/} & {\it ew\/} & {\it vel\/} & {\it fwhm\/} & {\it lum\/} & {\it ew\/} & {\it lum\/} & {\it lum\/} & {\it lum\/} \\
\hline\hline
23&\it NC\/&--- & 340 & 41.56 & 39 &--- & 344 & 40.79 & 5.0 & --- & --- & --- & --- & 41.0 & --- & 41.0\\
(1)$^{f}$&\it IC\/ & 130 & 1700 & 42.15 & 150 & 130 & 1700 & 41.66 & 37 & 31 & 256 & 40.59 & 3.2 & --- & --- & ---\\
&\it BC\/& 340 & 5970 & 41.92 & 89 & 340 & 5970 & 41.60 & 33 & -79 & 930 & 40.81 & 5.4 & --- & --- & ---\\
&\it I+B\/&--- & 1890 & 42.35 & 240 & --- & 1980 & 41.93 & 70 & --- & 340 & 41.01 & 8.7 & --- & --- & ---\\
\hline
24&\it NC\/&--- & 576 & 41.37 & 3.9 &--- & 575 & 41.04 & 1.6 & --- & --- & --- & --- & 42.3 & 42.2 & ---\\
(1)$^{f}$&\it IC\/ & 150 & 4700 & 42.74 & 91 & 150 & 4700 & 41.59 & 5.5 & 44 & 327 & 41.00 & 1.4 & --- & --- & ---\\
&\it BC\/& 1400 & 15400 & 43.19 & 260 & 1400 & 15400 & 42.91 & 120 & -110 & 1090 & 41.52 & 4.8 & --- & --- & ---\\
&\it I+B\/&--- & 7280 & 43.32 & 350 & --- & 13900 & 42.93 & 120 & --- & 577 & 41.64 & 6.2 & --- & --- & ---\\
\hline
25&\it NC\/&--- & 400 & 42.49 & 47 &--- & 395 & 41.79 & 5.6 & --- & --- & --- & --- & 42.1 & --- & ---\\
(1)$^{f}$&\it IC\/ & -74 & 3820 & 43.36 & 350 & -74 & 3820 & 42.72 & 47 & 29 & 365 & 42.51 & 31 & --- & --- & ---\\
&\it BC\/& -87 & 12700 & 43.22 & 250 & -87 & 12700 & 42.92 & 75 & -99 & 943 & 42.13 & 13 & --- & --- & ---\\
&\it I+B\/&--- & 4350 & 43.59 & 590 & --- & 4980 & 43.13 & 120 & --- & 395 & 42.66 & 43 & --- & --- & ---\\
\hline
26&\it NC\/&--- & 543 & 42.16 & 88 &--- & 544 & 41.63 & 16 & --- & --- & --- & --- & 41.5 & --- & ---\\
(2)$^{f}$&\it IC\/ & 44 & 1540 & 42.53 & 210 & 44 & 1540 & 42.05 & 43 & -40 & 545 & 41.71 & 21 & --- & --- & ---\\
&\it BC\/& 490 & 6770 & 42.42 & 160 & 490 & 6770 & 41.92 & 32 & -330 & 1630 & 41.21 & 6.6 & --- & --- & ---\\
&\it I+B\/&--- & 1730 & 42.78 & 370 & --- & 1720 & 42.29 & 76 & --- & 577 & 41.83 & 27 & --- & --- & ---\\
\hline
27$^{*}$&\it NC\/&--- & 255 & 40.98 & 7.6 &--- & 256 & 40.48 & 1.8 & 49 & 254 & 41.53 & 20 & 41.5 & 40.9 & 40.9\\
(2)$^{f}$&\it IC\/ & -150 & 2500 & 42.40 & 200 & -150 & 2500 & 41.64 & 26 & -130 & 504 & 41.15 & 8.6 & --- & --- & ---\\
&\it BC\/& 780 & 7780 & 42.43 & 220 & 780 & 7780 & 42.17 & 86 & -350 & 1240 & 41.00 & 6.1 & --- & --- & ---\\
&\it I+B\/&--- & 3030 & 42.72 & 420 & --- & 4310 & 42.28 & 110 & --- & 291 & 41.76 & 35 & --- & --- & ---\\
\hline
28&\it NC\/&--- & 368 & 42.31 & 55 &--- & 369 & 41.74 & 11 & --- & --- & --- & --- & 42.7 & --- & ---\\
(1)$^{f}$&\it IC\/ & 0 & 3260 & 42.99 & 260 & 0 & 3260 & 42.32 & 43 & -9.0 & 356 & 42.62 & 88 & --- & --- & ---\\
&\it BC\/& -690 & 10900 & 42.75 & 150 & -690 & 10900 & 42.52 & 68 & -21 & 1530 & 41.94 & 18 & --- & --- & ---\\
&\it I+B\/&--- & 3610 & 43.19 & 420 & --- & 4240 & 42.74 & 110 & --- & 365 & 42.71 & 110 & --- & --- & ---\\
\hline
29&\it NC\/&--- & 497 & 42.03 & 34 &--- & 494 & 41.40 & 5.6 & --- & --- & --- & --- & 41.8 & 41.5 & 41.0\\
(1)$^{f}$&\it IC\/ & 12 & 2810 & 42.87 & 240 & 12 & 2810 & 42.30 & 44 & 39 & 385 & 41.69 & 11 & --- & --- & ---\\
&\it BC\/& 810 & 9770 & 42.71 & 170 & 810 & 9770 & 42.46 & 64 & -160 & 789 & 41.75 & 13 & --- & --- & ---\\
&\it I+B\/&--- & 3170 & 43.10 & 400 & --- & 3560 & 42.69 & 110 & --- & 498 & 42.02 & 24 & --- & --- & ---\\
\hline
30&\it NC\/&--- & 386 & 41.31 & 39 &--- & 388 & 40.66 & 5.1 & --- & --- & --- & --- & 41.1 & 39.8 & 40.3\\
(1)$^{f}$&\it IC\/ & 110 & 808 & 41.43 & 52 & 110 & 808 & 41.09 & 14 & -19 & 336 & 40.97 & 11 & --- & --- & ---\\
&\it BC\/& 110 & 3040 & 41.41 & 49 & 110 & 3040 & 41.10 & 14 & -210 & 703 & 40.74 & 6.5 & --- & --- & ---\\
&\it I+B\/&--- & 943 & 41.72 & 100 & --- & 953 & 41.39 & 28 & --- & 389 & 41.17 & 17 & --- & --- & ---\\
\hline
31&\it NC\/&--- & 382 & 41.61 & 8.0 &--- & 382 & 40.54 & 0.51 & --- & --- & --- & --- & 41.2 & --- & ---\\
(3)$^{f}$&\it IC\/ & 610 & 5730 & 43.14 & 270 & 610 & 5730 & 42.59 & 57 & 20 & 333 & 41.29 & 3.0 & --- & --- & ---\\
&\it BC\/& 850 & 21500 & 42.75 & 110 & 850 & 21500 & 42.62 & 61 & -460 & 814 & 41.17 & 2.3 & --- & --- & ---\\
&\it I+B\/&--- & 6140 & 43.29 & 380 & --- & 6810 & 42.90 & 120 & --- & 377 & 41.54 & 5.2 & --- & --- & ---\\
\hline
32&\it NC\/&--- & 425 & 41.24 & 23 &--- & 424 & 40.61 & 5.3 & --- & --- & --- & --- & 41.1 & 41.0 & 40.3\\
(1)$^{f}$&\it IC\/ & 140 & 2550 & 42.17 & 200 & 140 & 2550 & 41.48 & 39 & -58 & 388 & 41.51 & 42 & --- & --- & ---\\
&\it BC\/& 1000 & 8070 & 41.68 & 63 & 1000 & 8070 & 41.52 & 43 & -130 & 1170 & 41.09 & 16 & --- & --- & ---\\
&\it I+B\/&--- & 2710 & 42.30 & 260 & --- & 3100 & 41.80 & 82 & --- & 419 & 41.65 & 58 & --- & --- & ---\\
\hline
33&\it NC\/&--- & 423 & 41.68 & 16 &--- & 426 & 41.03 & 2.7 & --- & --- & --- & --- & 41.6 & --- & ---\\
(1)$^{f}$&\it IC\/ & 17 & 4700 & 42.92 & 290 & 17 & 4700 & 42.37 & 59 & 100 & 412 & 41.99 & 25 & --- & --- & ---\\
&\it BC\/& 23 & 16700 & 42.66 & 160 & 23 & 16700 & 42.44 & 68 & -610 & 1370 & 41.42 & 6.8 & --- & --- & ---\\
&\it I+B\/&--- & 5170 & 43.11 & 450 & --- & 5690 & 42.71 & 130 & --- & 425 & 42.09 & 32 & --- & --- & ---\\
\hline
34&\it NC\/&--- & 838 & 42.57 & 44 &--- & 836 & 41.70 & 4.0 & --- & --- & --- & --- & 41.8 & --- & ---\\
(3)$^{f}$&\it IC\/ & 260 & 2470 & 43.06 & 130 & 260 & 2470 & 42.52 & 26 & -150 & 1440 & 42.48 & 25 & --- & --- & ---\\
&\it BC\/& 1200 & 6570 & 42.97 & 110 & 1200 & 6570 & 42.72 & 41 & -240 & 634 & 42.30 & 16 & --- & --- & ---\\
&\it I+B\/&--- & 2910 & 43.32 & 250 & --- & 3310 & 42.93 & 67 & --- & 829 & 42.70 & 41 & --- & --- & ---\\
\hline
35&\it NC\/&--- & 572 & 42.67 & 130 &--- & 576 & 42.01 & 22 & --- & --- & --- & --- & 42.0 & 41.8 & 42.1\\
(1)$^{f}$&\it IC\/ & 270 & 2380 & 43.31 & 560 & 270 & 2380 & 42.65 & 96 & -120 & 1250 & 42.28 & 43 & --- & --- & ---\\
&\it BC\/& 1300 & 5880 & 43.01 & 280 & 1300 & 5880 & 42.54 & 76 & -140 & 512 & 42.55 & 79 & --- & --- & ---\\
&\it I+B\/&--- & 2640 & 43.49 & 840 & --- & 2790 & 42.90 & 170 & --- & 571 & 42.74 & 120 & --- & --- & ---\\
\hline\hline
   \end{tabular}
 \end{minipage}
\end{table*}

\addtocounter{table}{-1}
\begin{table*}
 \centering
  \begin{minipage}{175mm}
   \caption{continued}
     \begin{tabular}{@{}ccccccccccccccccc@{}}
\hline\hline
   ID & &\multicolumn{4}{c}{H$\alpha$}& \multicolumn{4}{c}{H$\beta$}& \multicolumn{4}{c}{[OIII] 5007} &HeII&FeVII&FeX\\
  & & {\it vel\/} & {\it fwhm\/} & {\it lum\/} & {\it ew\/} & {\it vel\/} & {\it fwhm\/} & {\it lum\/} & {\it ew\/} & {\it vel\/} & {\it fwhm\/} & {\it lum\/} & {\it ew\/} & {\it lum\/} & {\it lum\/} & {\it lum\/} \\
\hline\hline
36$^{*}$&\it NC\/&--- & 274 & 41.15 & 36 &--- & 271 & 40.32 & 5.1 & 54 & 175 & 40.54 & 8.6 & 41.1 & 40.2 & 39.6\\
(1)$^{f}$&\it IC\/ & 210 & 1540 & 41.29 & 49 & 210 & 1540 & 40.69 & 12 & -62 & 448 & 40.62 & 10. & --- & --- & ---\\
&\it BC\/& -160 & 4450 & 40.98 & 24 & -160 & 4450 & 40.75 & 14 & -290 & 1150 & 40.42 & 6.5 & --- & --- & ---\\
&\it I+B\/&--- & 1690 & 41.47 & 73 & --- & 1890 & 41.02 & 26 & --- & 248 & 41.01 & 25 & --- & --- & ---\\
\hline
37&\it NC\/&--- & 567 & 41.92 & 13 &--- & 566 & 41.17 & 2.2 & --- & --- & --- & --- & 42.5 & 41.7 & 41.0\\
(1)$^{f}$&\it IC\/ & 220 & 3440 & 43.17 & 230 & 220 & 3440 & 42.68 & 72 & 72 & 294 & 41.37 & 3.7 & --- & --- & ---\\
&\it BC\/& 870 & 10000 & 42.88 & 120 & 870 & 10000 & 42.55 & 53 & -170 & 824 & 41.79 & 9.8 & --- & --- & ---\\
&\it I+B\/&--- & 3790 & 43.35 & 350 & --- & 3960 & 42.92 & 130 & --- & 558 & 41.93 & 13 & --- & --- & ---\\
\hline
38$^{*}$&\it NC\/&--- & 326 & 41.61 & 140 &--- & 325 & 41.05 & 38 & 38 & 635 & 41.22 & 56 & 41.1 & --- & ---\\
(1)$^{f}$&\it IC\/ & -100 & 4830 & 42.23 & 580 & -100 & 4830 & 41.25 & 60 & 60 & 286 & 41.51 & 110 & --- & --- & ---\\
&\it BC\/& -390 & 12200 & 41.83 & 230 & -390 & 12200 & 41.51 & 110 & -140 & 1280 & 40.77 & 20 & --- & --- & ---\\
&\it I+B\/&--- & 5250 & 42.38 & 800 & --- & 6630 & 41.70 & 170 & --- & 328 & 41.74 & 190 & --- & --- & ---\\
\hline
39&\it NC\/&--- & 212 & 41.00 & 32 &--- & 217 & 40.44 & 6.9 & --- & --- & --- & --- & 40.8 & 40.0 & 40.1\\
(1)$^{f}$&\it IC\/ & 29 & 829 & 41.47 & 93 & 29 & 829 & 40.87 & 19 & 33 & 193 & 41.06 & 29 & --- & --- & ---\\
&\it BC\/& 50 & 3000 & 41.25 & 57 & 50 & 3000 & 40.90 & 20 & -20 & 427 & 40.73 & 14 & --- & --- & ---\\
&\it I+B\/&--- & 925 & 41.67 & 150 & --- & 990 & 41.19 & 38 & --- & 216 & 41.22 & 43 & --- & --- & ---\\
\hline
40$^{*}$&\it NC\/&--- & 446 & 41.34 & 8.0 &--- & 451 & 40.45 & 0.88 & -130 & 465 & 40.73 & 1.7 & --- & --- & 41.0\\
(3)$^{f}$&\it IC\/ & 210 & 1930 & 42.64 & 160 & 210 & 1930 & 41.84 & 22 & 130 & 123 & 40.59 & 1.2 & --- & --- & ---\\
&\it BC\/& 76 & 6350 & 42.47 & 110 & 76 & 6350 & 42.21 & 51 & -160 & 1590 & 41.27 & 6.0 & --- & --- & ---\\
&\it I+B\/&--- & 2180 & 42.86 & 270 & --- & 2790 & 42.37 & 73 & --- & 450 & 41.45 & 8.9 & --- & --- & ---\\
\hline
41&\it NC\/&--- & 400 & 42.10 & 23 &--- & 401 & 41.47 & 3.5 & --- & --- & --- & --- & 42.3 & --- & ---\\
(2)$^{f}$&\it IC\/ & 170 & 2300 & 43.38 & 440 & 170 & 2300 & 42.85 & 84 & -52 & 401 & 42.32 & 25 & --- & --- & ---\\
&\it BC\/& 890 & 8300 & 43.09 & 220 & 890 & 8300 & 42.73 & 62 & -260 & 908 & 42.25 & 22 & --- & --- & ---\\
&\it I+B\/&--- & 2510 & 43.56 & 660 & --- & 2610 & 43.10 & 150 & --- & 486 & 42.59 & 47 & --- & --- & ---\\
\hline
42&\it NC\/&--- & 307 & 40.74 & 14 &--- & 306 & 39.90 & 1.6 & --- & --- & --- & --- & 41.1 & 40.6 & 39.9\\
(1)$^{f}$&\it IC\/ & 120 & 3760 & 41.89 & 190 & 120 & 3760 & 41.27 & 38 & 26 & 721 & 40.43 & 5.6 & --- & --- & ---\\
&\it BC\/& -300 & 9930 & 41.66 & 110 & -300 & 9930 & 41.44 & 56 & -3.0 & 287 & 41.02 & 22 & --- & --- & ---\\
&\it I+B\/&--- & 4240 & 42.09 & 300 & --- & 4920 & 41.67 & 94 & --- & 303 & 41.12 & 27 & --- & --- & ---\\
\hline
43&\it NC\/&--- & 414 & 42.10 & 29 &--- & 419 & 41.60 & 6.5 & --- & --- & --- & --- & 42.0 & --- & 41.0\\
(1)$^{f}$&\it IC\/ & 12 & 3550 & 43.23 & 380 & 12 & 3550 & 42.51 & 53 & 79 & 324 & 42.07 & 20 & --- & --- & ---\\
&\it BC\/& 1400 & 10900 & 42.95 & 200 & 1400 & 10900 & 42.67 & 75 & -310 & 993 & 42.22 & 28 & --- & --- & ---\\
&\it I+B\/&--- & 3920 & 43.41 & 590 & --- & 4550 & 42.90 & 130 & --- & 419 & 42.45 & 47 & --- & --- & ---\\
\hline
44&\it NC\/&--- & 265 & 41.86 & 26 &--- & 262 & 41.34 & 5.1 & --- & --- & --- & --- & 41.8 & 40.3 & 40.8\\
(1)$^{f}$&\it IC\/ & 73 & 852 & 42.41 & 91 & 73 & 852 & 41.80 & 15 & 110 & 220 & 41.67 & 11 & --- & --- & ---\\
&\it BC\/& 97 & 4190 & 42.32 & 74 & 97 & 4190 & 42.04 & 26 & -98 & 852 & 41.72 & 13 & --- & --- & ---\\
&\it I+B\/&--- & 952 & 42.67 & 170 & --- & 1070 & 42.24 & 40 & --- & 260 & 41.99 & 24 & --- & --- & ---\\
\hline
45&\it NC\/&--- & 345 & 42.82 & 33 &--- & 344 & 42.22 & 5.2 & --- & --- & --- & --- & --- & --- & ---\\
(1)$^{f}$&\it IC\/ & -870 & 6210 & 43.81 & 330 & -870 & 6210 & 43.01 & 32 & 15 & 267 & 42.79 & 20 & --- & --- & ---\\
&\it BC\/& 1400 & 17200 & 43.89 & 390 & 1400 & 17200 & 43.56 & 110 & -62 & 880 & 42.93 & 28 & --- & --- & ---\\
&\it I+B\/&--- & 7740 & 44.16 & 720 & --- & 10900 & 43.66 & 140 & --- & 346 & 43.17 & 48 & --- & --- & ---\\
\hline
46$^{*}$&\it NC\/&--- & 514 & 42.22 & 51 &--- & 516 & 41.66 & 13 & -11 & 432 & 42.32 & 63 & 41.4 & 41.9 & ---\\
(3)$^{f}$&\it IC\/ & 3000 & 6960 & 42.54 & 110 & 3000 & 6960 & 41.99 & 29 & 150 & 223 & 41.66 & 14 & --- & --- & ---\\
&\it BC\/& -1400 & 8410 & 42.93 & 260 & -1400 & 8410 & 42.09 & 36 & -150 & 1010 & 42.30 & 60 & --- & --- & ---\\
&\it I+B\/&--- & 9970 & 43.08 & 370 & --- & 9930 & 42.34 & 64 & --- & 534 & 42.66 & 140 & --- & --- & ---\\
\hline
47&\it NC\/&--- & 391 & 41.58 & 27 &--- & 391 & 40.93 & 5.7 & --- & --- & --- & --- & 41.7 & --- & ---\\
(1)$^{f}$&\it IC\/ & 160 & 3550 & 42.70 & 360 & 160 & 3550 & 41.94 & 58 & 50 & 1080 & 41.19 & 11 & --- & --- & ---\\
&\it BC\/& 670 & 7400 & 42.02 & 74 & 670 & 7400 & 41.78 & 40 & -24 & 355 & 41.70 & 34 & --- & --- & ---\\
&\it I+B\/&--- & 3730 & 42.78 & 430 & --- & 4100 & 42.17 & 98 & --- & 383 & 41.82 & 45 & --- & --- & ---\\
\hline
48&\it NC\/&--- & 302 & 41.27 & 51 &--- & 306 & 40.67 & 8.7 & --- & --- & --- & --- & 40.4 & --- & 40.2\\
(2)$^{f}$&\it IC\/ & 110 & 1040 & 41.53 & 92 & 110 & 1040 & 41.03 & 20 & 28 & 306 & 40.37 & 4.6 & --- & --- & ---\\
&\it BC\/& 33 & 4300 & 41.38 & 66 & 33 & 4300 & 41.00 & 19 & -130 & 719 & 40.30 & 3.9 & --- & --- & ---\\
&\it I+B\/&--- & 1160 & 41.77 & 160 & --- & 1190 & 41.32 & 39 & --- & 365 & 40.64 & 8.4 & --- & --- & ---\\
\hline\hline
   \end{tabular}
 \end{minipage}
\end{table*}

\clearpage

\addtocounter{table}{-1}
\begin{table*}
 \centering
  \begin{minipage}{175mm}
   \caption{continued}
     \begin{tabular}{@{}ccccccccccccccccc@{}}
\hline\hline
   ID & &\multicolumn{4}{c}{H$\alpha$}& \multicolumn{4}{c}{H$\beta$}& \multicolumn{4}{c}{[OIII] 5007} &HeII&FeVII&FeX\\
  & & {\it vel\/} & {\it fwhm\/} & {\it lum\/} & {\it ew\/} & {\it vel\/} & {\it fwhm\/} & {\it lum\/} & {\it ew\/} & {\it vel\/} & {\it fwhm\/} & {\it lum\/} & {\it ew\/} & {\it lum\/} & {\it lum\/} & {\it lum\/} \\
\hline\hline
49&\it NC\/&--- & 284 & 41.27 & 19 &--- & 281 & 40.65 & 3.8 & --- & --- & --- & --- & 41.5 & --- & 40.5\\
(1)$^{f}$&\it IC\/ & 16 & 1010 & 41.65 & 45 & 16 & 1010 & 41.05 & 9.4 & 44 & 268 & 40.98 & 8.1 & --- & --- & ---\\
&\it BC\/& 160 & 4040 & 41.49 & 31 & 160 & 4040 & 41.33 & 18 & -260 & 578 & 40.37 & 2.0 & --- & --- & ---\\
&\it I+B\/&--- & 1120 & 41.88 & 76 & --- & 1340 & 41.51 & 27 & --- & 285 & 41.07 & 10. & --- & --- & ---\\
\hline
50$^{*}$&\it NC\/&--- & 274 & 42.84 & 36 &--- & 275 & 42.23 & 4.5 & -73 & 402 & 41.87 & 2.1 & 42.3 & 42.2 & 41.9\\
(1)$^{f}$&\it IC\/ & 120 & 1880 & 43.60 & 210 & 120 & 1880 & 43.06 & 30 & 130 & 224 & 42.34 & 6.2 & --- & --- & ---\\
&\it BC\/& 91 & 8960 & 43.50 & 170 & 91 & 8960 & 43.12 & 35 & -470 & 1770 & 42.42 & 7.4 & --- & --- & ---\\
&\it I+B\/&--- & 2090 & 43.86 & 370 & --- & 2200 & 43.39 & 65 & --- & 266 & 42.74 & 16 & --- & --- & ---\\
\hline
51$^{*}$&\it NC\/&--- & 451 & 41.73 & 65 &--- & 445 & 41.20 & 16 & 11 & 364 & 42.03 & 110 & --- & 41.2 & ---\\
(1)$^{f}$&\it IC\/ & -1300 & 6490 & 42.44 & 330 & -1300 & 6490 & 41.59 & 39 & 130 & 1150 & 41.57 & 39 & --- & --- & ---\\
&\it BC\/& 1400 & 17200 & 42.53 & 410 & 1400 & 17200 & 42.10 & 130 & 340 & 247 & 41.23 & 18 & --- & --- & ---\\
&\it I+B\/&--- & 8170 & 42.79 & 750 & --- & 11100 & 42.21 & 170 & --- & 450 & 42.21 & 170 & --- & --- & ---\\
\hline\hline
   \end{tabular}
   \\
$^{f}$ : The final fitting method used for each object. see subsection~\ref{emission line fitting} for detailed 
description of each fitting methods.\\
$^{*}$ : Three gaussian profiles are used for these objects, in this case (I+B) means the total of all three components.\\
$^{d}$ : UM 269, the only object in our sample showing double-peak feature in Balmer lines. Two gaussian profiles are 
used for fitting the two peaks, thus the velocity shift of each component related to the line centre is huge. \\
 \end{minipage}
\end{table*}

\end{document}